\documentclass[twocolumn,floatfix,prb,aps,showpacs]{revtex4}
\usepackage{graphicx,amsmath,amssymb,color}
\usepackage{nicefrac,subfigure}
\usepackage[titletoc,title]{appendix}

\newcommand{\be}{\begin{equation}}
\newcommand{\ee}{\end{equation}}

\newcommand{\ba}{\begin{eqnarray}}
\newcommand{\ea}{\end{eqnarray}}

\begin{document}

\title{Interacting composite fermions: Nature of the 4/5, 5/7, 6/7, and 6/17 fractional quantum Hall states}
\author{Ajit C. Balram$^{1,2}$}
\affiliation{$^{1}$Department of Physics, 104 Davey Lab, Pennsylvania State University, University Park, Pennsylvania 16802, USA}
\affiliation{$^{2}$Niels Bohr International Academy and the Center for Quantum Devices, Niels Bohr Institute, University of Copenhagen, 2100 Copenhagen, Denmark}

\begin{abstract} 
Numerical studies by W\'ojs, Yi and Quinn have suggested that an unconventional fractional quantum Hall effect is plausible at filling factors $\nu=$ 1/3 and 1/5, provided the interparticle interaction has an unusual form for which the energy of two fermions in the relative angular momentum three channel dominates. The interaction between composite fermions in the second $\Lambda$ level (composite fermion analog of the electronic Landau level) satisfies this property, and recent studies have supported unconventional fractional quantum Hall effect of composite fermions at $\nu^*=$ 4/3 and 5/3, which manifests as fractional quantum Hall effect of electrons at $\nu=$ 4/11, 4/13, 5/13, and 5/17. I investigate in this article the nature of the fractional quantum Hall states at $\nu=$ 4/5, 5/7, 6/17, and 6/7, which correspond to composite fermions at $\nu^*=$ 4/3, 5/3 and 6/5, and find that all these fractional quantum Hall states are conventional. The underlying reason is that the interaction between composite fermions depends substantially on both the number and the direction of the vortices attached to the electrons. I also study in detail the states with different spin polarizations at 6/17 and 6/7 and predict the critical Zeeman energies for the spin phase transitions between them. I calculate the excitation gaps at 4/5, 5/7, 6/7, and 6/17 and compare them against recent experiments.
\end{abstract}
\maketitle

\section{Introduction}
The fractional quantum Hall effect (FQHE) \cite{Tsui82} is the most celebrated example of a state of matter possessing emergent topological order. It arises due to the formation of composite fermions \cite{Jain89,Jain07} (CFs), where a composite fermion is the bound state of an electron and an even number ($2p$) of quantized vortices. CFs experience a reduced magnetic field compared to the external magnetic field and form their own Landau-like levels, called $\Lambda$ levels ($\Lambda$Ls), in the reduced field. The filling factor of composite fermions $\nu^{*}$ is related to the electron filling factor as $\nu=\nu^{*}/(2p\nu^{*}\pm1)$. Various mechanisms have been invoked to explain the various observed fractional quantum Hall states (FQHSs). The $\nu^*=n$ integer quantum Hall effect (IQHE) of composite fermions produces FQHE at the Jain filling fractions $\nu=n/(2pn\pm 1)$, which explains a vast majority of the FQHSs in the lowest Landau level (LLL). The CF theory also successfully predicts at these fractions states with various spin polarizations \cite{Wu93,Park98,Park98b,Park01,Balram15a}, as observed in experiments \cite{Eisenstein89,Engel92,Du95,Kang97,Yeh99,Kukushkin99,Kukushkin00,Melinte00,Freytag01,Tiemann12,Feldman13,Liu14}. A second mechanism is $p$-wave pairing of composite fermions \cite{Moore91,Read00}, which is thought to be responsible for the FQHS at $\nu=5/2$ \cite{Willett87} and the closely related $\nu=7/2$. Pairing of composite fermions may also underlie FQHE at $\nu=3/8$ and $3/10$ \cite{Mukherjee12,Mukherjee14c,Mukherjee15b} (experimental evidence for which exists \cite{Pan03,Bellani10,Pan15,Samkharadze15b} but is not yet conclusive).\\

\begin{figure}
\begin{center}
\includegraphics[width=0.48\textwidth]{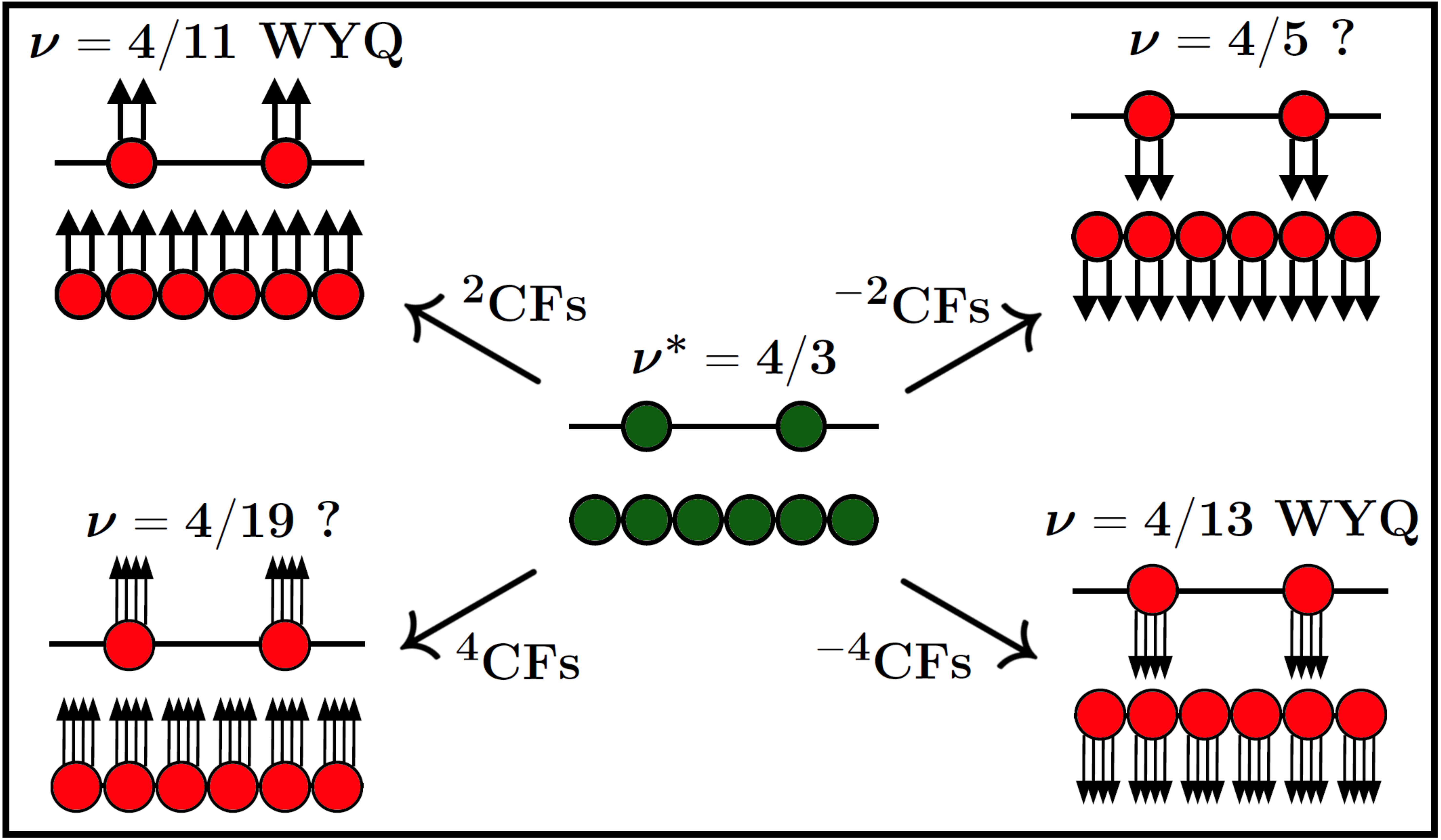}
\includegraphics[width=0.48\textwidth]{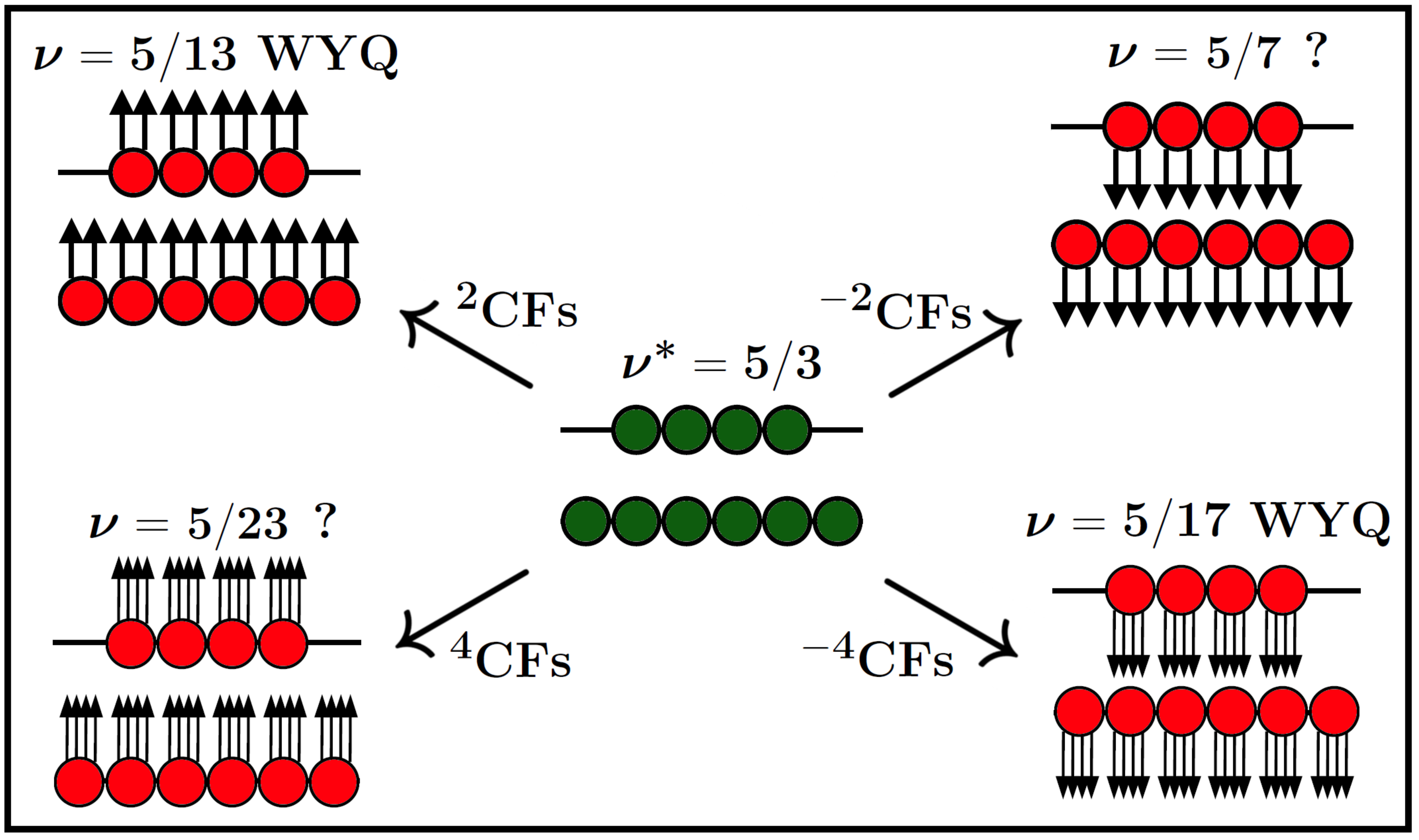}
\includegraphics[width=0.48\textwidth]{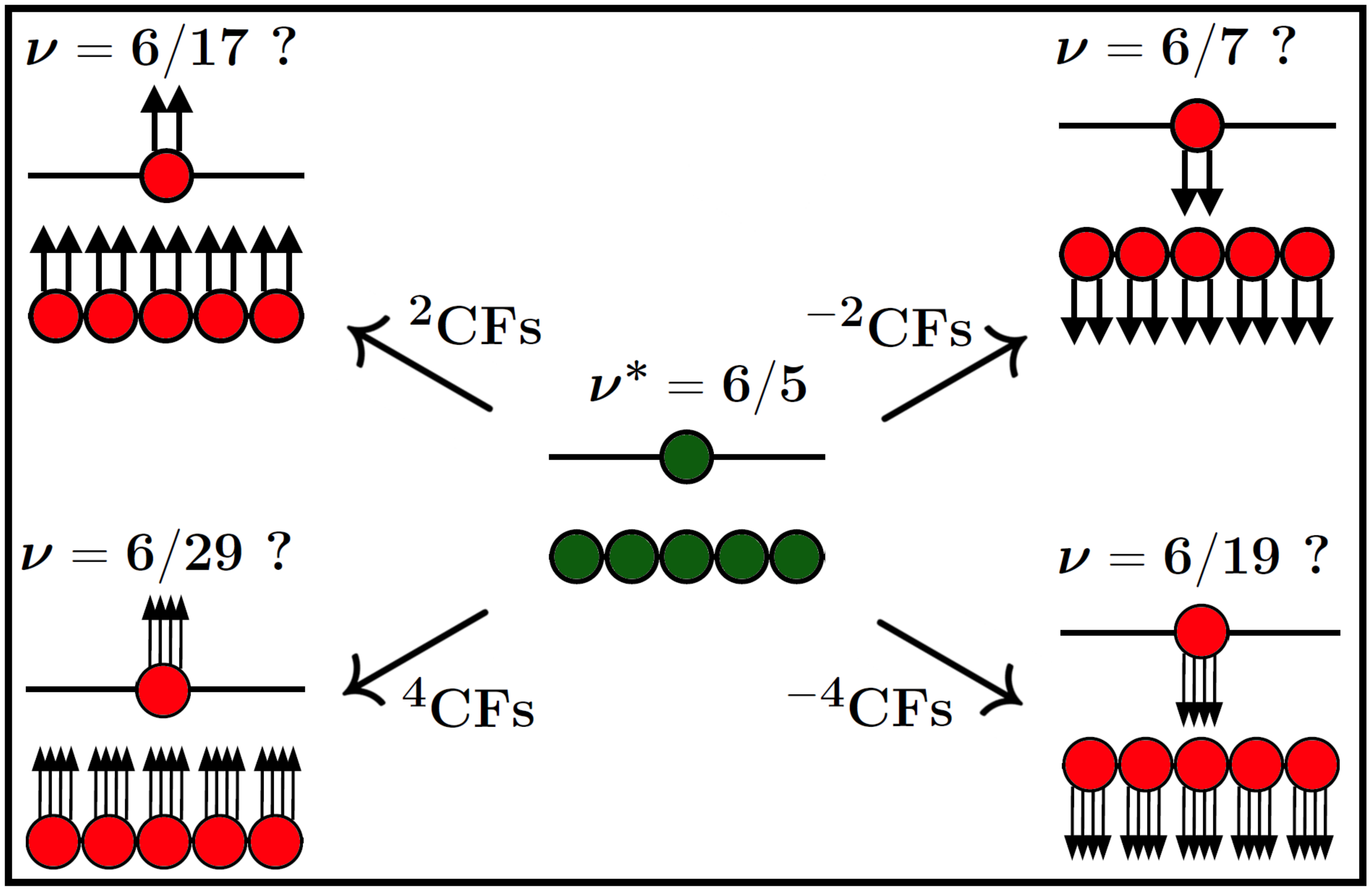}
\caption{(color online) Schematic of the various fractional quantum Hall states obtained from a filling $\nu^{*}=4/3$ (top panel), $\nu^{*}=5/3$ (middle panel), and $\nu^{*}=6/5$ (bottom panel) of composite fermions. The green solid dots represent electrons while the red solid dots with $2p$ arrows are composite fermions carrying $2p$ vortices in the same (arrows pointing up) or opposite (arrows pointing down) direction to the external magnetic field.}
\label{fig:schematic}
\end{center}
\end{figure}

These two mechanisms, namely the IQHE and pairing of composite fermions, do not explain all of the observed FQHSs, such as those at $\nu=~4/11,~4/13,~5/13,~5/17,~6/17,~4/5,~5/7$ and $6/7$. For some of these states FQHE has been definitively established in experiments while for others signatures have been observed \cite{Pan03,Pan15,Samkharadze15b,Yeh99,Liu14a}. Given detailed tests of the CF theory for filling factors in this range, there is little doubt that these are FQHSs of interacting composite fermions, with incompressibility arising from the weak residual interaction between CFs. Building on the work of W\'ojs \emph{et al}. \cite{Wojs00,Wojs04}, Mukherjee \emph{et al}. \cite{Mukherjee14} made a strong theoretical case that the mechanism of FQHE at the fully spin polarized $\nu=4/11$ and $\nu=5/13$ is \emph{unconventional}. These states arise from a filling $\nu^{*}=4/3=1+1/3$ and $\nu^{*}=5/3=1+2/3$ of composite fermions carrying two vortices in the same direction as the external magnetic field (parallel vortex attachment) respectively (see Fig. \ref{fig:schematic}), but the $1/3$ and $2/3$ states in the second $\Lambda$L of composite fermions is of the W\'ojs-Yi-Quinn (WYQ) \cite{Wojs04} type, which is topologically distinct from the conventional Laughlin state \cite{Laughlin83}. Subsequently Mukherjee and Mandal \cite{Mukherjee15b} suggested that the FQHE at the fully spin polarized $\nu=4/13$ and $\nu=5/17$, which arises from $\nu^{*}=4/3$ and $\nu^{*}=5/3$ of composite fermions carrying four vortices in the opposite direction as the external magnetic field (reverse vortex attachment) respectively (see Fig. \ref{fig:schematic}), could also owe their origin to the WYQ type $1/3$ and $2/3$ states, respectively. Following these works Ref. \cite{Balram15} provided a detailed spin phase diagram for the fractional quantum Hall states (FQHSs) of composite fermions which agrees reasonably well with the experimental observations of the spin transitions at these filling factors \cite{Yeh99,Liu14a}.\\

The aim of this work is to ask if the WYQ mechanism could be responsible for other FQHSs. The idea is illustrated in Fig.~\ref{fig:schematic}. Some of the fractions where composite fermions can possibly exhibit unconventional WYQ FQHSs are $\nu^*=$ 4/3, 5/3, and 6/5. These produce FQHE of electrons at several filling factors, depending on the number and the direction of the vortices attached to the electrons, as shown in Fig.~\ref{fig:schematic}. Previous works mentioned above have suggested that some of these belong to the WYQ class. I will consider below some of the others. \\

Whether CFs form the unconventional WYQ state depends on their interaction pseudopotentials \cite{Sitko96,Lee01,Lee02} $V_m^{\rm CF}$, which are the energies of two composite fermions in relative angular momentum $m$ states. Composite fermions form a conventional state if the interaction pseudopotential $V_1^{\rm CF}$ is dominant, but an unconventional state if $V_3^{\rm CF}$ is dominant. (Only odd $m$ is relevant for fully spin polarized composite fermions.) The CF pseudopotentials derive from the Coulomb interaction between electrons, and depend on the $\Lambda$L index, the value of $2p$, and the direction of the attached vortices. In this work I evaluate the CF pseudopotentials for spinful composite fermions in various $\Lambda$Ls in detail. Following this I consider the fully spin polarized states at 4/5, 5/7, 6/17, and 6/7 and show that these are all conventional. I have also considered various non-fully spin polarized states at 6/17 and 6/7 and present evidence in favor of them being conventional. Evidence for FQHE at 6/17 was reported by Pan \emph{et al.} \cite{Pan03} in 2003, but it has not yet been definitively confirmed \cite{Samkharadze15b,Pan15}. \\

The plan of the paper is as follows: In Sec. \ref{sec:sphere_CF} I list some facts on the spherical geometry, which is used for all calculations presented in this paper, and  of the CF theory. Following this, in Sec. \ref{sec:int_CFs} I look at the interaction between two composite fermions on top of the filled lowest $\Lambda$L. By calculating these interaction energies one can ascertain the plausibility of unconventional order for FQHSs with various spin polarizations. These results suggest that unconventional WYQ states are only possible for fully spin polarized composite fermions while for partially spin polarized states conventional order is expected. In Sec. \ref{sec:4_5_and_5_7_fp} I look at two candidate states, namely the fully spin polarized FQHSs at $\nu=4/5$ and $\nu=5/7$ and demonstrate that they support only conventional order. In Sec. \ref{sec:CF_6_5} I look at two other candidate states, namely at $\nu=6/17$ and $\nu=6/7$, and find that for all spin polarizations only conventional order prevails. Finally, in Sec. \ref{sec:conclusions} I conclude the paper with experimentally verifiable predictions for these states and a summary of the results. \\

\section{Background}
\label{sec:sphere_CF}
For all calculations in this paper I use Haldane's spherical geometry \cite{Haldane83,Greiter11} in which $N$ electrons reside on the surface of a sphere in the presence of a radial magnetic field of flux $2Q\phi_{0}$ ($\phi_{0}=hc/e$ is a quantum of flux) generated by a monopole sitting at the center of the sphere ($2Q$ according to Dirac's quantization condition is an integer). Appropriate to this geometry the total orbital angular momentum $L$ and its $z$-component $L_{z}$ are good quantum numbers. FQHSs are uniform incompressible states and hence have $L=0$. These states occur at $2Q=\nu^{-1}N-\mathcal{S}$, where $\mathcal{S}$ is a rational number called the shift \cite{Wen92}. Topologically distinct candidate wave functions often occur at different values of the shift. A note is in order regarding the usage of the spherical geometry. A major advantage of using the spherical geometry is that it is devoid of edges, which makes it appropriate for an investigation of the bulk properties of a state. In contrast to the planar geometry (without an external confining potential), the degeneracy of a LL is finite on the sphere, which makes it useful in identifying incompressible ground states for finite systems. The spherical geometry has been quite instrumental in establishing the validity of the theory of the FQHE, not just qualitatively, but also quantitatively \cite{Jain07}. Wrapping the plane on the spherical surface is just a choice of the boundary conditions, which should not affect the bulk properties of a state. Periodic boundary conditions, which are equivalent to a toroidal geometry (which too does not have any edges and is thus useful to investigate bulk properties of a state), has also been considered in the literature \cite{Haldane85,Haldane85b}; but to date it has not yet been possible to evaluate composite fermion wave functions in a simple manner on the torus (although attempts have been made \cite{Hermanns13,Fremling13,Fremling14}). \\

A composite fermion is defined as a bound state of an electron and $2p$ quantized vortices. Due to the binding of vortices, CFs see a reduced flux given by $2Q^{*}=2Q-2p(N-1)$, which could be either positive or negative. Composite fermions carrying $2p$ vortices and experiencing an effective magnetic field in the same (opposite) direction as the external magnetic field, are said to arise from parallel (reverse) vortex attachment and denoted as $^{2p}$CFs ($^{-2p}$CFs). The basic postulate of CF theory is that in this reduced field, to the first approximation, CFs can be considered weakly interacting and hence form their own Landau-like levels which are termed $\Lambda$ levels ($\Lambda$Ls). The FQHS at $\nu=n/(2pn\pm 1)$ maps into an IQHE of CFs with $n$ filled $\Lambda$Ls. For composite fermions carrying a spin degree of freedom, we can write the CF filling as $n=n_{\uparrow}+n_{\downarrow}$, where $n_{\uparrow}$ and $n_{\downarrow}$, respectively, denote the number of filled spin up and spin down $\Lambda$Ls. At a given value of $n$ states with various spin polarizations [defined by $\gamma=(n_{\uparrow}-n_{\downarrow})/(n_{\uparrow}+n_{\downarrow})$] can occur, each of which has a different value of $\mathcal{S}$, which the CF theory identifies. The Jain wave functions for the ground state at $n/(2pn\pm 1)$ are:
\begin{equation}
\Psi_{n/(2pn+1)}=\mathcal{P}_{\rm LLL}\Phi_{n}J^{2p}=\mathcal{P}_{\rm LLL}\Phi_{n_{\uparrow}}\Phi_{n_{\downarrow}}J^{2p}, 
\label{Jainwf_paral}
\end{equation}
\begin{equation}
\Psi_{n/(2pn-1)}=\mathcal{P}_{\rm LLL}[\Phi_{n}]^{*}J^{2p}=\mathcal{P}_{\rm LLL}[\Phi_{n_{\uparrow}}\Phi_{n_{\downarrow}}]^{*}J^{2p},
\label{Jainwf_rev}
\end{equation}
where the Jastrow factor $J$ is given by:
\begin{equation}
 J=\prod_{1\leq i<j \leq N}(u_{i}v_{j}-u_{j}v_{i}).
\end{equation}
Here $u=\cos(\theta/2)e^{i\phi/2}$ and $v=\sin(\theta/2)e^{-i\phi/2}$ are the spinor coordinates, and $\theta$ and $\phi$ are the polar and azimuthal angles on the sphere, respectively. $\Phi_{n}$ is the Slater determinant wave function of $n$ filled LLs of electrons and $\mathcal{P}_{\rm LLL}$ denotes LLL projection which in this work is carried out using the Jain-Kamilla method, details of which can be found in the literature \cite{Jain97,Jain97b,Jain07,Moller05,Davenport12,Balram15a,Mukherjee15b}. For spinful CFs, besides $L$, we also use the total spin angular momentum quantum number $S$, which equals half the difference of number of up and down spins, to characterize the states (without loss of generality I will assume that the number of up spins is greater than or equal to the number of down spins.). In terms of $S$, the spin polarization of a state $\gamma=\lim_{N\rightarrow \infty}2S/N$. In this work I shall only be concerned with states occurring in the following three $\Lambda$Ls: lowest $\Lambda$L spin up (L$\Lambda$L $\uparrow$), lowest $\Lambda$L spin down (L$\Lambda$L $\downarrow$), and second $\Lambda$L spin up (S$\Lambda$L $\uparrow$).\\

To describe the FQHSs of composite fermions I use the notation of Ref. \cite{Balram15} to denote the rather complex wave functions. In this notation, the symbol $[u,d]_{\pm 2p}$ denotes a state where $u$ and $d$ are states in the spin up and spin down sectors, which can themselves be integer or fractional quantum Hall states of composite fermions. Furthermore, the symbols $[~]_{2p}$ and $[~]_{-2p}$ denote composite fermionization with parallel and reverse vortex attachment, respectively. 
To give an example, the Jain states of Eqs.~(\ref{Jainwf_paral}) and (\ref{Jainwf_rev}) are denoted as $[n_{\uparrow},n_{\downarrow}]_{2p}$ and $[n_{\uparrow},n_{\downarrow}]_{-2p}$ respectively. I will also denote the state $[u,d]_{\pm 2p}$ by $(\nu_{\uparrow},\nu_{\downarrow})$, where $\nu_{\uparrow}$ ($\nu_{\downarrow}$) is the filling factor in the spin up (spin down) sector, which gives the spin polarization $\gamma=(\nu_{\uparrow}-\nu_{\downarrow})/(\nu_{\uparrow}+\nu_{\downarrow})$.\\

The state $[n+[k]_{\pm 2r},[m]_{\pm 2q}]_{\pm 2p}$ has a CF filling factor of:
\begin{equation}
 \nu^{*}=\nu_{\uparrow}^{*}+\nu_{\downarrow}^{*},~\nu_{\uparrow}^{*}=n+\frac{k}{2kr\pm1}, \nu_{\downarrow}^{*}=\frac{m}{2qm\pm1},
\end{equation}
which gives an electron filling factor of:
\begin{equation}
 \nu=\frac{n+\frac{k}{2kr\pm1}+\frac{m}{2qm\pm1}}{2p\Big(n+\frac{k}{2kr\pm1}+\frac{m}{2qm\pm1}\Big)\pm 1}.
\end{equation}
The explicit wave functions for these states are given by:
\begin{eqnarray}
 \Psi_{[n+[k]_{\pm 2r},[m]_{\pm 2q}]_{2p}}&=&\mathcal{P}_{\rm LLL} \Phi_{n+k/(2kr\pm1)} \psi_{[m]_{\pm 2q}}J^{2p}, \\
 \Psi_{[n+[k]_{\pm 2r},[m]_{\pm 2q}]_{-2p}}&=&\mathcal{P}_{\rm LLL} [\Phi_{n+k/(2kr\pm1)} \psi_{[m]_{\pm 2q}}]^{*} J^{2p}, \nonumber
\end{eqnarray}
where $\Phi_{n+k/(2kr\pm1)}$ denotes the state in which the lowest $n$ LLs are fully occupied and the electrons in the $(n+1)$th LL form a $k/(2kr\pm1)$ state in the spin up component (for $k=0$ this is just the Slater determinant of $n$ filled LLs for electrons) and $\psi_{[m]_{\pm 2q}}$ is the Jain wave function of spin down electrons at $m/(2mq\pm1)$. Finally, the hole-conjugate of the state $[\cdots]$ will be denoted by $\overline{[\cdots]}$, which takes the state at $(\nu_{\uparrow},\nu_{\downarrow})$ to a state at $\overline{(\nu_{\uparrow},\nu_{\downarrow})}\equiv (1-\nu_{\uparrow},1-\nu_{\downarrow})$. \\

For systems with small values of $N$ and $2Q$ we can do a brute force exact diagonalization of the Coulomb Hamiltonian in the LLL Hilbert space. In cases where the Hilbert space is beyond reach of exact diagonalization, I calculate ground state energies and gaps using the the accurate approximation method called composite fermion diagonalization (CFD), details of which can be found in the literature \cite{Mandal02,Jain07,Mukherjee14}. Briefly, this method involves diagonalization of the $1/r$ Coulomb Hamiltonian within a small set of CF basis functions using the Metropolis Monte Carlo method as follows: First construct simultaneous eigenstates of the $L^{2}$ and $S^{2}$ operators at filling factor $\nu^{*}$ in the corresponding IQHE system. Then composite-fermionize these states, i.e., multiply by $J^{2p}$ and project the state to the LLL, to obtain a set of basis states at filling factor $\nu$. Finally, diagonalize the Coulomb interaction by evaluating the resulting multi-dimensional integrals using the Monte Carlo method. As a technical point I note that for all states considered in this work, except the 6/17 spin-singlet state, the $L^{2}$ eigenstates designed in CFD satisfy the Fock conditions \cite{Hamermesh62,Jain07,Balram15} and hence are by construction also eigenstates of the $S^{2}$ operator. \\

Besides the ground state energy I shall also calculate gaps for the various incompressible states considered in this work. I consider two kinds of spin-conserving gaps. The neutral gap is defined as the minimum energy gap, i.e., it corresponds to the difference in energy between the ground state and the lowest lying excited state at a given value of $N$, $2Q$, and $S$. In the exciton dispersion this low-lying excited state appears as roton minima and its energy can be measured in inelastic light \cite{Pinczuk93} or phonon \cite{Mellor95} scattering experiments.
The second gap, called the charge gap, is defined, for an incompressible state occurring at flux $2Q$, as: 
 \begin{equation}
 \Delta_{2Q}= E_{2Q+1}+E_{2Q-1}-2E_{2Q}
 \end{equation}
 where $E_{N_{\phi}}$ is the ground state energy of a system of $N$ electrons at flux $N_{\phi}$. The quantity $\Delta_{2Q}/n$ is the energy required to create a quasihole-quasiparticle pair, where $n$ is the number of fundamental quasiholes produced per flux quantum [$n=1$ for the Laughlin states at $\nu=1/(2p+1)$]. Since this involves calculation of the spectra at fluxes either side of the incompressible state (which is time and resource consuming), I have used an alternate definition of the charge gap, namely, the long wave vector limit of the neutral exciton gap (corresponds to a far separated quasihole-quasiparticle pair) at a given $N$, $2Q$, and $S$. This gap appears in transport experiments as the activation energy.\\

I work in the ideal limit of zero thickness and neglect effects of LL mixing and disorder. The ground state energies quoted below include contributions from the background-background and electron-background interactions. The density for a finite system in the spherical geometry depends on the number of electrons $N$ and differs slightly from its thermodynamic value. To reduce the effect of the $N$-dependence we use the density corrected energy \cite{Morf86} $E_{N}'=\sqrt{2Q\nu/N}E_{N}$, which is then extrapolated to the thermodynamic limit $N^{-1}\rightarrow 0$. All the ground state energies quoted in this work are the thermodynamic limits of the density corrected per-particle energies, namely $\lim_{N\rightarrow \infty} E_{N}'/N$. The error bars shown in the following CFD spectra are obtained from the statistical uncertainity of the Monte Carlo sampling. \\

\section{Interaction pseudopotentials of composite fermions}
\label{sec:int_CFs}
Some insight into what FQHE is possible can be gained by looking at the two-particle interaction pseudopotentials $V_{m}$ \cite{Haldane83}, where $m$ is the relative angular momentum of the two particles. If $V_{1}$ is dominant then the 1/3 and 1/5 FQHSs have correlations of the conventional Laughlin kind, while if $V_{3}$ is dominant the 1/3 and 1/5 states are likely to be of the unconventional WYQ type. Of course for electrons in the LLL, $V_{1}$ is dominant, but for CFs other relative angular momentum pseudopotentials may be dominant, as shown in Refs.~ \cite{Sitko96,Wojs00,Lee01,Lee02,Wojs04}.\\

To understand the nature of differently spin polarized FQHSs of composite fermions, I first evaluate their pseudopotentials, which depend on the background state as well as the number and the direction of the vortices they carry. For this purpose, I consider states with filled L$\Lambda$L $\uparrow$ (background state) with two additional composite fermions in either the S$\Lambda$L $\uparrow$ or L$\Lambda$L $\downarrow$. For the sake of completeness I have also evaluated the interaction energy of two composite fermions one each placed in the S$\Lambda$L $\uparrow$ and L$\Lambda$L $\downarrow$. In analogy to the electron pseudopotentials \cite{Haldane83} this interaction energy of two composite fermions is termed CF pseudopotentials \cite{Sitko96} and is denoted by $V^{\rm CF}_{m}$, where $m$ is the relative angular momentum of the two CFs.\\

\begin{figure*}[htpb]
\begin{center}
\includegraphics[width=8cm,height=4.5cm]{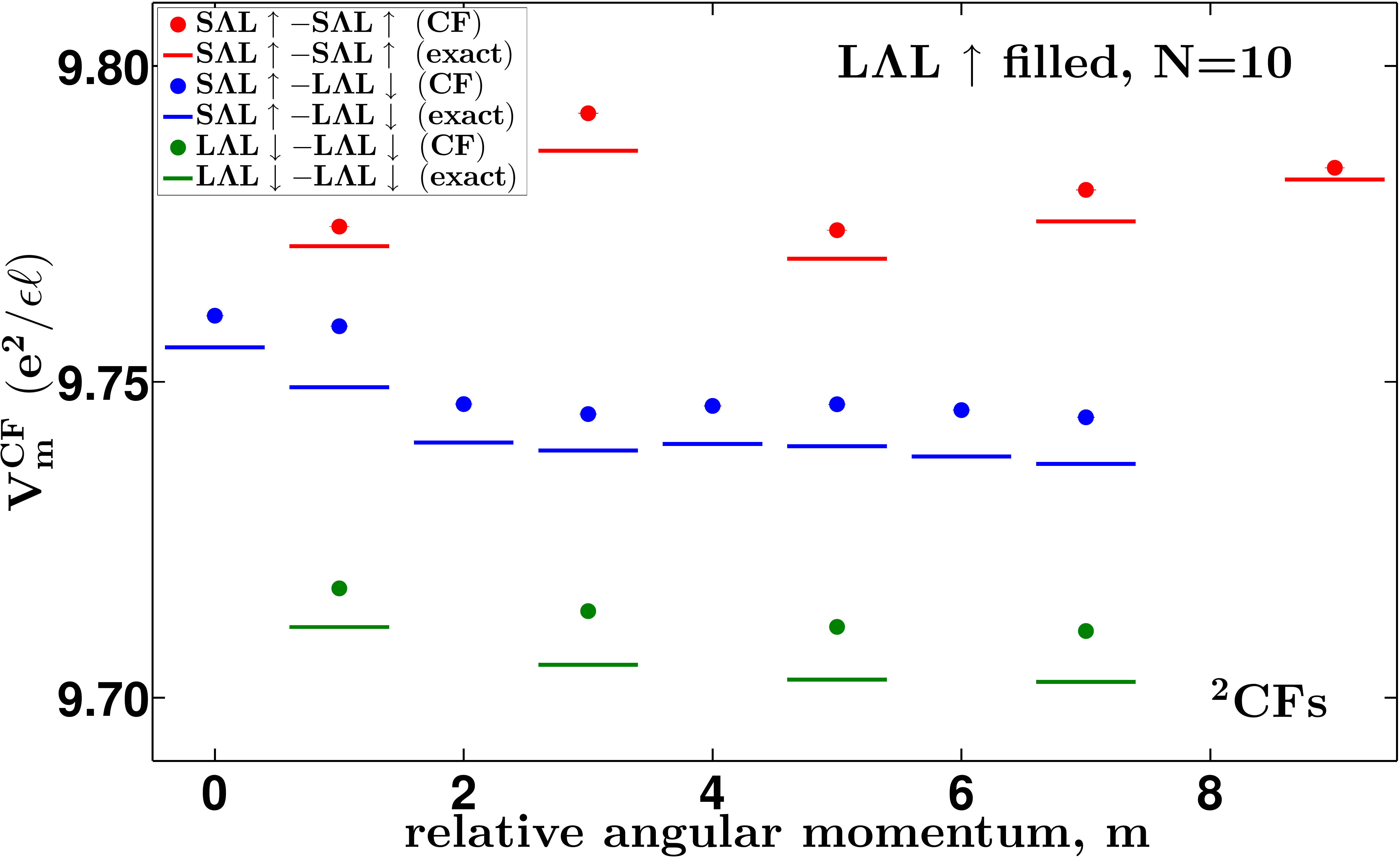}
\includegraphics[width=8cm,height=4.5cm]{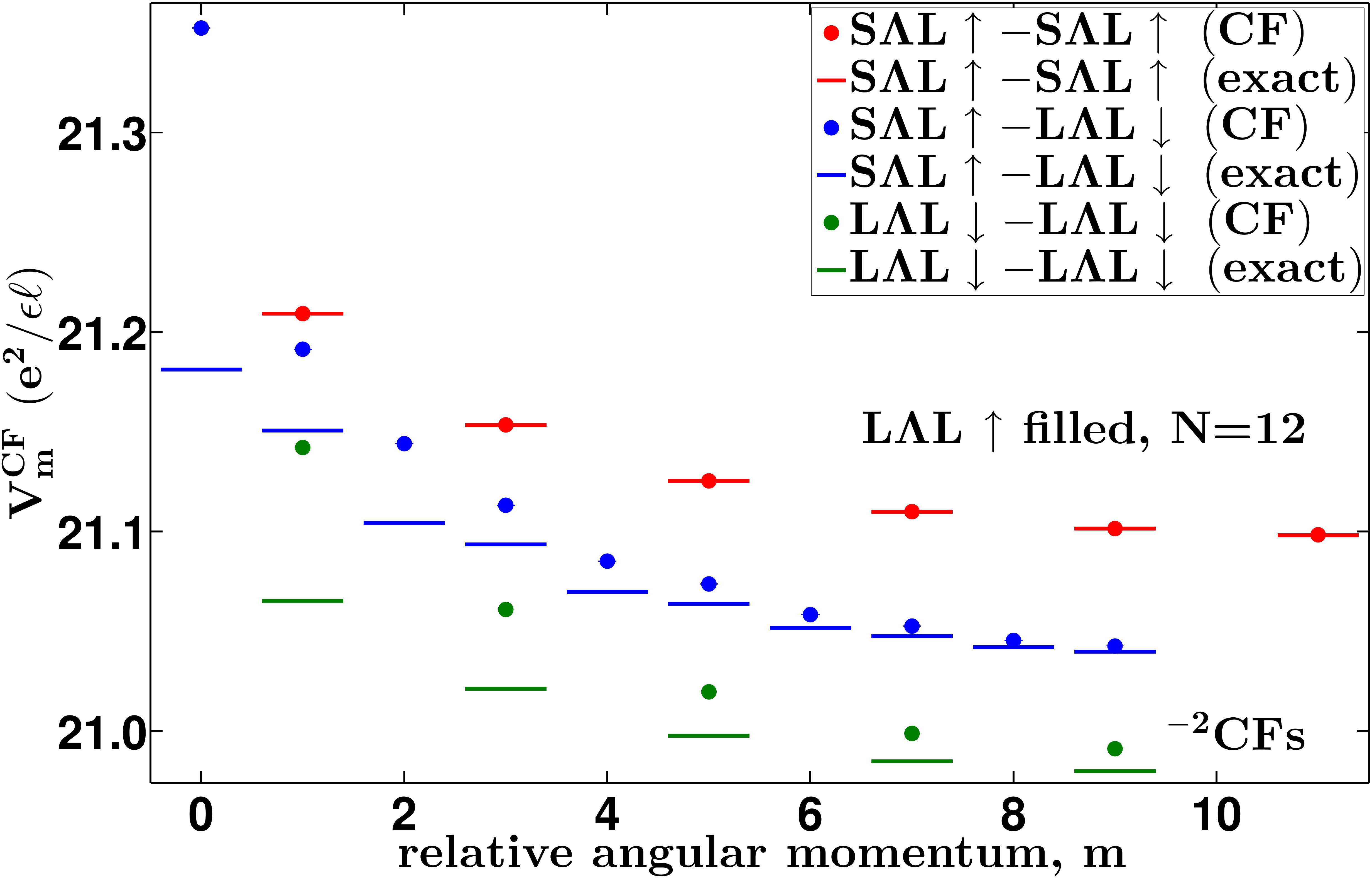}
\includegraphics[width=8cm,height=4.5cm]{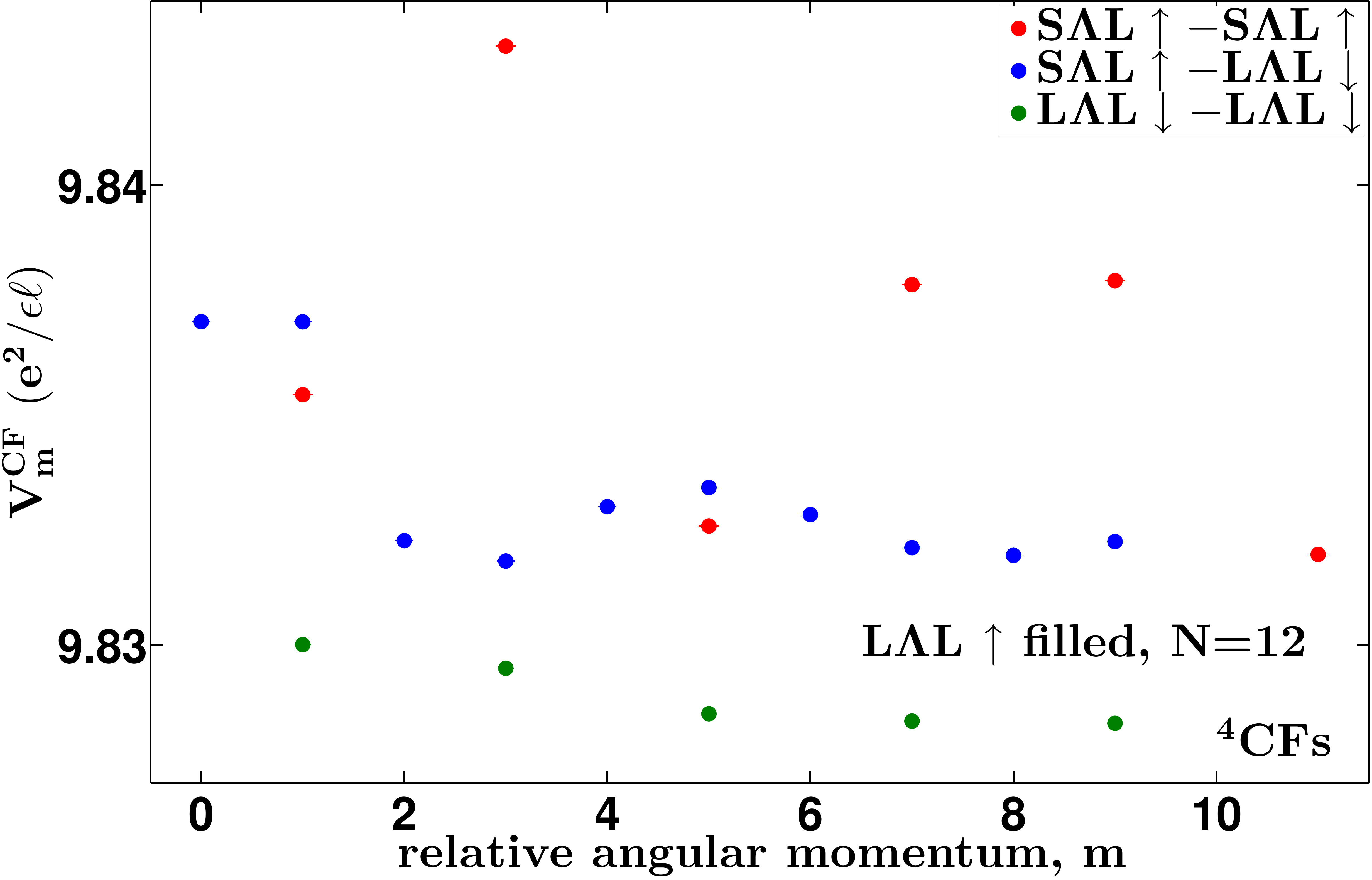}
\includegraphics[width=8cm,height=4.5cm]{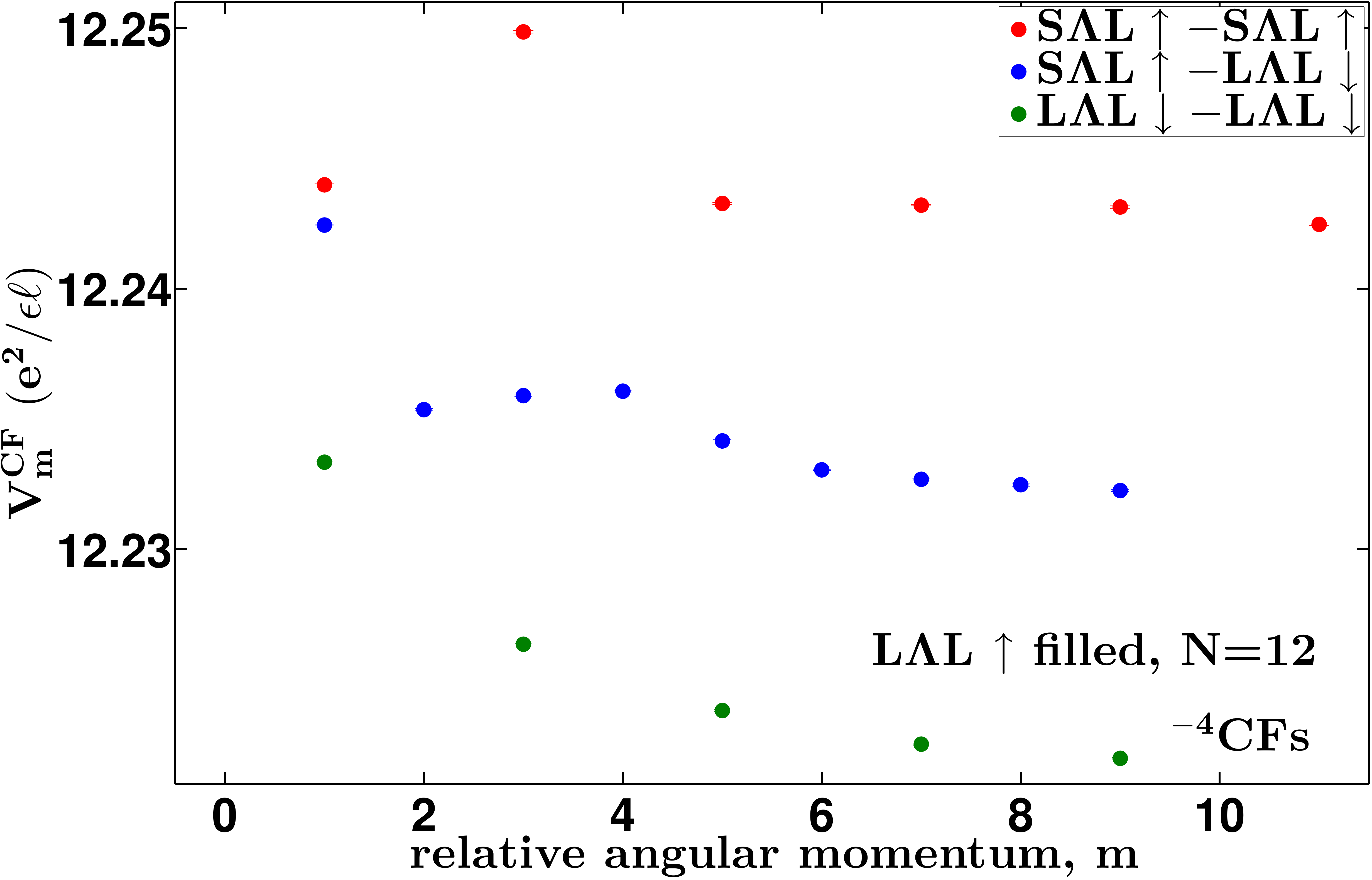}
\end{center}
\caption{\label{fig:interaction_CFs}Coulomb interaction energy of two composite fermions in the presence of a filled spin up lowest $\Lambda$ level carrying two (top panels) or four (bottom panels) vortices and experiencing an effective magnetic field in the same (left panels) or opposite  (right panels) direction as the external magnetic field calculated in the spherical geometry. The dashes (dots) show the energies obtained from exact (CF) diagonalization.}
\end{figure*}

In Fig. \ref{fig:interaction_CFs} I show these two-particle interaction energies for a representative system of $N$ electrons evaluated in the spherical geometry. These represent the interaction energy of
\begin{itemize}
 \item for $^{2}$CFs: two quasiparticles at $\nu=1/3$
 \item for $^{4}$CFs: two quasiparticles at $\nu=1/5$
 \item for $^{-2}$CFs: two quasiholes at $\nu=1$
 \item for $^{-4}$CFs: two quasiholes at $\nu=1/3$
\end{itemize}

Based on the CF pseudopotentials shown in Fig. \ref{fig:interaction_CFs} I make the following points:
\begin{itemize}
 \item The CF pseudopotential $V^{\rm CF}_{3}$ is maximum when two composite fermions are located in the S$\Lambda$L $\uparrow$ for all cases except for $^{-2}$CFs. As noted above this should be contrasted with the electronic pseudopotentials in the LLL where $V_{1}$ is maximum (as is seen for $^{-2}$CFs), which in turn provides sufficiently strong short-range interactions \cite{Haldane85a} to stabilize correlations of the Laughlin kind \cite{Laughlin83}. Refs. \cite{Sitko96,Wojs00,Lee01,Lee02} noticed this observation which inspired the search of a fully spin polarized incompressible state at $\nu=1/3$ and $\nu=1/5$ (and by particle-hole conjugation at $\nu=2/3$ and $\nu=4/5$) which was different from the Laughlin state. By considering the model electronic Hamiltonian $V_{m}=\delta_{m,3}$, Refs. \cite{Wojs04,Mukherjee14} found that incompressibility for this interaction occurs at $\nu=1/3$ and $\nu=1/5$ at shifts $\mathcal{S}=7$ and $\mathcal{S}=9$ respectively (shift for the Laughlin state at $1/3$ ($1/5$) is $\mathcal{S}=3$ ($\mathcal{S}=5$). The $\nu=1/5$ Laughlin state is by construction an exact zero energy eigenstate of the $V_{3}$-only model Hamiltonian (while the $\nu=1/3$ Laughlin state is not), so trivially there is incompressibility at $\nu=1/5$ at shift $\mathcal{S}=5$ for this model interaction.). These states are termed the WYQ states and Refs. \cite{Mukherjee14,Mukherjee15b} made a strong case that the 1/3 (2/3) WYQ state occurs in the S$\Lambda$L $\uparrow$ at fully spin polarized $\nu^{*}=4/3$ ($\nu^{*}=5/3$) which leads to FQHE at fully spin polarized $\nu=4/11$ and $\nu=4/13$ ($\nu=5/13$ and $\nu=5/17$).
 \item For the case of two CFs located in the L$\Lambda$L $\downarrow$ the pseudopotential ordering follows that of electrons in the LLL which suggests that partially spin polarized FQHSs of composite fermions within the L$\Lambda$L would support conventional Laughlin-type order.
 \item By comparing the CF pseudopotentials of $^{2}$CFs and $^{-4}$CFs we notice that the interaction between two quasiparticles at $\nu=1/3$ is qualitatively similar to that of two quasiholes at $\nu=1/3$. This sheds light into the observation of Mukherjee \emph{et al.} \cite{Mukherjee14,Mukherjee14c} who found that the physics of the the 4/11 and 4/13 states, which both arise from $\nu^{*}=4/3$ (the former with parallel vortex attachment with CFs carrying two vortices and the latter with reverse vortex attachment with CFs carrying four vortices as shown in Fig. \ref{fig:schematic}), is similar in that both likely support the WYQ type $1/3$ state in the S$\Lambda$L $\uparrow$. The same argument explains the qualitative similarity of the fully spin polarized 5/13 and 5/17 states (see Fig. \ref{fig:schematic}).
 \item The physics of parallel vortex attached states could be very different from that of reverse vortex attached states. 
 \item The vorticity of composite fermions plays a key role in determining the interaction between them which subsequently decides the nature of an FQHE state.
\end{itemize}

The above points suggest that WYQ states could occur in the S$\Lambda$L $\uparrow$ for the fully spin polarized states of the form $[1+1/(2p+1)]_{2}$, $[1+\overline{1/(2p+1)}]_{2}$, $[1+1/(2p+1)]_{\pm 4}$ or $[1+\overline{1/(2p+1)}]_{\pm 4}$ with $p=1,2$. Of these states, besides the aformentioned states at $\nu=4/11,~4/13,~5/13$ and $5/17$, the only state for which experimental signatures have been observed is the state at $\nu=6/17$ \cite{Pan03}. This state is obtained by parallel vortex attachment of two vortices to the state at $\nu^{*}=6/5$ (see Fig. \ref{fig:schematic}) and I shall consider it in detail in Sec. \ref{sec:CF_6_5}.\\

Since the CF pseudopotentials indicate only a possibility for certain states to occur and do not tell the complete story (which of course has to be determined by looking at the full problem of interacting electrons), I consider three other potential WYQ candidate states in this work, namely the ones at $\nu=4/5$, $\nu=5/7$ and $\nu=6/7$. The $6/7$ state is closely related to the $6/17$ state in that it too arises from $\nu^{*}=6/5$ of composite fermions by reverse vortex attachment with two vortices (see Fig. \ref{fig:schematic}). On the other hand, the fully spin polarized $4/5$ ($5/7$) states are related to the fully spin polarized $4/11$ and $4/13$ ($5/13$ and $5/17$) states, where WYQ order has been shown to be very plausible \cite{Mukherjee14,Mukherjee14c} (see Fig. \ref{fig:schematic}).   \\

In the next two sections I shall look at the fully spin polarized $\nu=4/5$ and $\nu=5/7$ states and the various spin polarized states at $\nu=6/17$ and $\nu=6/7$. Our approach would be to consider both the Laughlin and WYQ $1/5$ state in the S$\Lambda$L $\uparrow$ and see which of these is stabilized for the fully spin polarized $\nu=6/17$ state. I shall do the same for the fully spin polarized state at $\nu=4/5$, $\nu=5/7$ and $\nu=6/7$ even though the S$\Lambda$L $\uparrow$ $-$ S$\Lambda$L $\uparrow$ pseudopotentials of $^{-2}$CFs indicate that this state is likely to host a Laughlin state in the S$\Lambda$L $\uparrow$. For all these fully spin polarized states I find that conventional Laughlin order prevails over the unconventional WYQ order. Thus $V_{3}^{\rm CF}$ being maximum provides only a necessary but not sufficient condition for FQHSs to host the unconventional WYQ order. As for the partially spin polarized state, the CF pseudopotentials of two CFs placed in the L$\Lambda$L $\downarrow$, strongly suggest that Laughlin type correlations are stabilized in the L$\Lambda$L $\downarrow$ and hence I shall only consider this state. Finally for the spin-singlet state only standard CF states are expected to be stabilized and I shall only consider these in this work. For the non-fully spin polarized states at the conventional flux, all systems considered in this work have ground states with $L=0$. This suggests that the conventional states are the ground states, since otherwise the ground state at the conventional flux would be an excited state of the true ground state, in which case the ground state at the conventional flux can have $L\neq 0$.\\ 

Ref. \cite{Balram15} had already considered the differently spin polarized states at $\nu=4/5$ and $5/7$ supporting conventional order. I vindicate their assumption for the fully spin polarized $4/5$ and $5/7$ states and produce improved numbers for the ground state energy by considering larger system sizes which supersede their results. For the conventional non-fully spin polarized states at $\nu=4/5$ and $\nu=5/7$ I have not carried out any new calculations. Results on these can be found in Ref. \cite{Balram15} which I shall not reproduce here. \\

\section{Nature of the fully spin polarized states at $\nu=4/5$ and $\nu=5/7$}
\label{sec:4_5_and_5_7_fp}
The $\nu=4/5$ and $\nu=5/7$ states are obtained from $\nu^{*}=4/3$ and $\nu^{*}=5/3$, respectively, by reverse vortex attachment with two vortices (see Fig. \ref{fig:schematic}). Since the composite fermion wave functions projected in the manner as stated in Sec. \ref{sec:sphere_CF} are not very accurate when doing reverse vortex attachment \cite{Balram15a} (see for example the top right panel of Fig. \ref{fig:interaction_CFs}), I shall only consider results obtained from exact diagonalization in this section. 
\subsection{Fully spin polarized state at $\nu=4/5$}
\label{subsec:4_5_fp}
\begin{figure*}[htpb]
\centering
\subfigure[~Laughlin]{
\includegraphics[width=6cm,height=3.5cm]{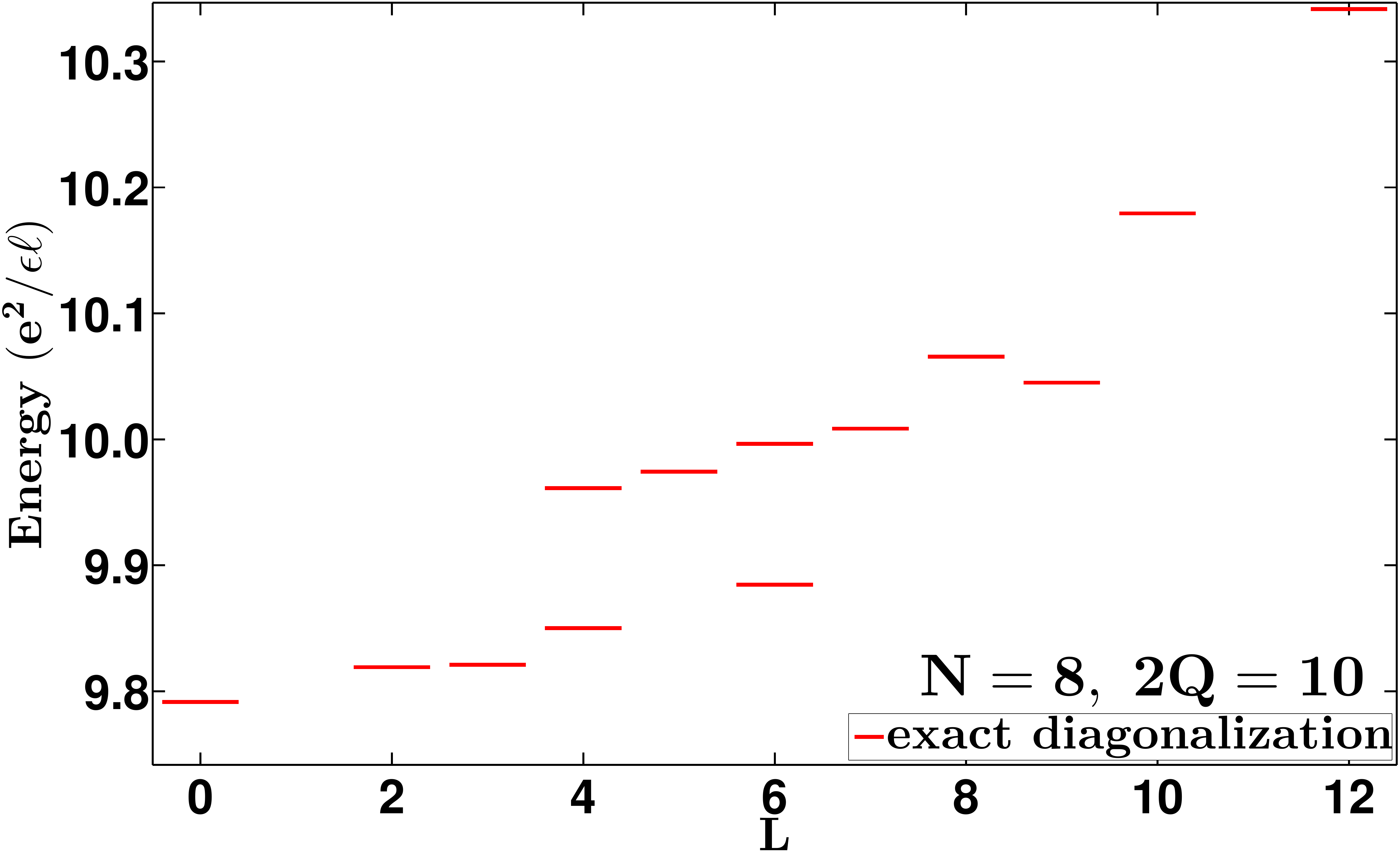}
\includegraphics[width=6cm,height=3.5cm]{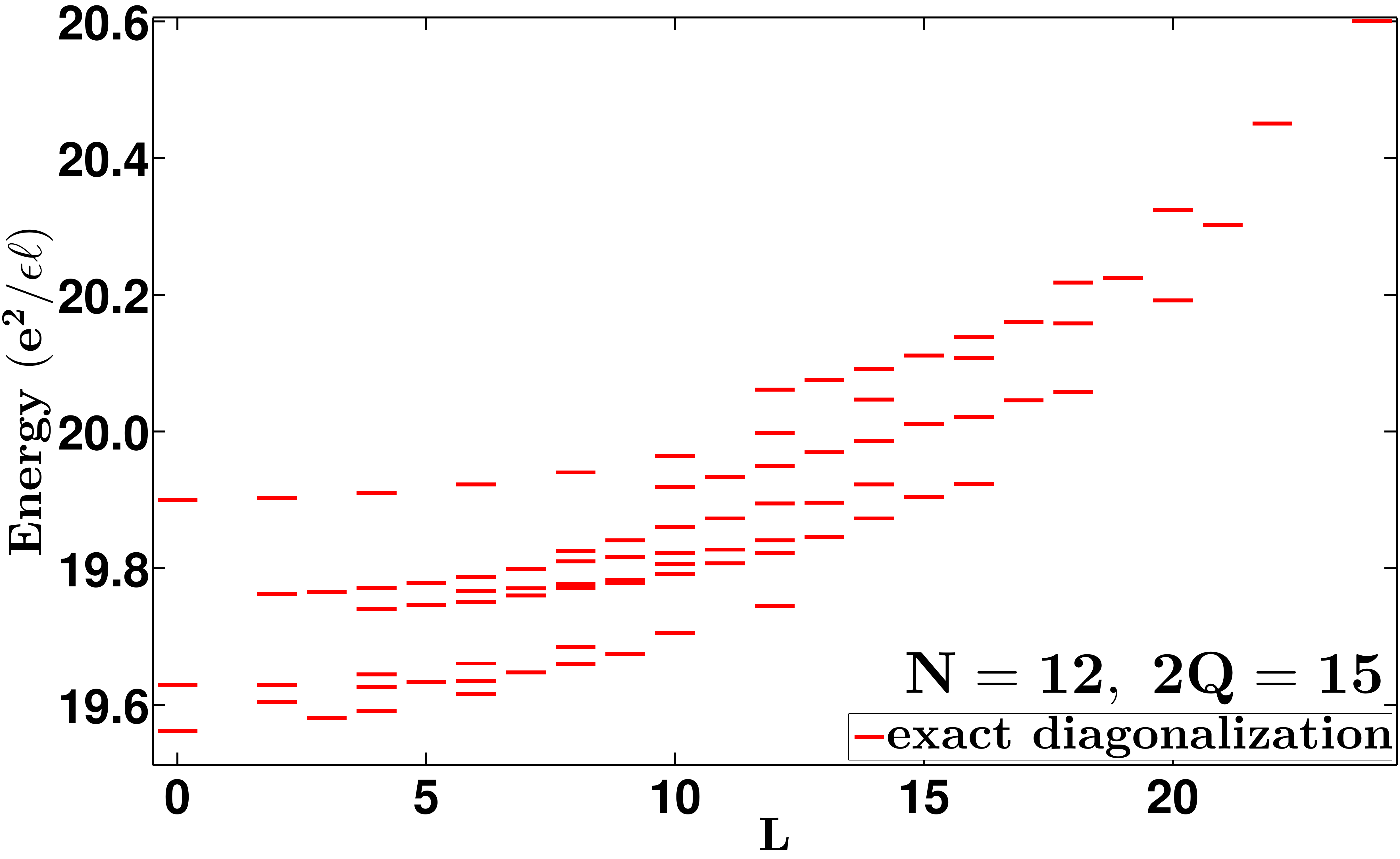}
\includegraphics[width=6cm,height=3.5cm]{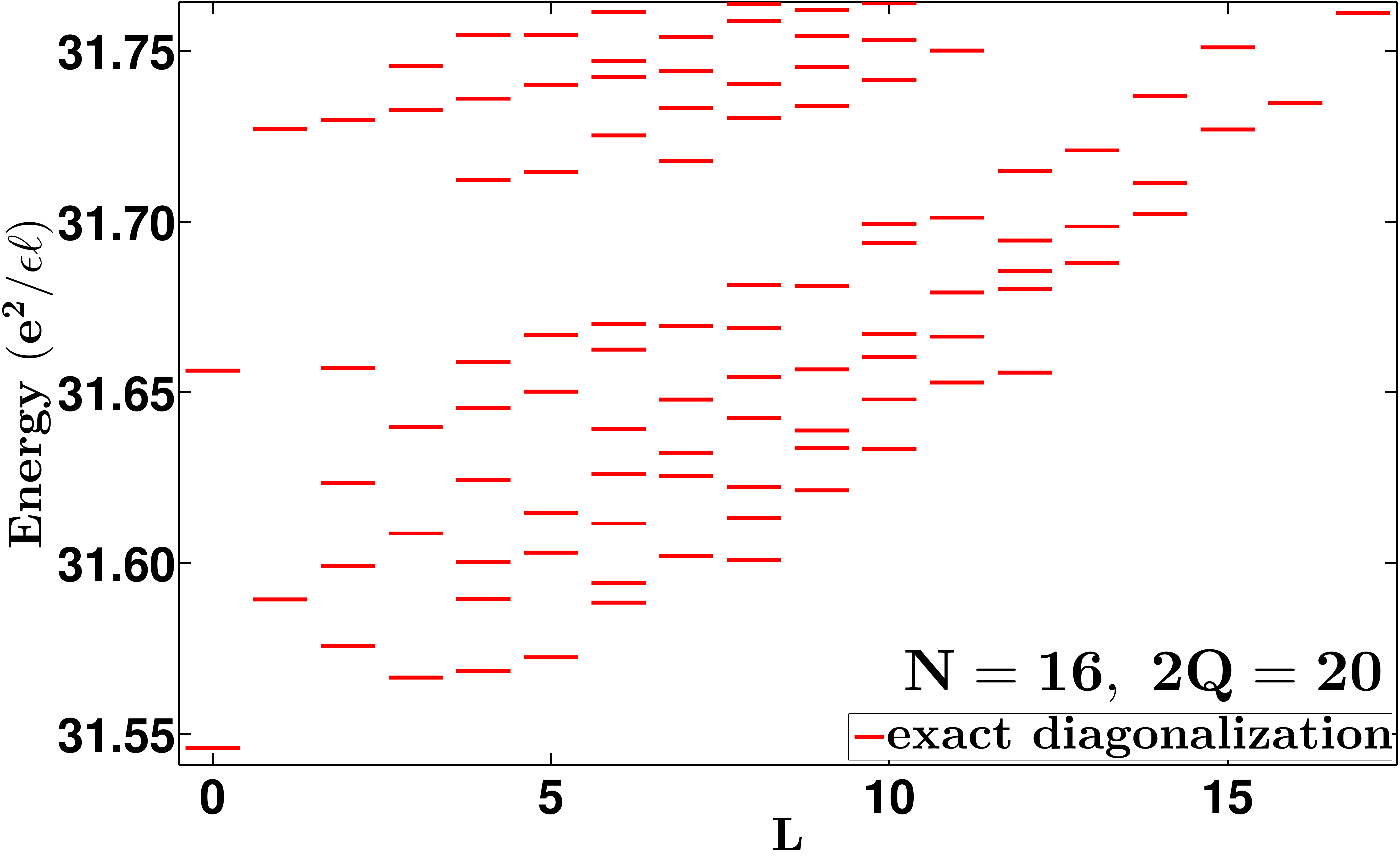}}
\subfigure[~Laughlin]{
\includegraphics[width=6cm,height=3.5cm]{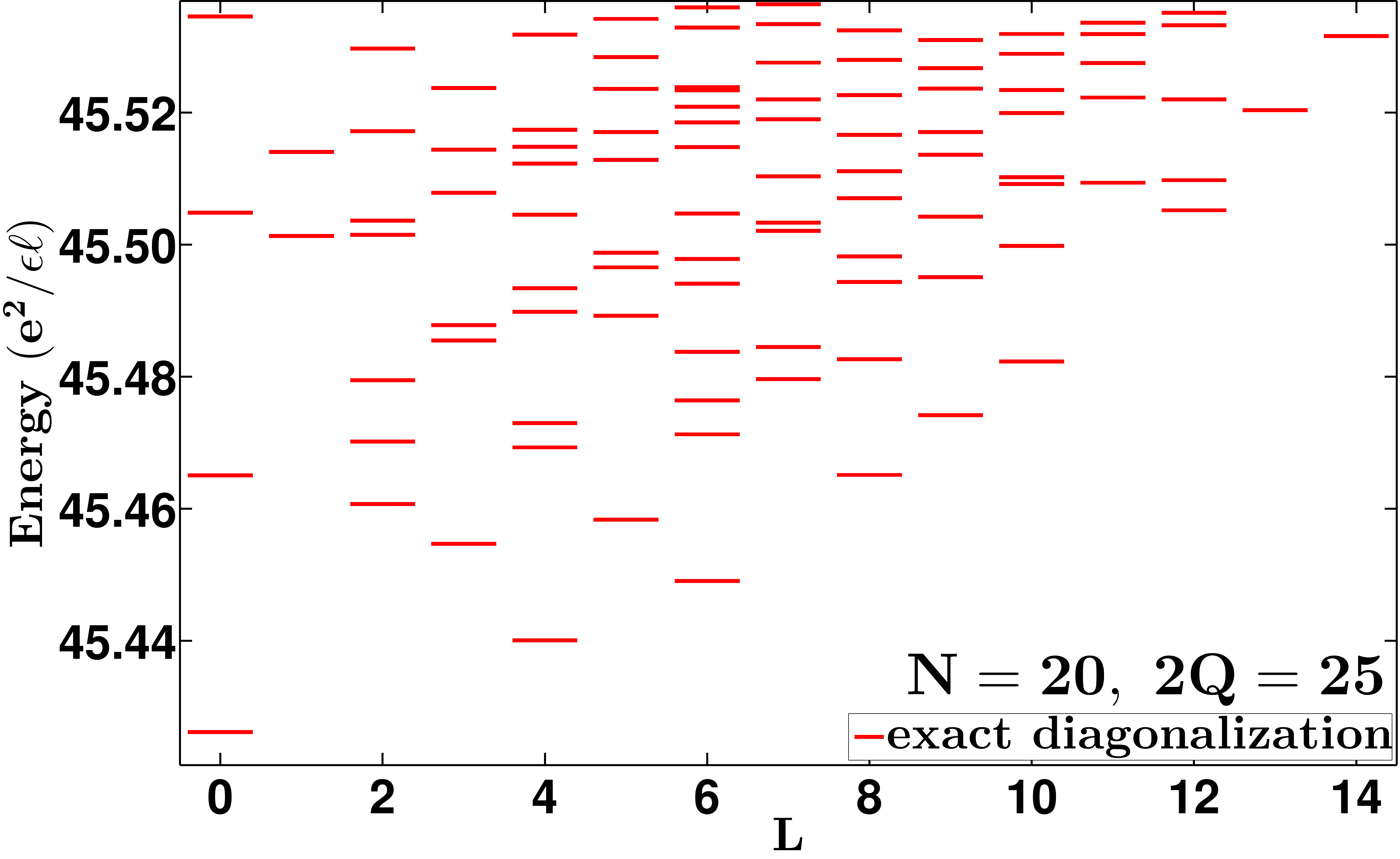}
\includegraphics[width=6cm,height=3.5cm]{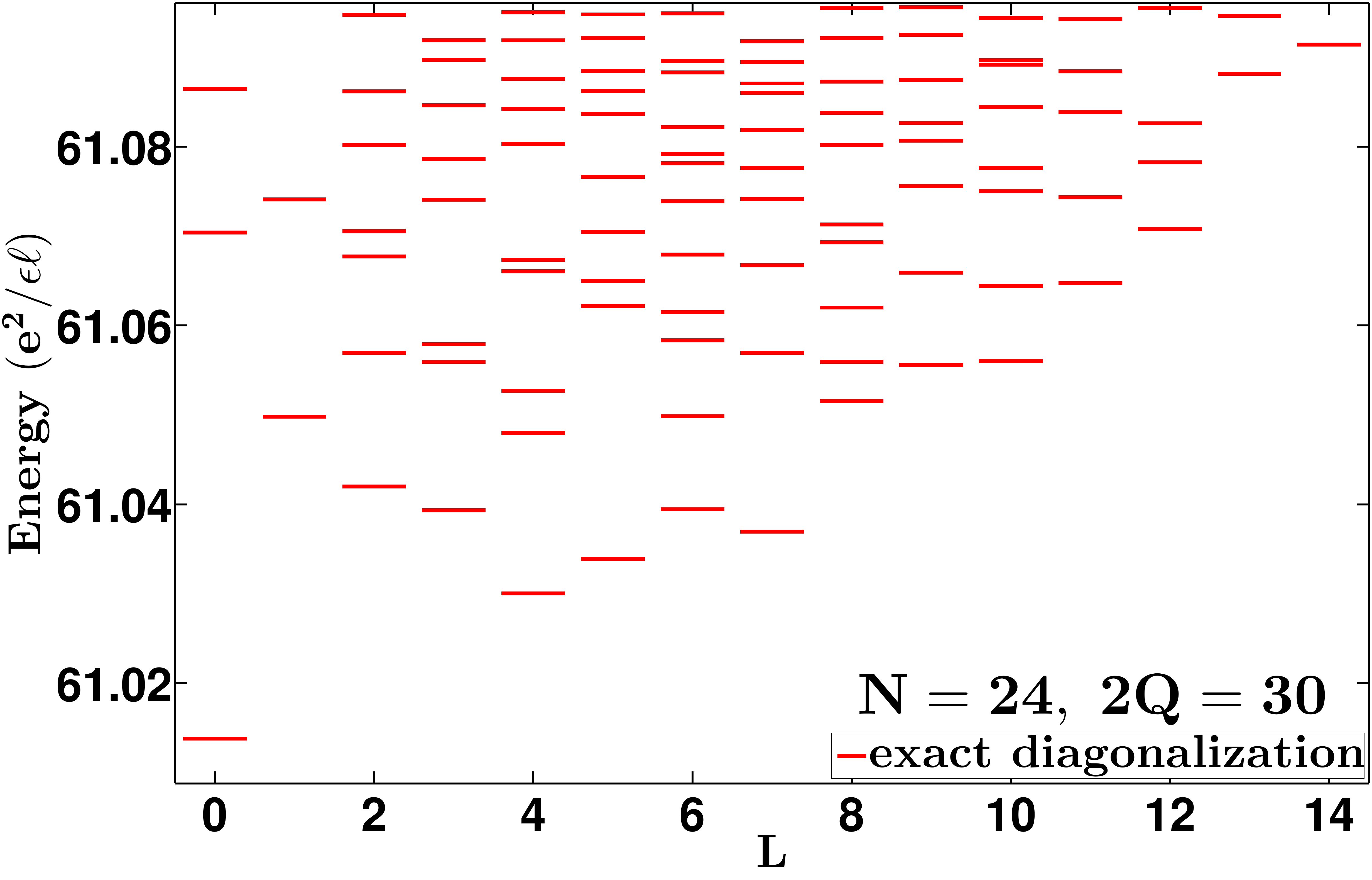}
\includegraphics[width=6cm,height=3.5cm]{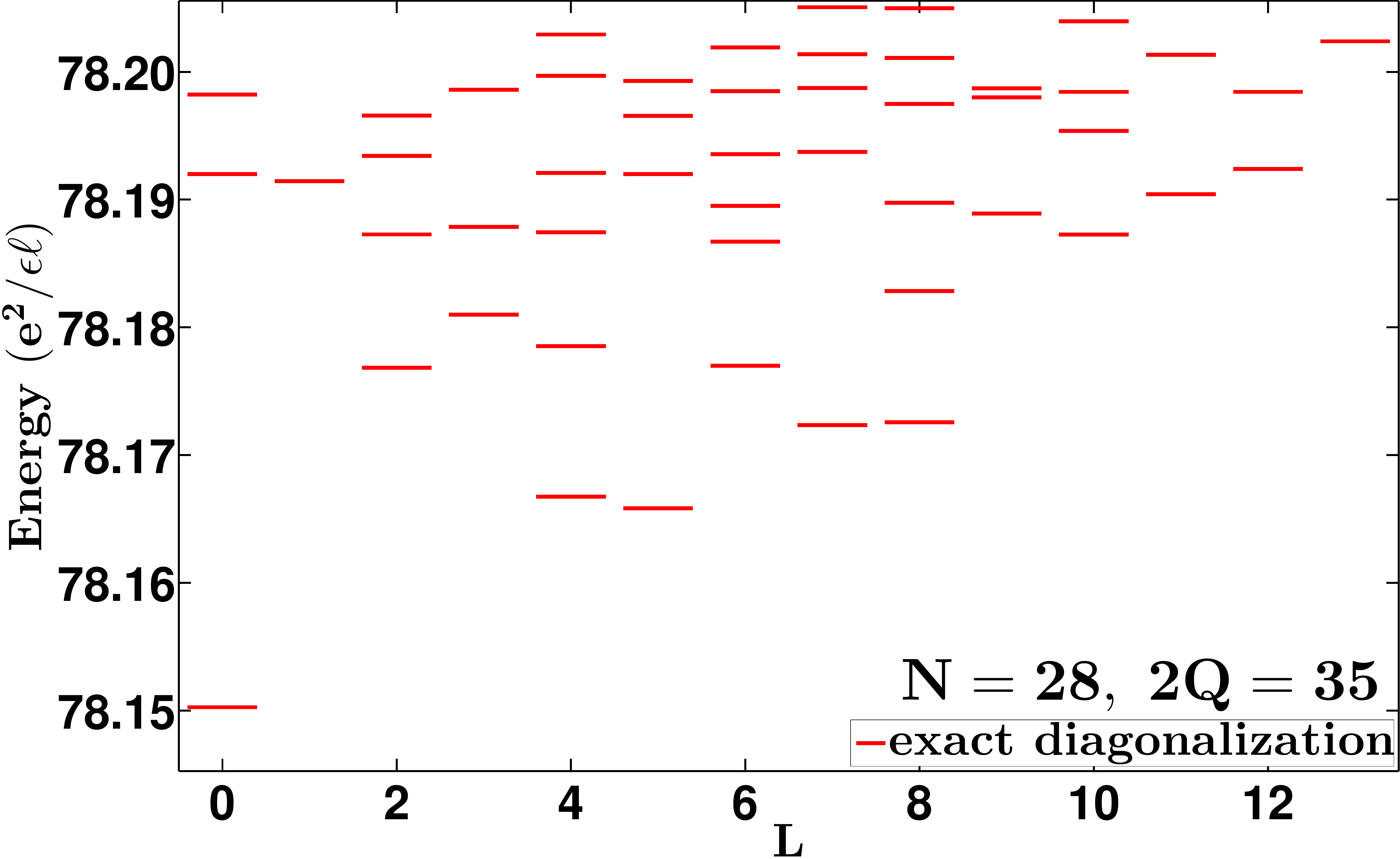}}
\quad
\subfigure[~WYQ]{
\includegraphics[width=6cm,height=3.5cm]{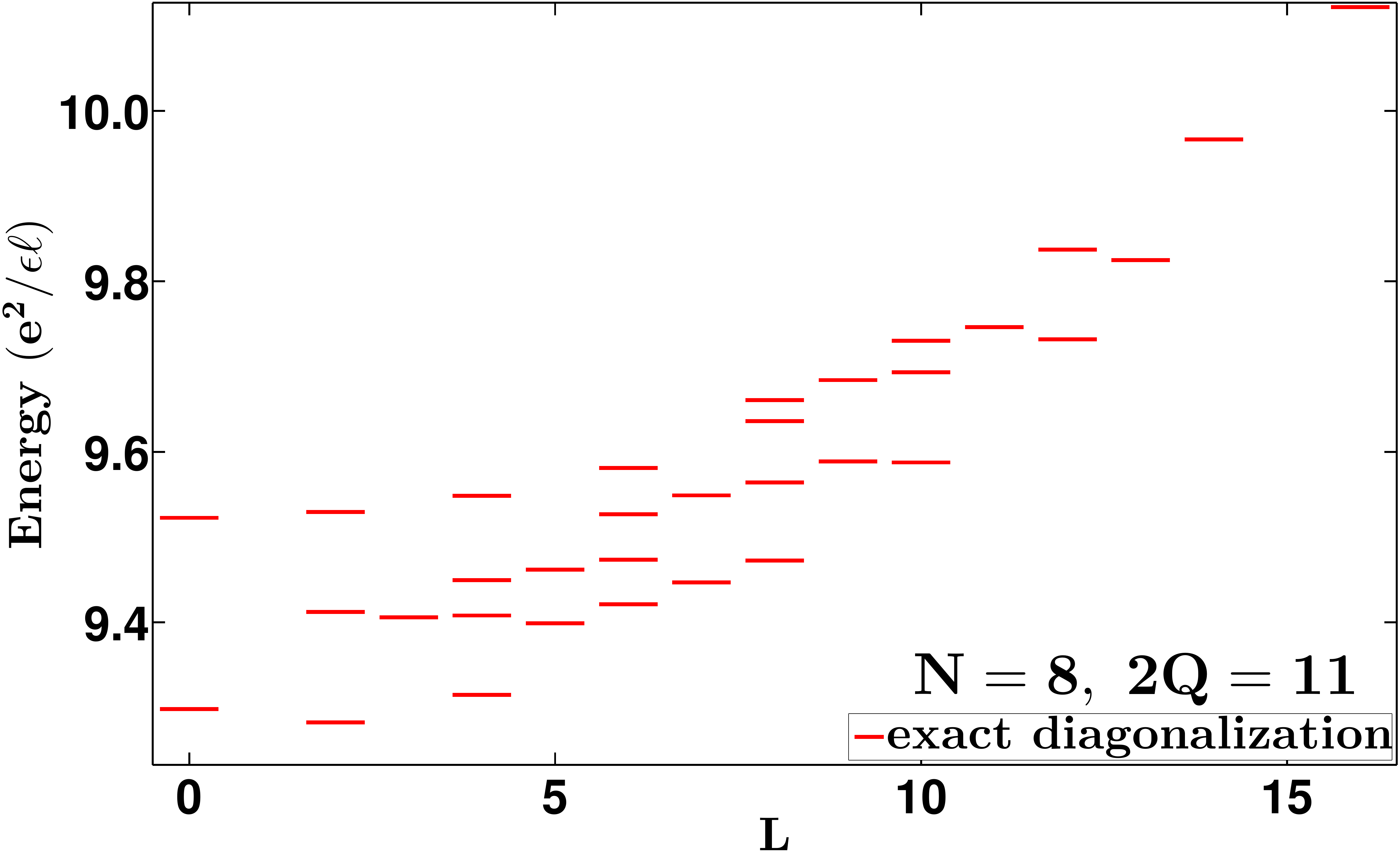}
\includegraphics[width=6cm,height=3.5cm]{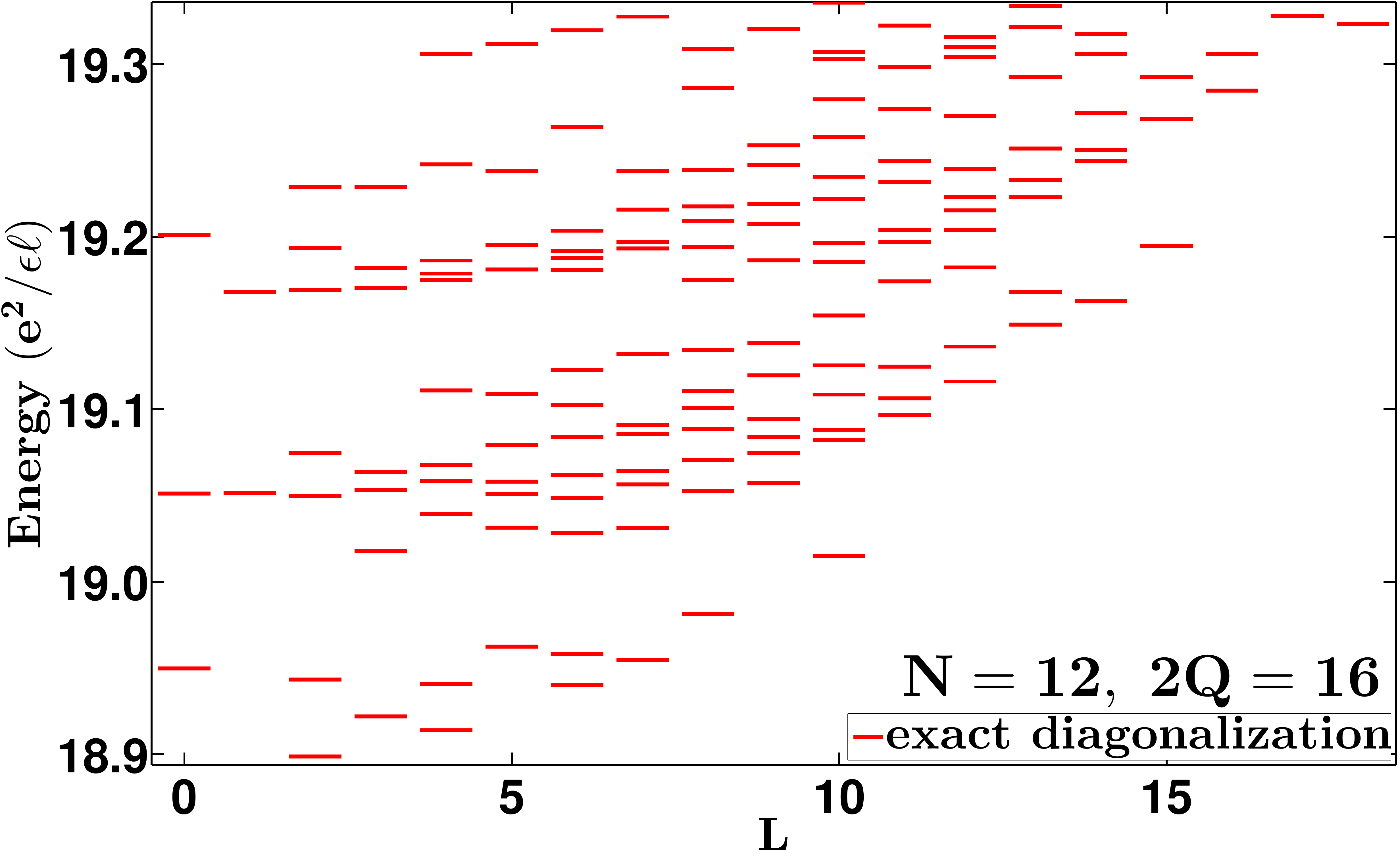}
\includegraphics[width=6cm,height=3.5cm]{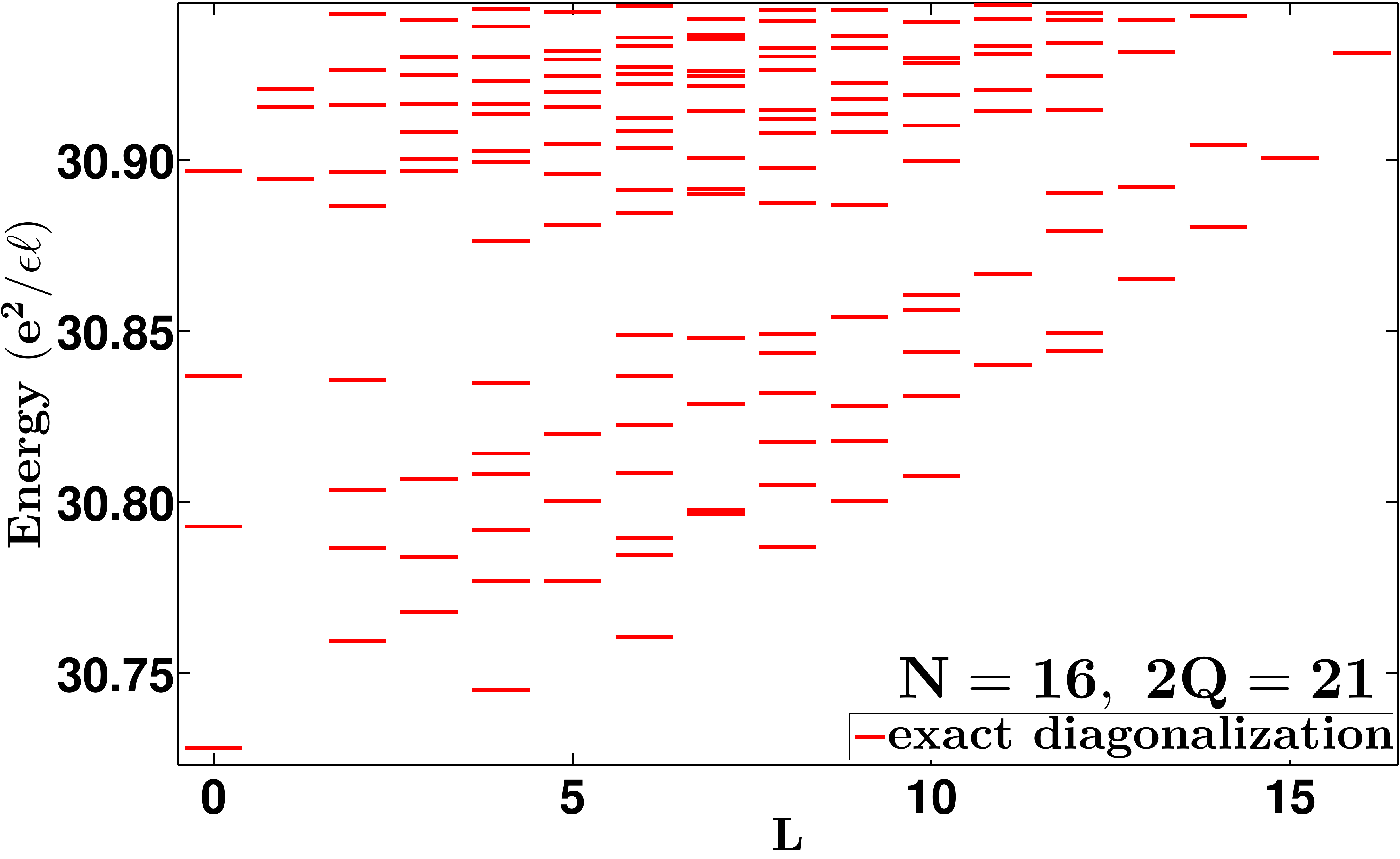}}
\subfigure[~WYQ]{
\includegraphics[width=6cm,height=3.5cm]{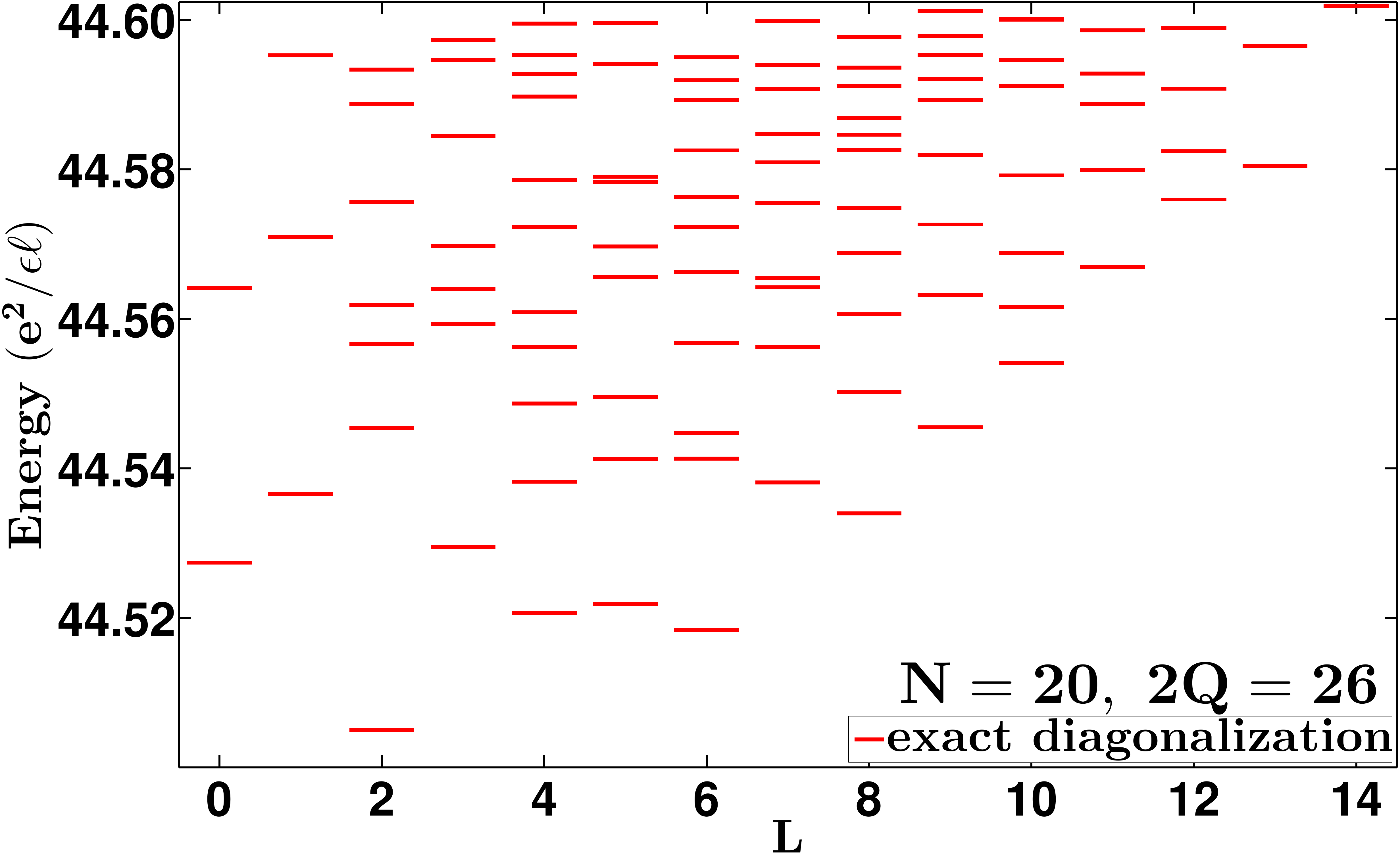}
\includegraphics[width=6cm,height=3.5cm]{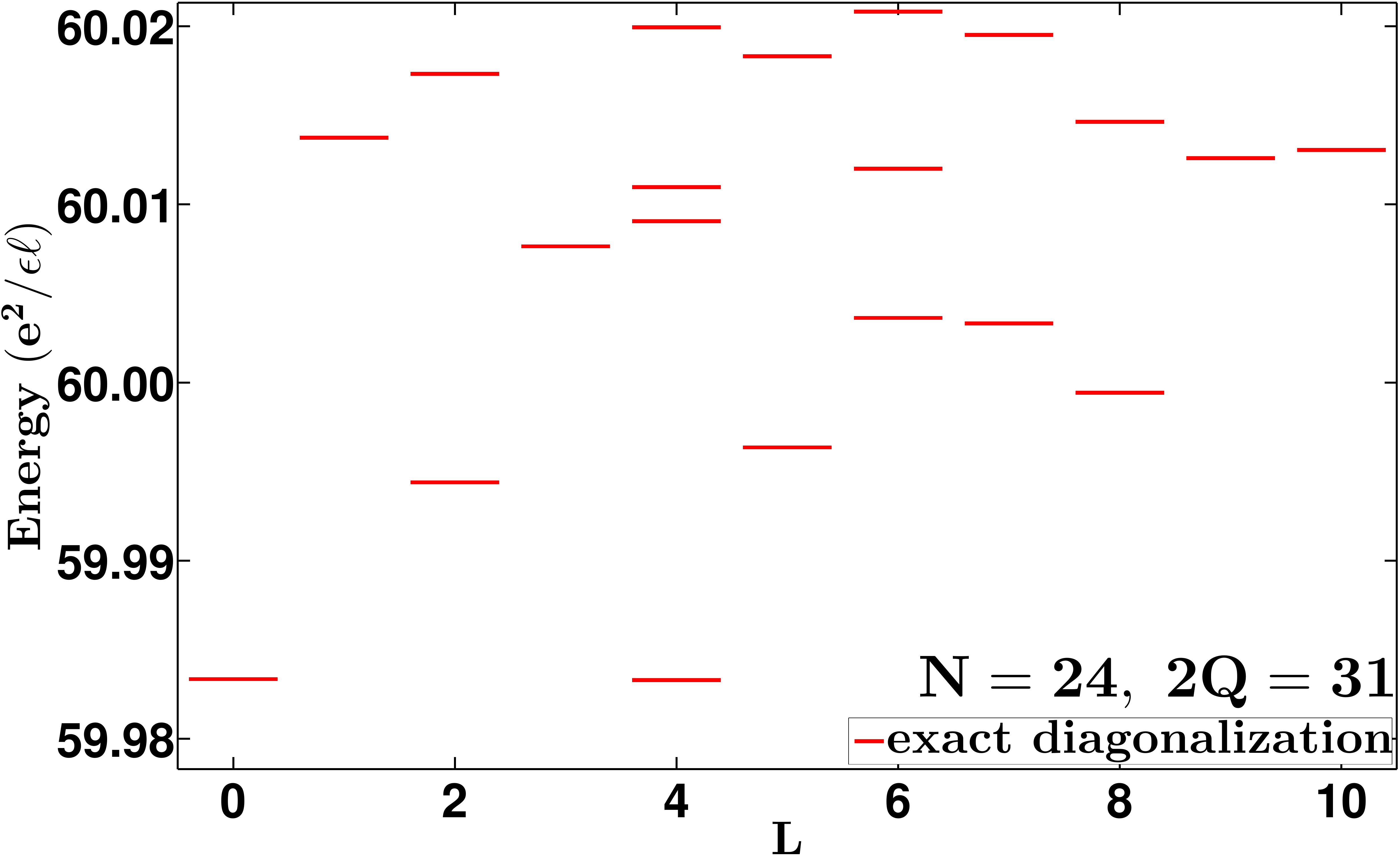}
\includegraphics[width=6cm,height=3.5cm]{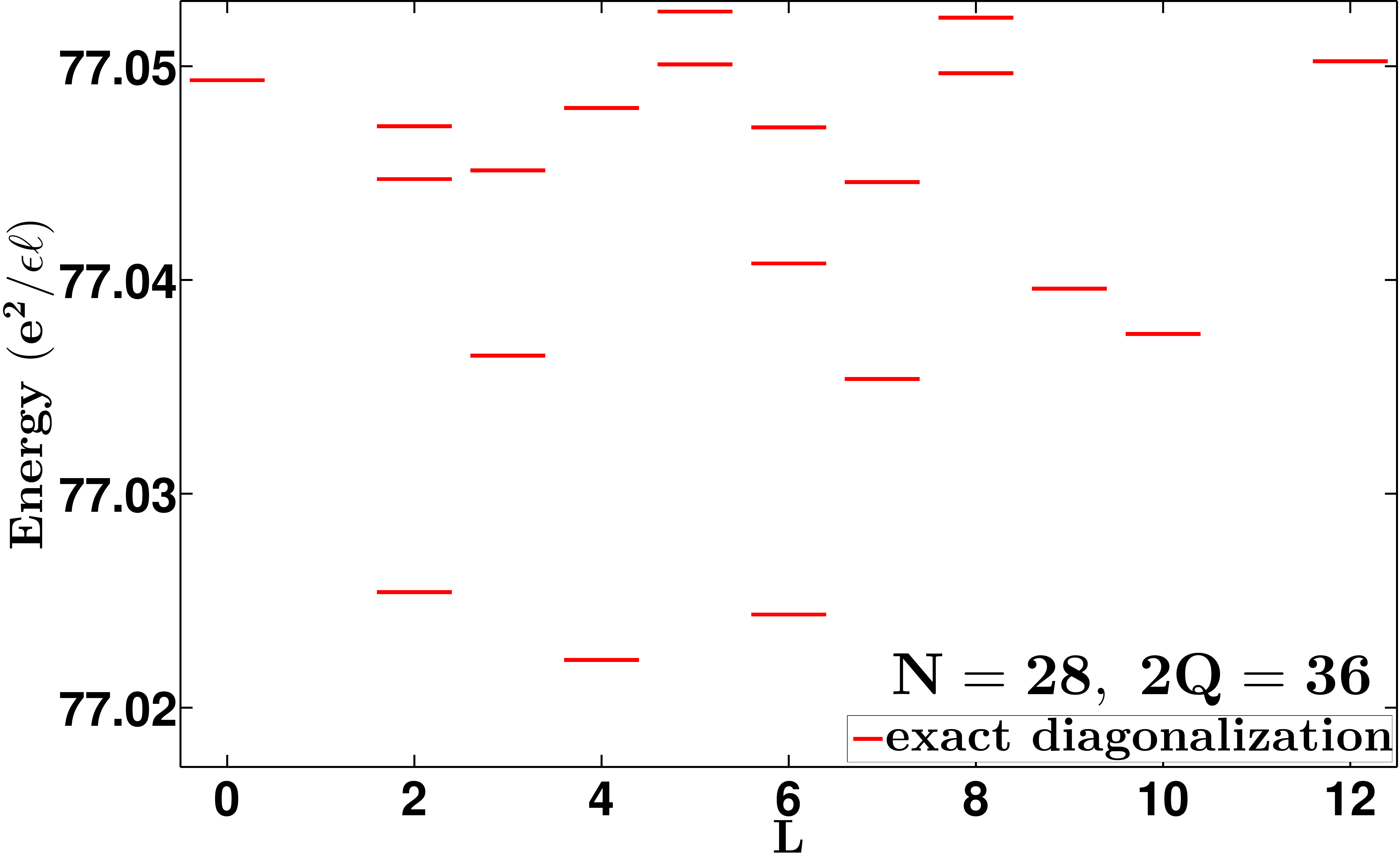}}
\caption{\label{spectra_fp_4_5}Coulomb spectra obtained from exact diagonalization in the spherical geometry for $N$ fully spin polarized electrons at $\nu=4/5$ at the total flux $2Q(hc/e)$ corresponding to the Laughlin (top two panels (a) and (b)) and WYQ state (bottom two panels (c) and (d)).}
\end{figure*}

\begin{figure}[htbp]
\begin{center}
\includegraphics[width=8cm,height=4.5cm]{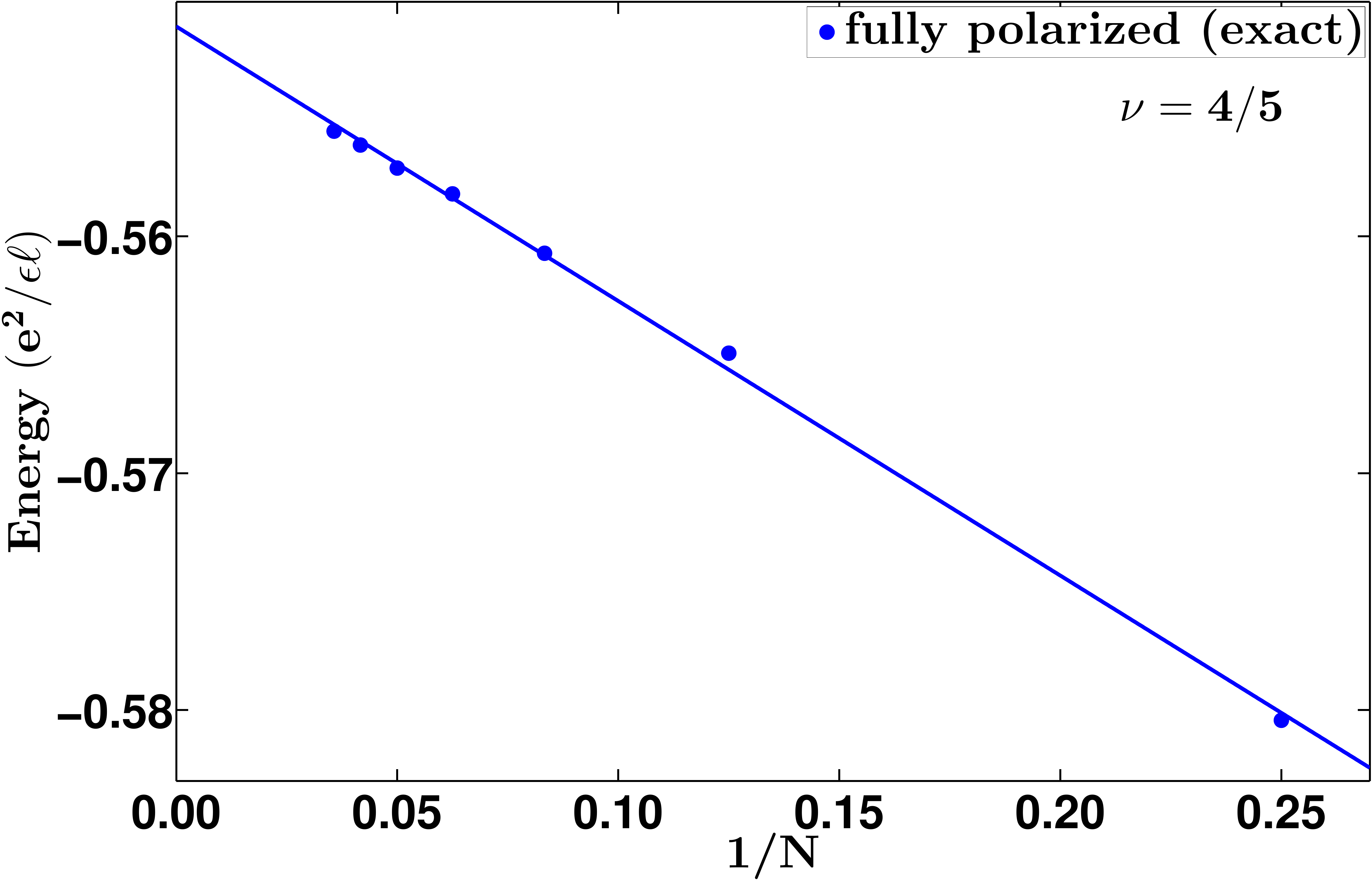}
\end{center}
\caption{\label{extrap_4_5_fp}
Thermodynamic extrapolation of the lowest Landau level Coulomb ground state energy for the fully spin polarized state at $\nu=4/5$.}
\end{figure}

The S$\Lambda$L $\uparrow$ $-$ S$\Lambda$L $\uparrow$ $^{-2}$CFs pseudopotentials shown in Fig. \ref{fig:interaction_CFs} indicate that the ground state at $\nu=4/5$ arises from a $1/3$ Laughlin like state in the S$\Lambda$L $\uparrow$. If this is the case, it would be in stark contrast to the ground state at the related $\nu=4/11$ which likely arises from a $1/3$ WYQ state in the S$\Lambda$L $\uparrow$ \cite{Mukherjee14}. To establish the identity of the fully spin polarized $\nu=4/5$ state I consider the following two states:
\begin{itemize}
 \item The state:
 \begin{equation}
  [1+[1]_{2}]_{-2} \leftrightarrow(4/5)~:\gamma=1
  \label{eq_CF_4_3_m2}
 \end{equation} 
corresponds to $\nu^*=4/3$, which is obtained by filling the L$\Lambda$L $\uparrow$ completely and forming a 1/3 Laughlin state in the S$\Lambda$L $\uparrow$. This state at $\nu=4/5$ occurs at $\mathcal{S}=0$ and has $S=N/2$.
 \item The state:
 $$[1+1/3^{\rm WYQ}]_{-2} \leftrightarrow(4/5)~:\gamma=1$$
corresponds to $\nu^*=4/3$, which is obtained by filling the L$\Lambda$L $\uparrow$ completely and forming a 1/3 WYQ state in the S$\Lambda$L $\uparrow$. This state at $\nu=4/5$ occurs at $\mathcal{S}=-1$ and has $S=N/2$.
\end{itemize}
The Coulomb spectra for both these cases obtained from exact diagonalization is shown in Fig. \ref{spectra_fp_4_5}. We notice that with a 1/3 Laughlin state in the S$\Lambda$L $\uparrow$ the ground state is incompressible for all system sizes while this is not the case for all systems hosting the 1/3 WYQ state in the S$\Lambda$L $\uparrow$. Thus the fully spin polarized 4/5 state is a FQHS of composite fermions with a filled L$\Lambda$L $\uparrow$ and a conventional 1/3 Laughlin state in the S$\Lambda$L $\uparrow$. Figure \ref{extrap_4_5_fp} shows the thermodynamic extrapolation of the LLL Coulomb ground state energy for the fully spin polarized $4/5$ state. The extrapolated ground state energy obtained is 0.5511(2) (the number in the parenthesis denotes the uncertainity in our linear extrapolation) which is consistent with previous results \cite{Fano86,Jain97,Balram15}. \\

Another plausible candidate state for the fully spin polarized FQHE at $4/5$ is 
\begin{equation}
\overline{[1]_{4}} \leftrightarrow(4/5)~:\gamma=1,
\label{eq_hole_1_5}
\end{equation}
which is the hole conjugate of the 1/5 Laughlin state. However, in spite of this superficial difference, the wave functions for the above two states of Eqs. (\ref{eq_CF_4_3_m2}) and (\ref{eq_hole_1_5}) are equivalent, i.e., they represent the same state, as can be seen by noting that they occur at the same shift in the spherical geometry and have an identical excitation spectrum (see the Supplemental Material of Ref. \cite{Balram15b} for details). This situation is analogous to the fully spin polarized $2/3$ state for which two candidate wave functions can be written: namely $[2]_{-2}$ and $\overline{[1]_2}$, but explicit calculations \cite{Wu93} have shown that in fact these wave functions are essentially identical to each other. \\

One advantage of using the form given in Eq. (\ref{eq_hole_1_5}) is that it allows us to quickly evaluate the quantum numbers of the far-separated quasiparticle-quasihole pair, whose energy I use to define the charge gap. This is done as follows: first note that the number of holes in LLL is $N_{h}=2Q+1-N$ and these holes bind four vortices to form composite fermions which completely fill their L$\Lambda$L $\uparrow$. Therefore, the effective flux seen by the holes $2Q_{h}^{*}$ is given by: $2Q_{h}^{*}+1=N_{h}$. The maximum total orbital angular momentum of the exciton state is thus: $L_{\rm max}=Q_{h}^{*}+(Q_{h}^{*}+1)=N_{h}=2Q+1-N$. \\

\subsection{Fully spin polarized state at $\nu=5/7$}
\label{subsec:5_7_fp}
\begin{figure*}[htpb]
\centering
\subfigure[~Laughlin]{
\includegraphics[width=6cm,height=3.5cm]{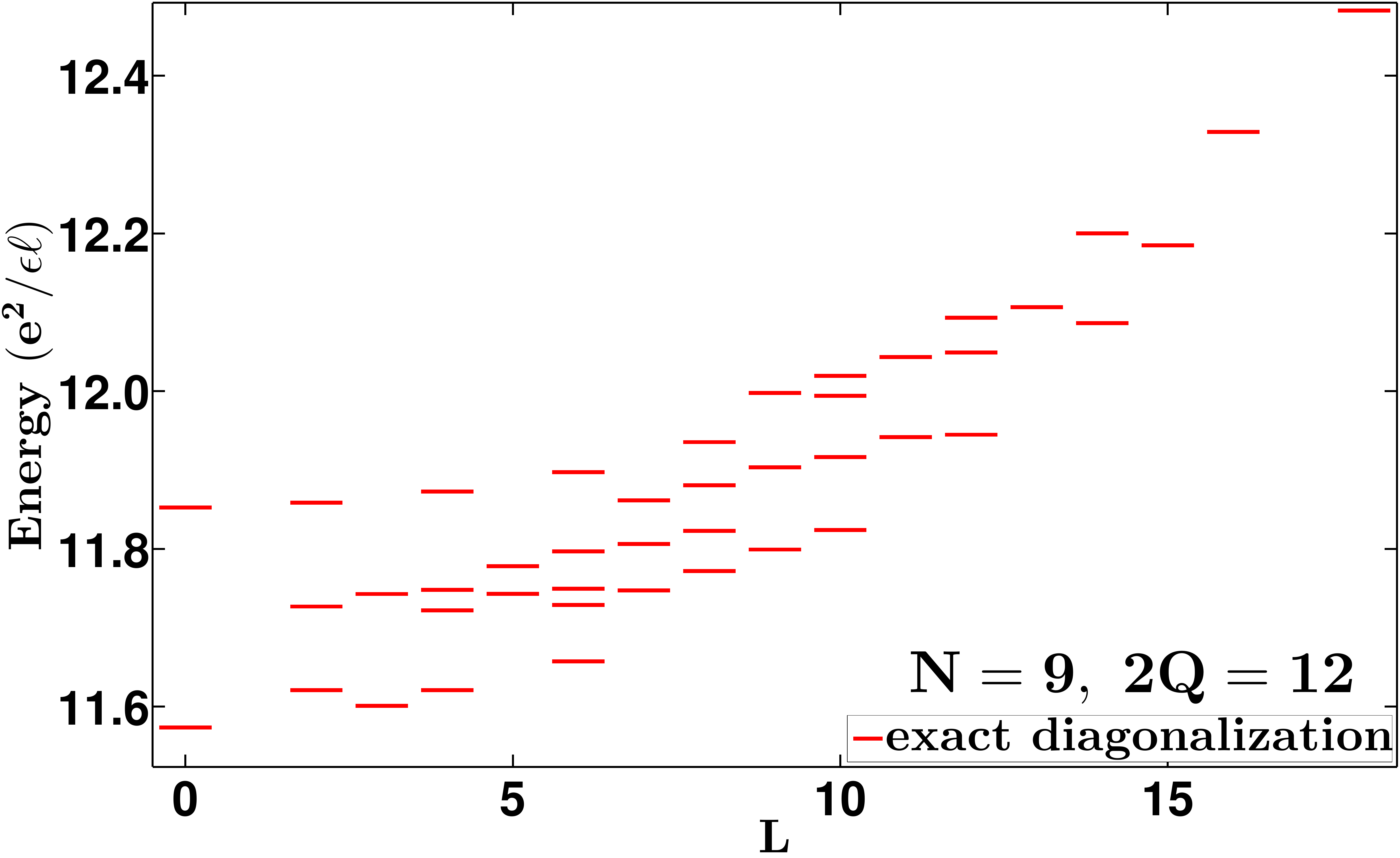}
\includegraphics[width=6cm,height=3.5cm]{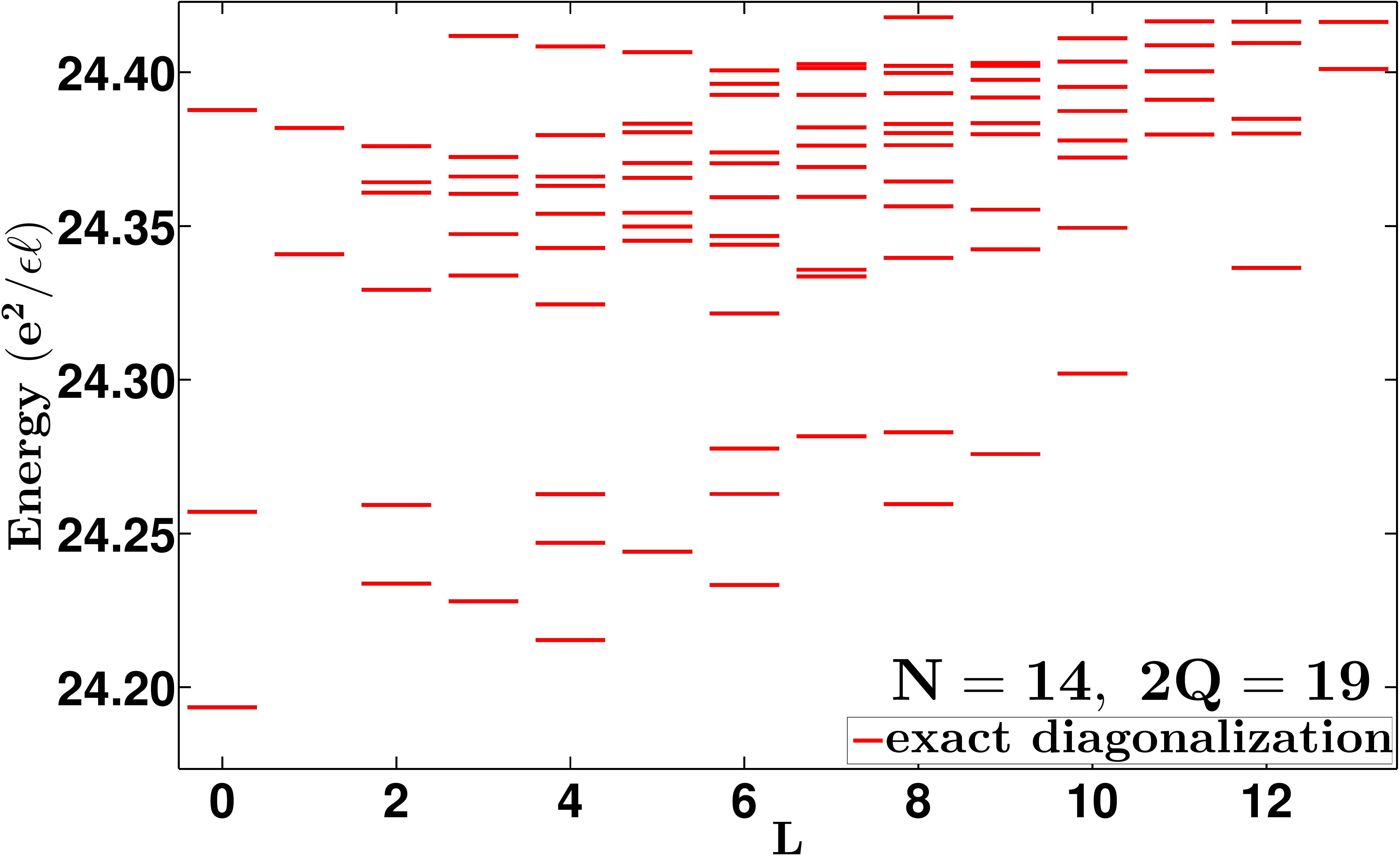}}
\subfigure[~Laughlin]{
\includegraphics[width=6cm,height=3.5cm]{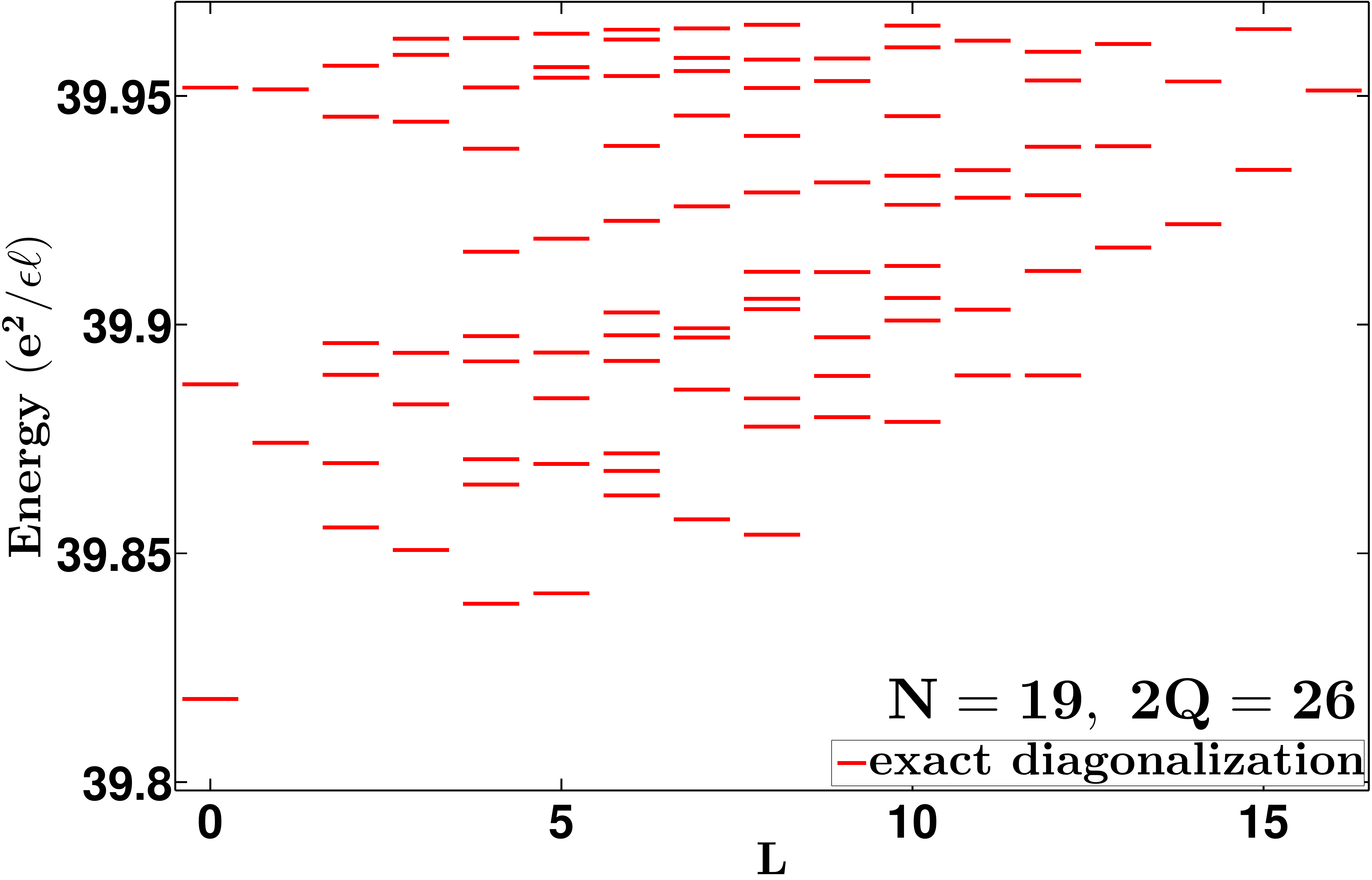}
\includegraphics[width=6cm,height=3.5cm]{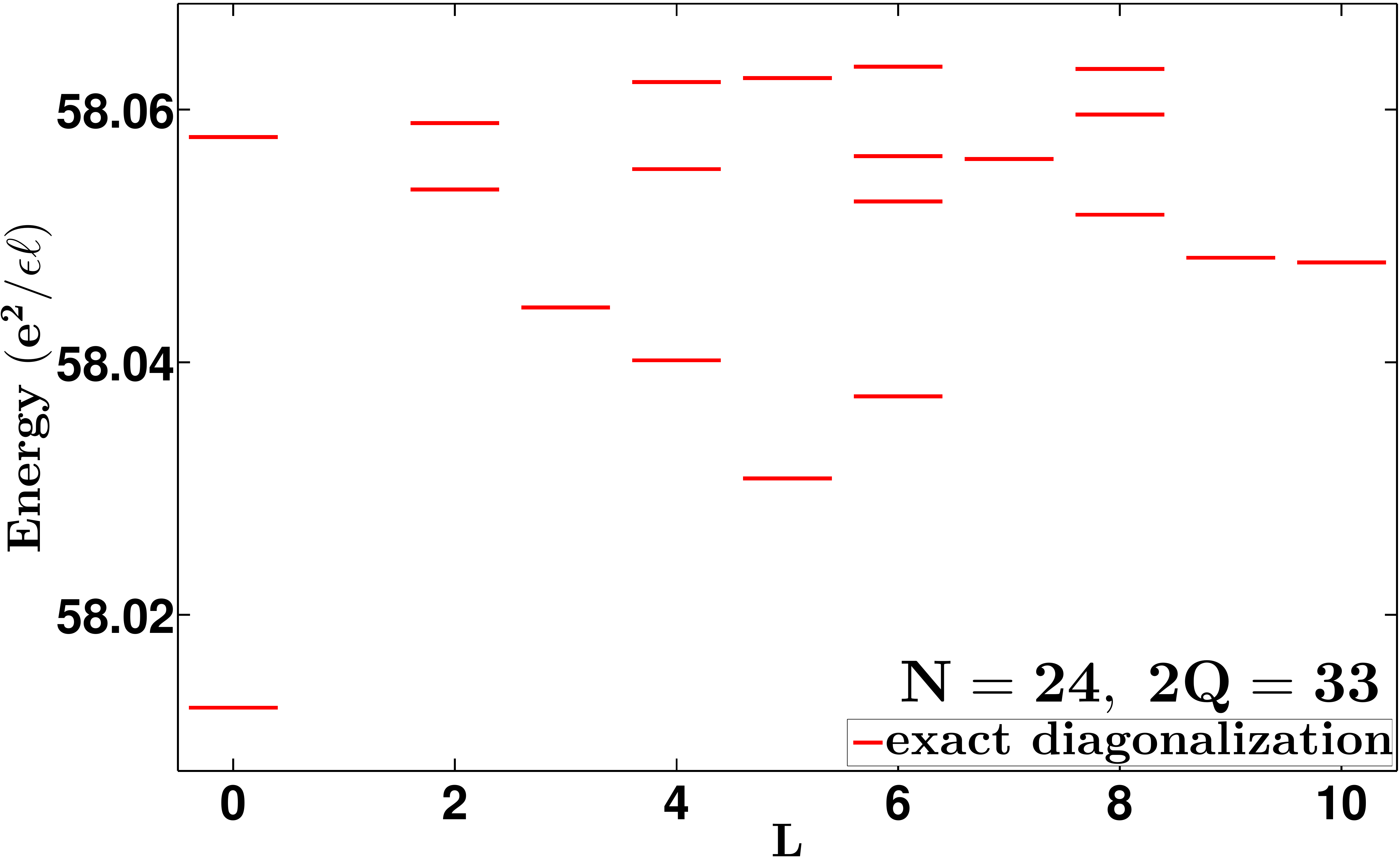}}
\quad
\subfigure[~WYQ]{
\includegraphics[width=6cm,height=3.5cm]{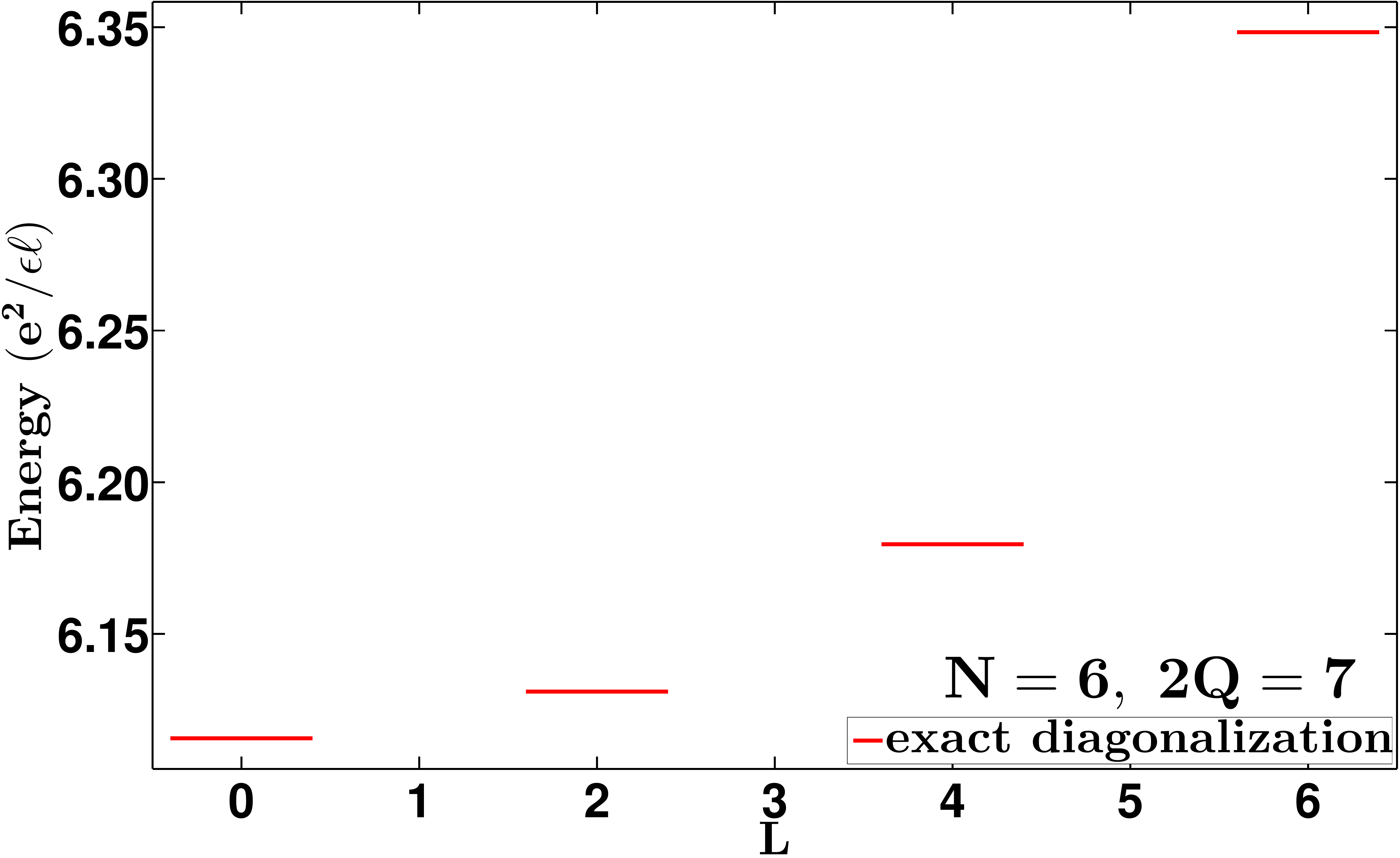}
\includegraphics[width=6cm,height=3.5cm]{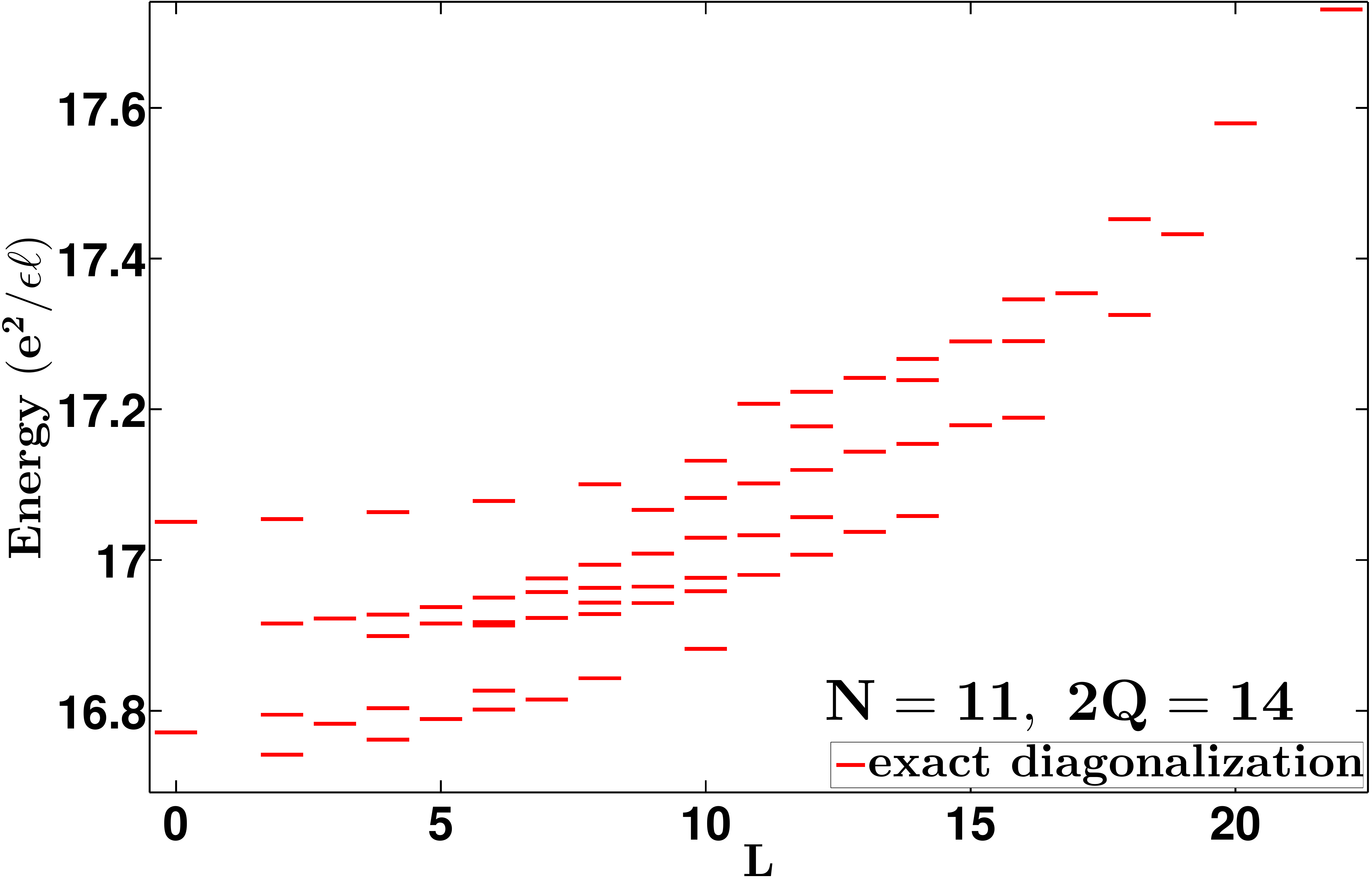}}
\subfigure[~WYQ]{
\includegraphics[width=6cm,height=3.5cm]{N_16_2Q_21.pdf}
\includegraphics[width=6cm,height=3.5cm]{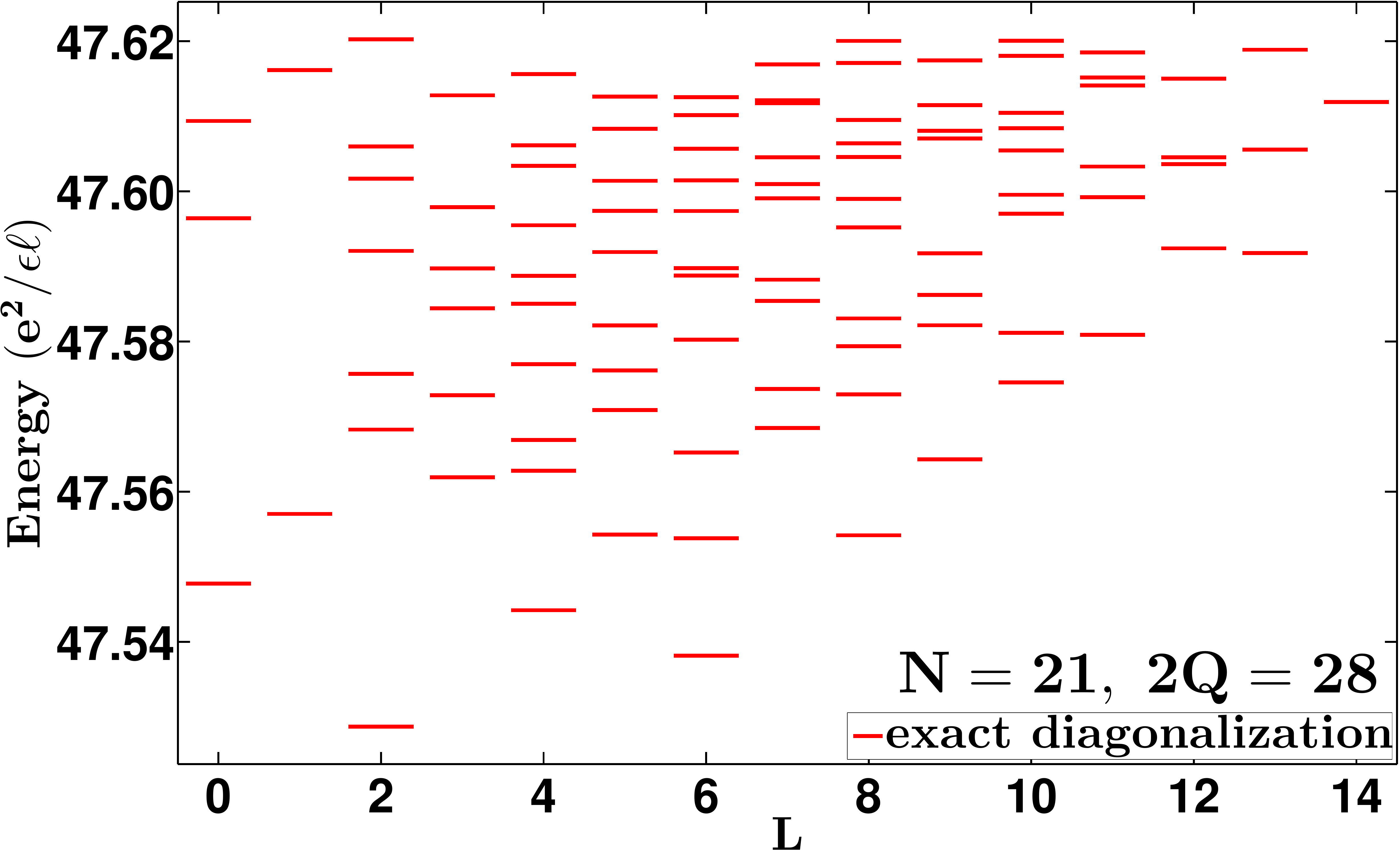}}
\caption{\label{spectra_fp_5_7}Coulomb spectra obtained from exact diagonalization in the spherical geometry for $N$ fully spin polarized electrons at $\nu=5/7$ at the total flux $2Q(hc/e)$  corresponding to the Laughlin (top two panels (a) and (b)) and WYQ state (bottom two panels (c) and (d)).}
\end{figure*}

\begin{figure}[htbp]
\begin{center}
\includegraphics[width=8cm,height=4.5cm]{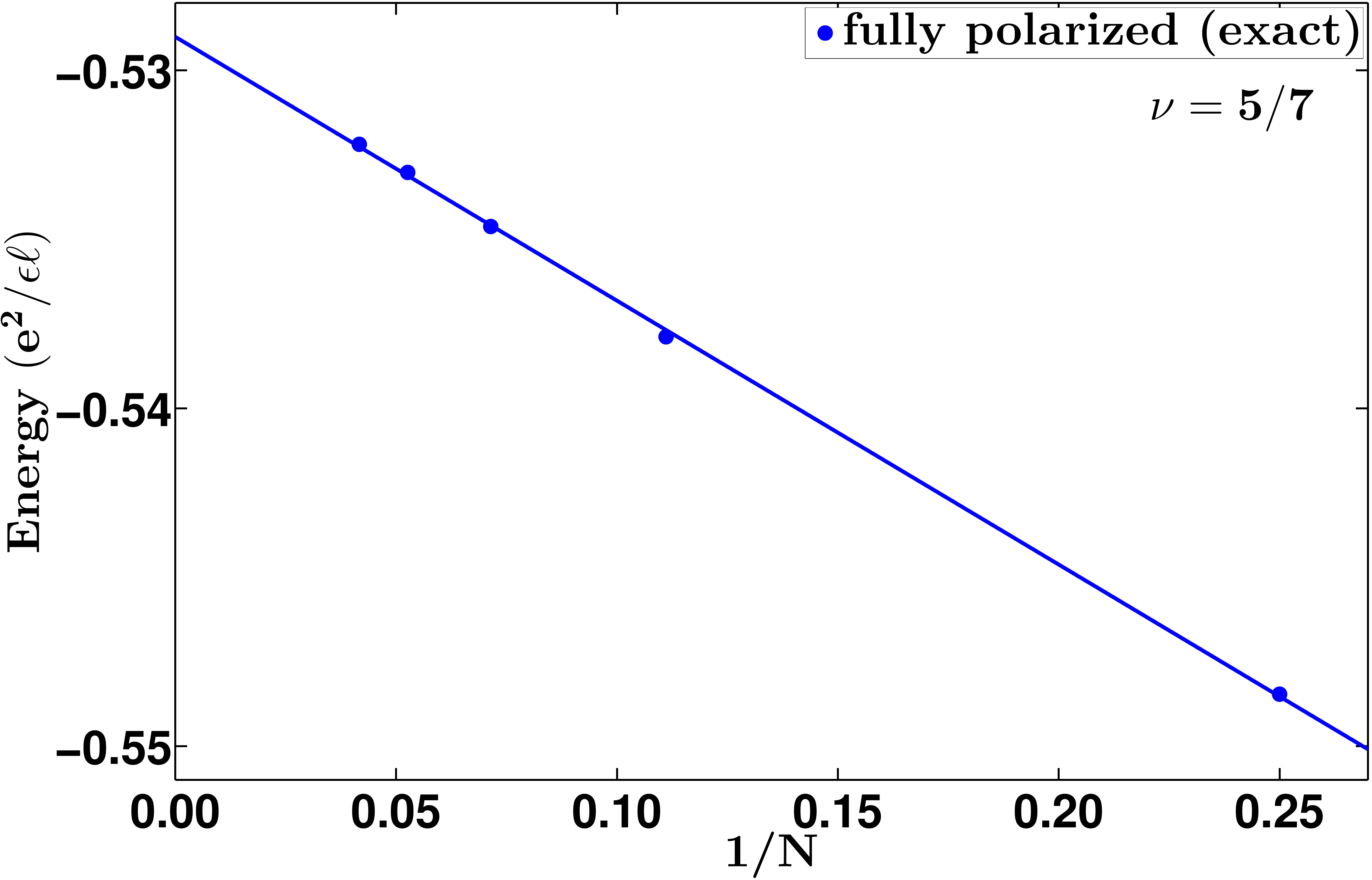}
\end{center}
\caption{\label{extrap_5_7_fp}
Thermodynamic extrapolation of the lowest Landau level Coulomb ground state energy for the fully spin polarized state at $\nu=5/7$.}
\end{figure}

The S$\Lambda$L $\uparrow$ $-$ S$\Lambda$L $\uparrow$ $^{-2}$CFs pseudopotentials shown in Fig. \ref{fig:interaction_CFs} suggests that the ground state at $\nu=5/7$ arises from a $2/3$ Laughlin like state in the S$\Lambda$L $\uparrow$. If this turns out to be the case, it would be in stark contrast to the ground state at the related $\nu=5/13$ which likely supports a $2/3$ WYQ state in the S$\Lambda$L $\uparrow$ \cite{Mukherjee14}. To unambiguously identify the nature of the fully spin polarized $\nu=5/7$ state I consider the following two cases:
\begin{itemize}
 \item The state:
 \begin{equation}
 [1+[2]_{-2}]_{-2} \leftrightarrow(5/7)~:\gamma=1
 \label{eq_CF_5_3_m2}
 \end{equation}
corresponds to $\nu^*=5/3$, which is obtained by filling the L$\Lambda$L $\uparrow$ completely and forming a 2/3 Laughlin state in the S$\Lambda$L $\uparrow$. This state at $\nu=5/7$ occurs at $\mathcal{S}=3/5$ and has $S=N/2$.
 \item The state:
 $$[1+2/3^{\rm WYQ}]_{-2} \leftrightarrow(5/7)~:\gamma=1$$
corresponds to $\nu^*=5/3$, which is obtained by filling the L$\Lambda$L $\uparrow$ completely and forming a 2/3 WYQ state in the S$\Lambda$L $\uparrow$. This state at $\nu=5/7$ occurs at $\mathcal{S}=7/5$ and has $S=N/2$.
\end{itemize}
The Coulomb spectra for both these cases obtained from exact diagonalization is shown in Fig. \ref{spectra_fp_5_7}. We notice that with a 2/3 Laughlin state in the S$\Lambda$L $\uparrow$ the ground state is incompressible for all system sizes while this is not the case for all systems hosting the 2/3 WYQ state in the S$\Lambda$L $\uparrow$. Thus the fully spin polarized 5/7 state is an FQHE state of composite fermions with a filled L$\Lambda$L $\uparrow$ and a conventional 2/3 Laughlin state in the S$\Lambda$L $\uparrow$. Figure \ref{extrap_5_7_fp} shows the thermodynamic extrapolation of the LLL Coulomb ground state energy for the fully spin polarized $5/7$ state. The extrapolated ground state energy obtained is 0.5290(1), which is consistent with previous results \cite{Balram15}. \\

Just as at the fully spin polarized $\nu=4/5$, there exists another plausible candidate state for the fully spin polarized FQHE at $5/7$. This state is 
\begin{equation}
\overline{[2]_{-4}} \leftrightarrow(5/7)~:\gamma=1,
\label{eq_hole_2_7}
\end{equation}
which is the hole conjugate of the 2/7 Jain state. Again this difference is merely artificial. The wave functions for the above two states of Eqs. (\ref{eq_CF_5_3_m2}) and (\ref{eq_hole_2_7}) are completely equivalent to each other. One can use the form of Eq. (\ref{eq_hole_2_7}) to evaluate the quantum number of the far-separated quasiparticle-quasihole pair at 5/7. This is done as follows: first note that the number of holes in LLL is $N_{h}=2Q+1-N$ and these holes bind four vortices to form composite fermions which completely fill their lowest two $\Lambda$Ls namely, L$\Lambda$L $\uparrow$ and S$\Lambda$L $\uparrow$. Therefore, the effective flux seen by the holes $2Q_{h}^{*}$ is given by: $2Q_{h}^{*}+1+2Q_{h}^{*}+3=N_{h}$. The maximum total orbital angular momentum of the exciton state is thus: $L_{\rm max}=(Q_{h}^{*}+1)+(Q_{h}^{*}+2)=(N_{h}-4)/2+3=((2Q+1-N)-4)/2+3$. \\

\section{States at $\boldsymbol{\nu=6/17}$ and $\boldsymbol{\nu=6/7}$}
\label{sec:CF_6_5}

\subsection{$\nu=6/17$ (parent state $\nu^{*}=6/5$)}
The $\nu=6/17$ state is obtained from $\nu^{*}=6/5$ by parallel vortex attachment with two vortices. We have the following three differently spin polarized states at this filling factor:
\subsubsection{Fully spin polarized $6/17$}
\begin{figure*}[htpb]
\centering
\subfigure[~Laughlin]{
\includegraphics[width=6cm,height=3.5cm]{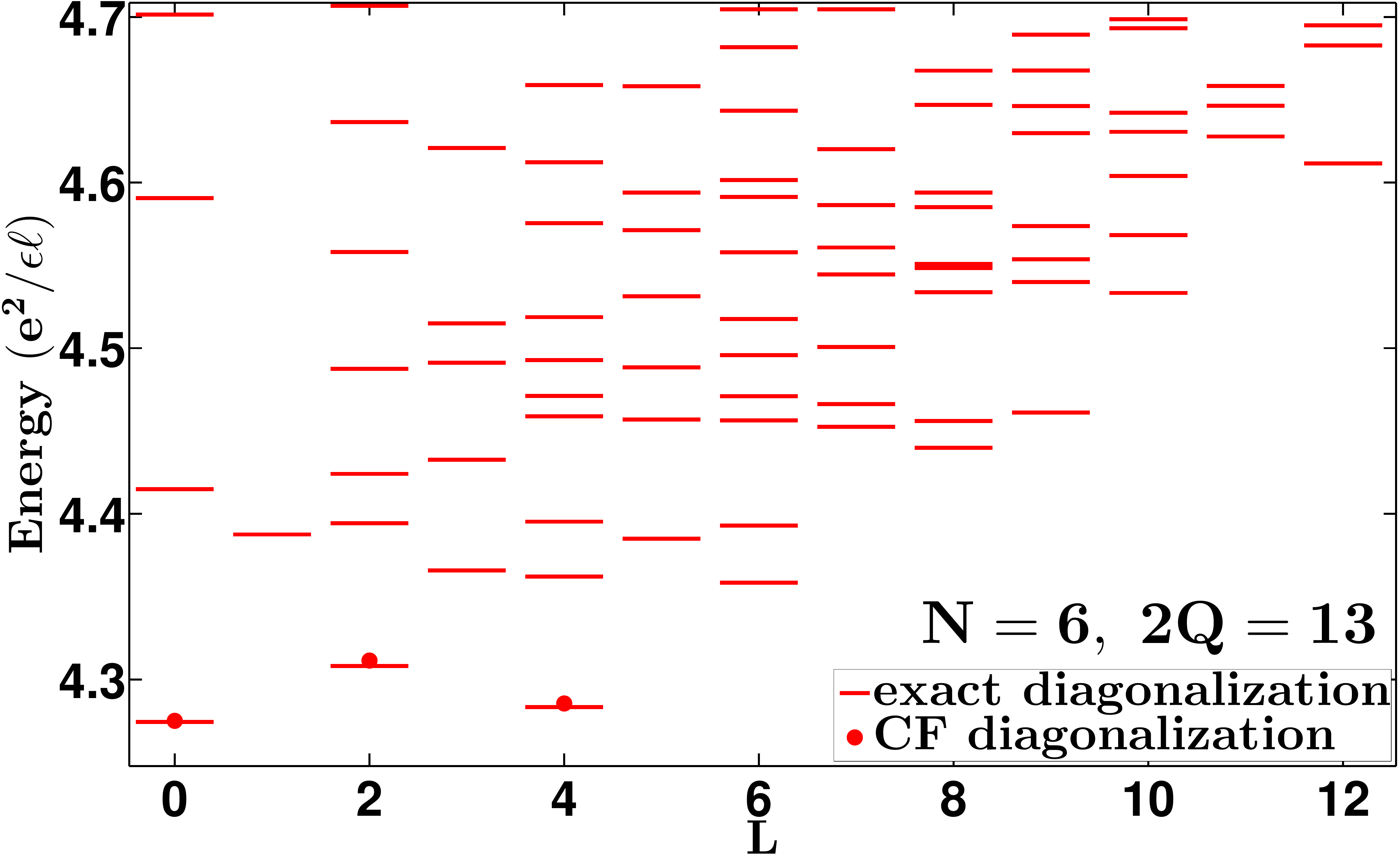}
\includegraphics[width=6cm,height=3.5cm]{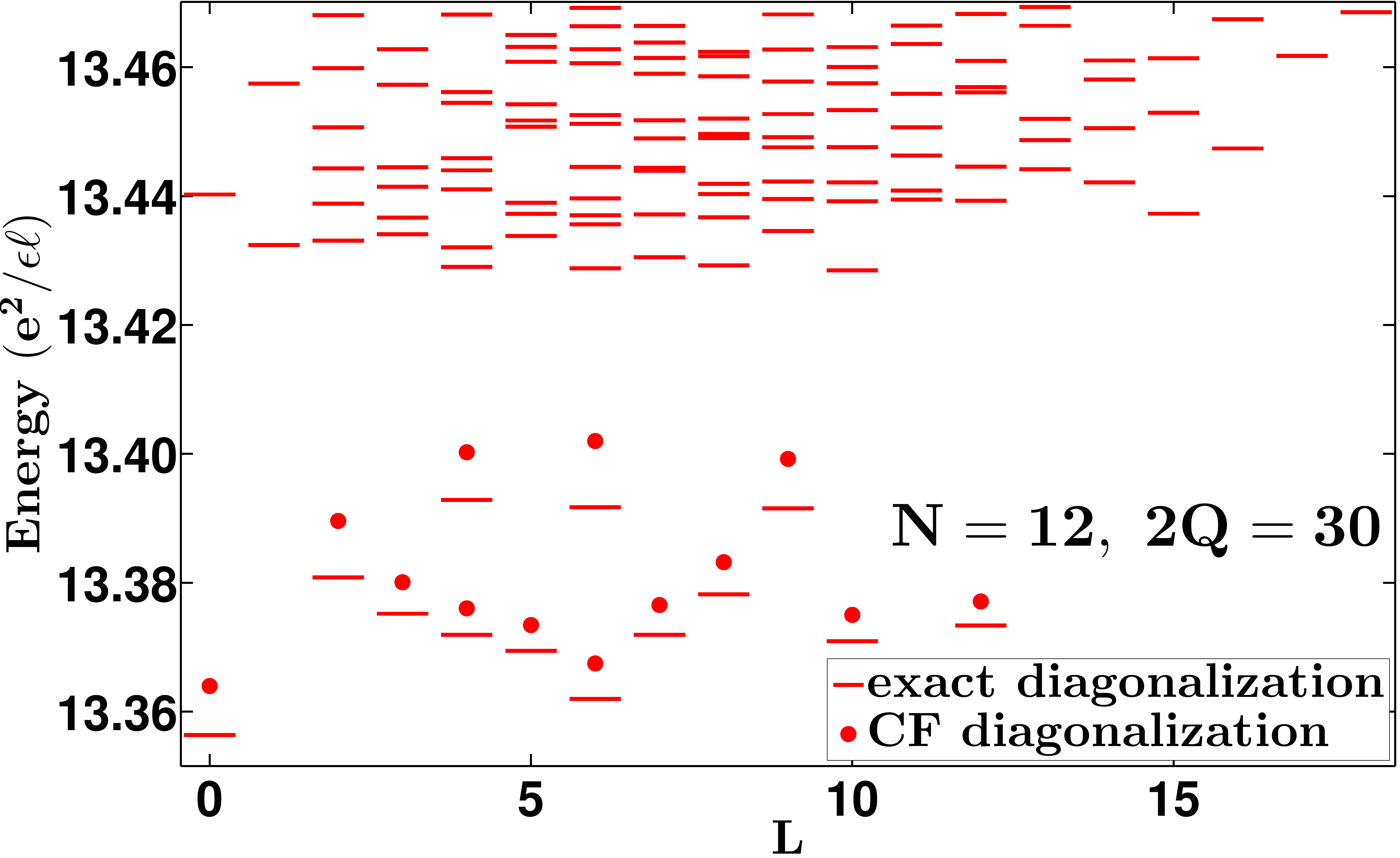}
\includegraphics[width=6cm,height=3.5cm]{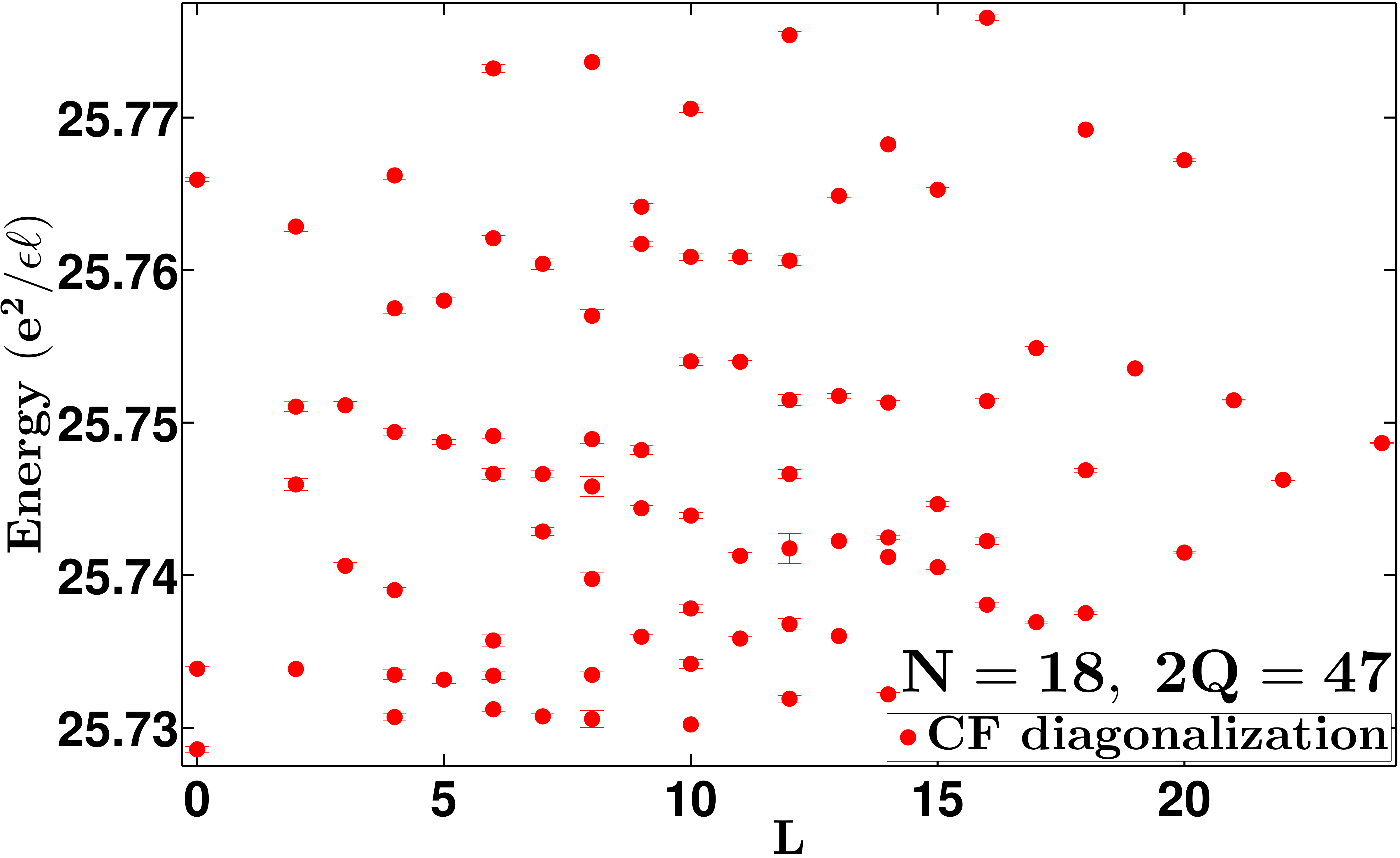}}
\quad
\subfigure[~WYQ]{
\includegraphics[width=6cm,height=3.5cm]{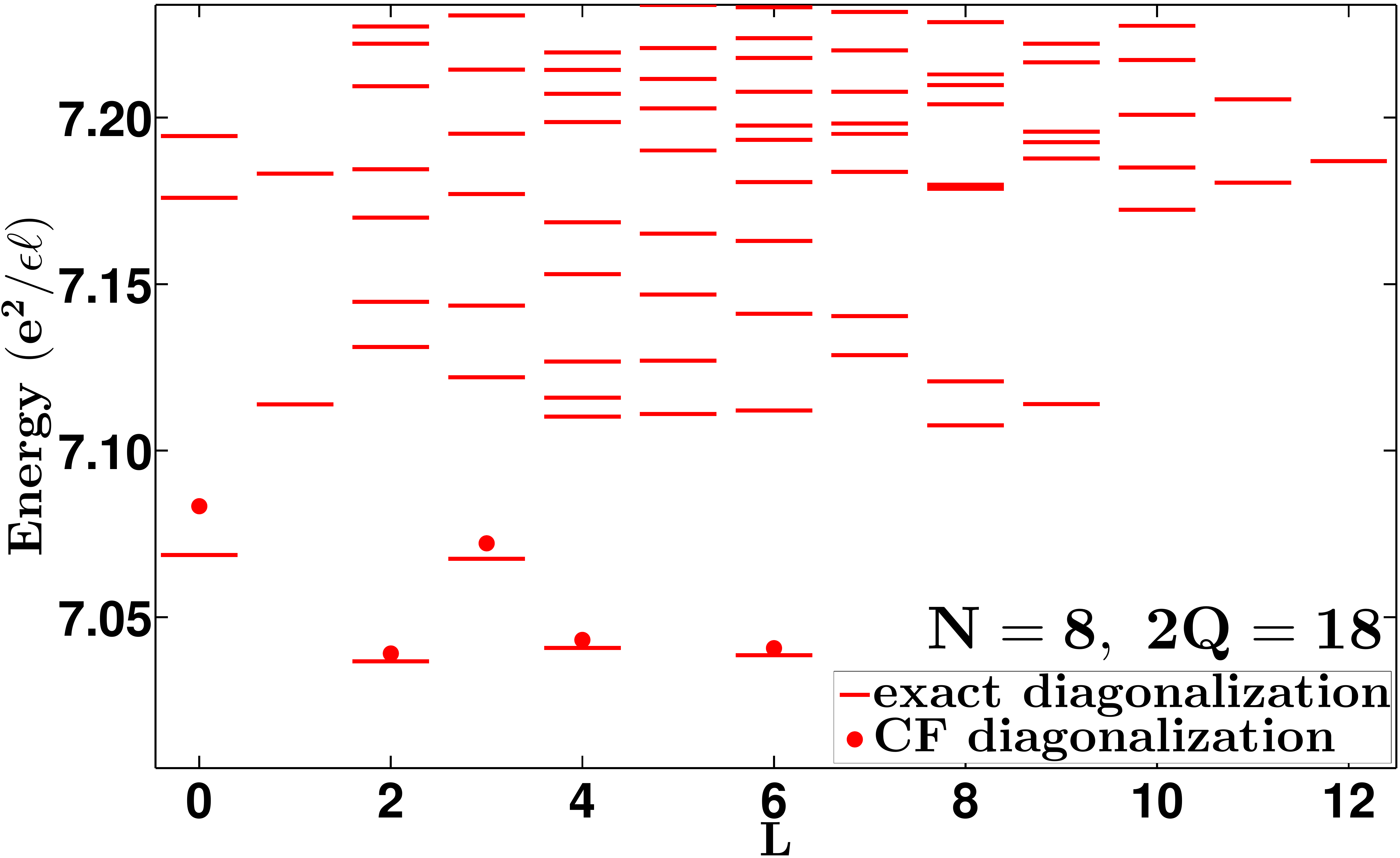}
\includegraphics[width=6cm,height=3.5cm]{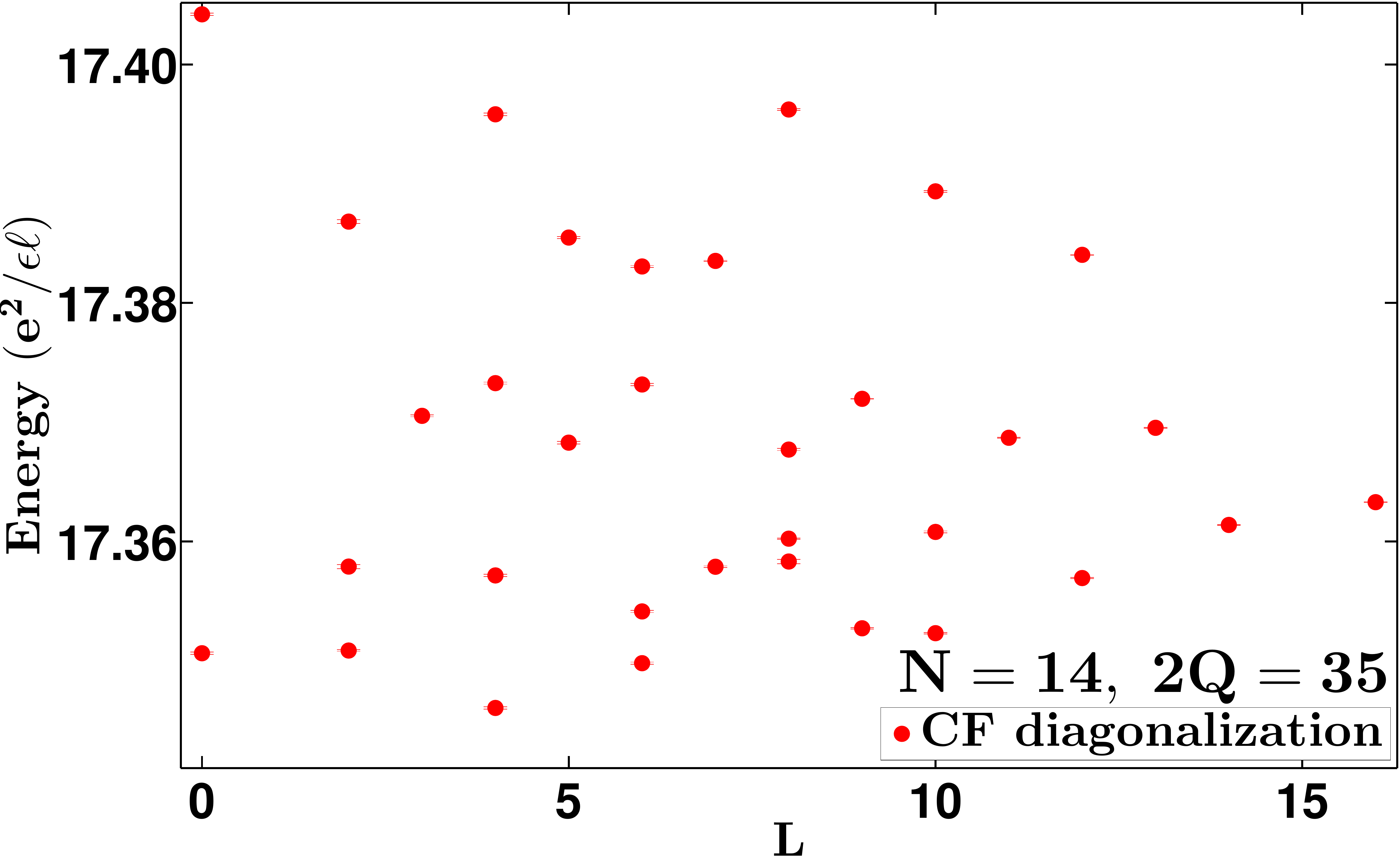}
\includegraphics[width=6cm,height=3.5cm]{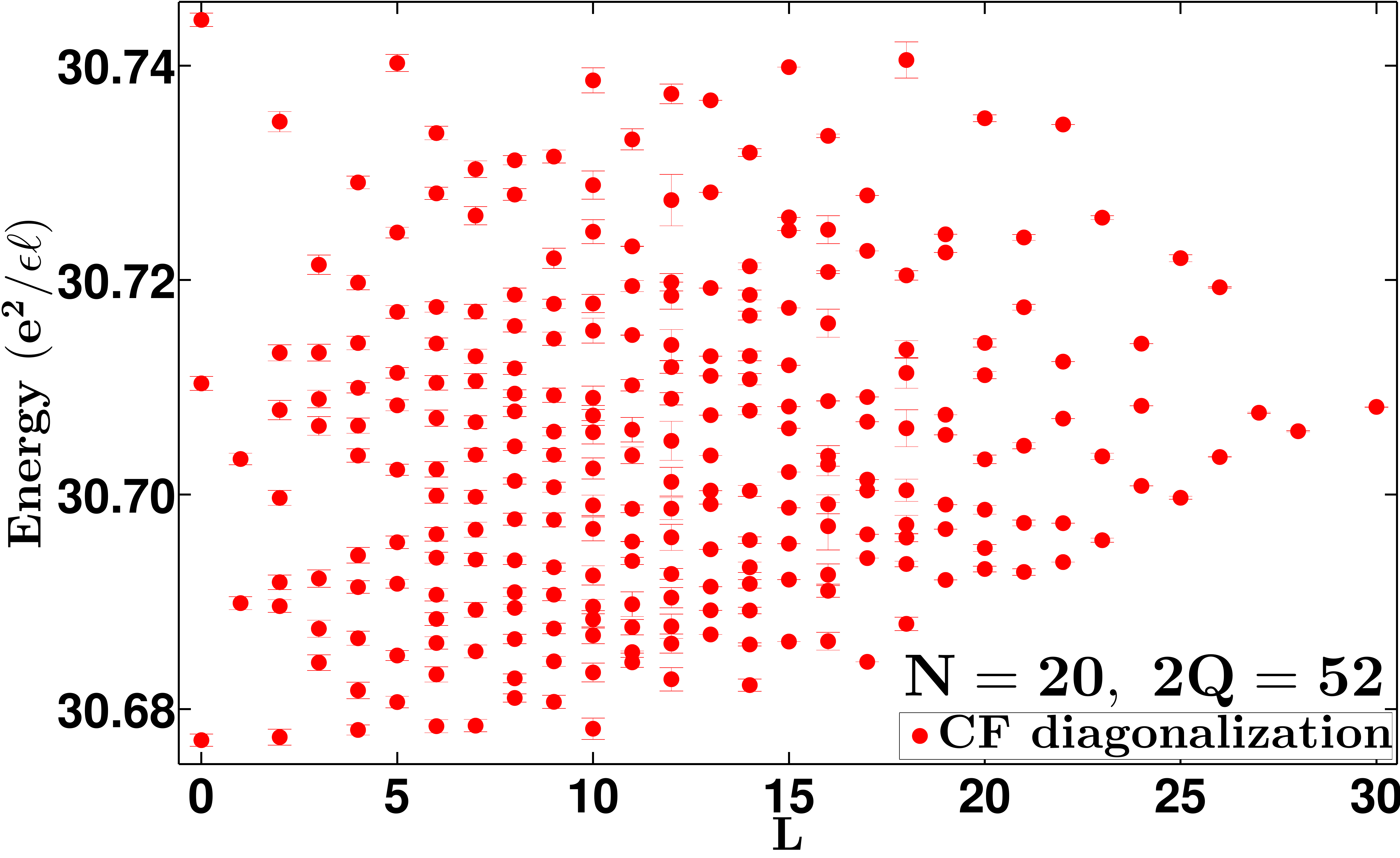}}
\caption{\label{spectra_fp_6_17}Coulomb spectra in the spherical geometry for $N$ fully spin polarized electrons at total flux $2Q(hc/e)$ at $\nu=6/17$. The red dashes (dots) show the energies obtained from exact (CF) diagonalization at the flux corresponding to the Laughlin (top panel (a)) and WYQ state (bottom panel (b)).}
\end{figure*}

For the fully spin polarized 6/17 state two possibilities arise:
\begin{itemize}
 \item The state:
 $$[1+[1]_{4}]_2 \leftrightarrow(6/17)~:\gamma=1$$
corresponds to $\nu^*=6/5$, which is obtained by filling the L$\Lambda$L $\uparrow$ completely and forming a 1/5 Laughlin state in the S$\Lambda$L $\uparrow$. This state at $\nu=6/17$ occurs at $\mathcal{S}=4$ and has $S=N/2$.
 \item The state:
 $$[1+1/5^{\rm WYQ}]_2 \leftrightarrow(6/17)~:\gamma=1$$
corresponds to $\nu^*=6/5$, which is obtained by filling the L$\Lambda$L $\uparrow$ completely and forming a 1/5 WYQ state in the S$\Lambda$L $\uparrow$. This state at $\nu=6/17$ occurs at $\mathcal{S}=14/3$ and has $S=N/2$.
\end{itemize}
The Coulomb spectra for both these cases obtained from exact and CF diagonalization is shown in Fig. \ref{spectra_fp_6_17}. We find that when there is a 1/5 Laughlin state in the S$\Lambda$L $\uparrow$ the ground state is incompressible for all system sizes while this is not the case for all systems hosting the 1/5 WYQ state in the S$\Lambda$L $\uparrow$. The largest system size for which I could carry out CF diagonalization at $\nu=6/17$ at the corresponding WYQ flux is $N=20$. Within error bars this system has an $L=0$ ground state, but here the gap is quite small of the order of $0.0003$ $e^2/\epsilon\ell$ ($\epsilon$ is the dielectric constant of the background host material and $\ell=\sqrt{\hbar c/eB}$ is the magnetic length), which indicates that it is unlikely to be stabilized. Thus the fully spin polarized 6/17 state is an FQHE state of composite fermions with a filled L$\Lambda$L $\uparrow$ and a conventional 1/5 Laughlin state in the S$\Lambda$L $\uparrow$. This supports the assertion made in Ref. \cite{Wojs04} who argued that the fully spin polarized $\nu=6/17$ might arise from a standard $1/5$ Laughlin state in the S$\Lambda$L $\uparrow$.\\

I mention here that Jolicoeur \cite{Jolicoeur07} has constructed a candidate state of the form:
\begin{equation}
 \Psi_{\nu=\frac{k}{(2p+1)k\pm2}}=\mathcal{P}_{\text{LLL}} \prod_{i<j}(z_{i}-z_{j})^{2p+1} \Phi^{\text{RR}}_{\nu^{*}=\pm k/2}e^{-\frac{1}{4\ell^2}\sum_{i}|z_{i}|^2}
\end{equation}
where $z_{i}$ is the planar coordinate of the $i$th electron and $\Phi^{\text{RR}}_{\nu^{*}=k/2}$ ($\Phi^{\text{RR}}_{\nu^{*}=-k/2}=[\Phi^{\text{RR}}_{\nu^{*}=k/2}]^{*}$) is the bosonic Read-Rezayi wave function with clusters of $k$ particles \cite{Read99,Rezayi09}:
\begin{equation}
\Phi^{\text{RR}}_{\nu^{*}=k/2}=\mathcal{S}[\prod_{i_{1}<j_{1}}(z_{i_{1}}-z_{j_{1}})^2\cdots \prod_{i_{k}<j_{k}}(z_{i_{k}}-z_{j_{k}})^2],
\end{equation}
where $\mathcal{S}$ is the symmetrization over all partitions of $N$ particles into subsets of $N/k$ particles. A candidate state of this kind at $\nu=6/17$ state is obtained with the value $k=12$, $p=1$ and $-$ sign in the above equation. This state at $\nu=6/17$ occurs at $\mathcal{S}=1$ and has $S=N/2$. Implementing the LLL projection \cite{Girvin84b,Dev92a,Jain07} by bringing all the factors of $\bar{z}$'s to the left of $z$'s in each term and replacing $\bar{z}$ by $2\partial_{z}\equiv 2\partial/\partial z$ (which acts on all terms except the Gaussian factor), we can write down the wave function as:
\begin{equation}
 \Psi_{\nu=\frac{k}{(2p+1)k-2}}=e^{-\frac{1}{4\ell^2}\sum_{i}|z_{i}|^2}\Phi^{\text{RR}}_{\nu^{*}=k/2}(\{\partial_{z_{i}}\}) \prod_{i<j}(z_{i}-z_{j})^{2p+1} 
\end{equation}
(analogous wave functions on the sphere can be obtained by making the transformations: $z_{i}-z_{j}\Rightarrow u_{i}v_{j}-u_{j}v_{i},~~~\partial_{z_{i}}-\partial_{z_{j}}\Rightarrow \partial_{u_{i}}\partial_{v_{j}}- \partial_{v_{i}}\partial_{u_{j}}$)
Firstly this wave function is not readily amenable to evaluation as it involves taking a large number of derivatives of the Jastrow factor. Secondly CF states are known to be remarkably stable for the Coulomb interaction in the LLL, so it is not anticipated that a state with such a high value of $k$ would compete with it. Hence I have not pursued this candidate state further.\\

\subsubsection{Partially spin polarized $6/17$}
\begin{figure}[htpb]
\begin{center}
\includegraphics[width=8cm,height=4.5cm]{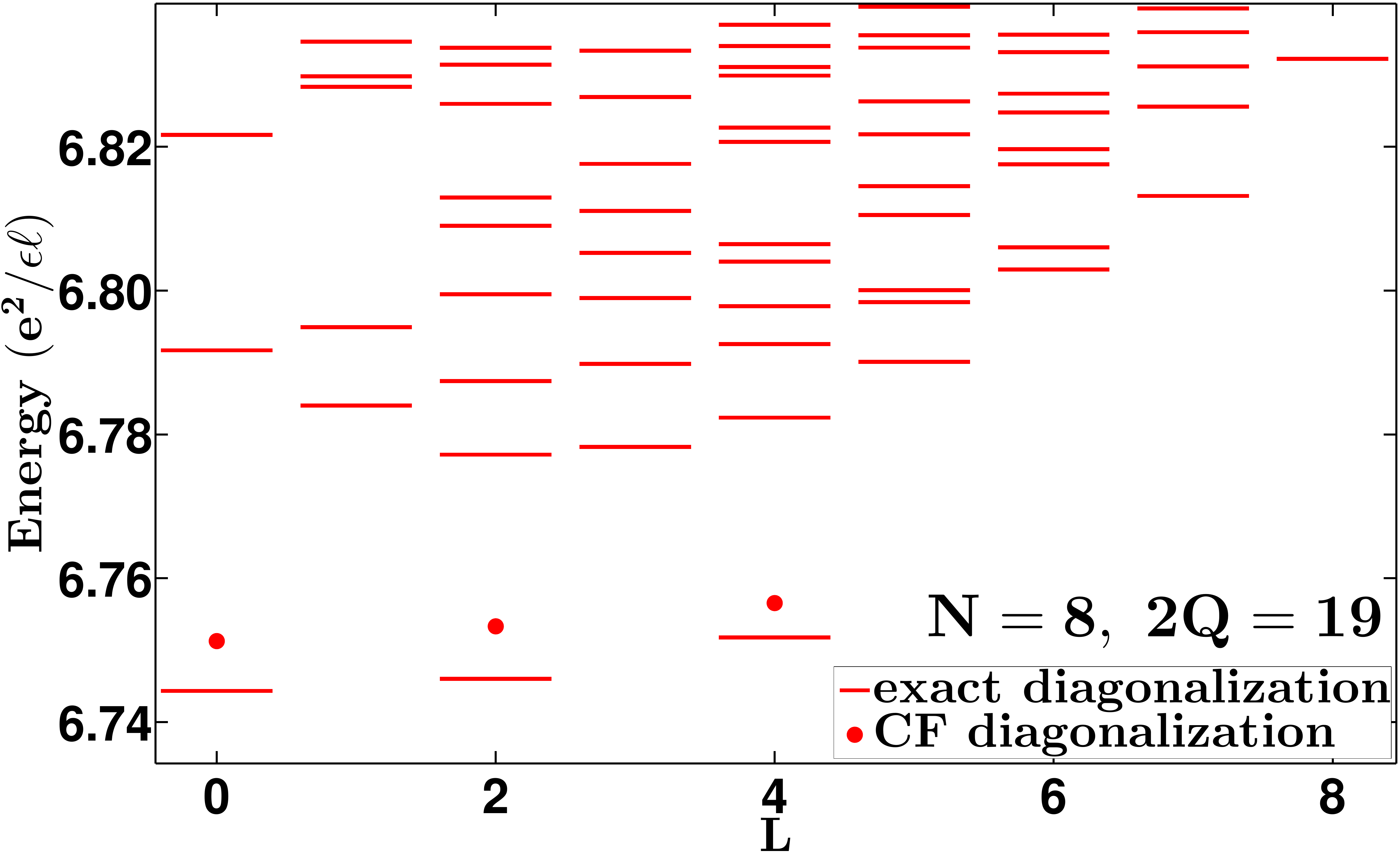}
\includegraphics[width=8cm,height=4.5cm]{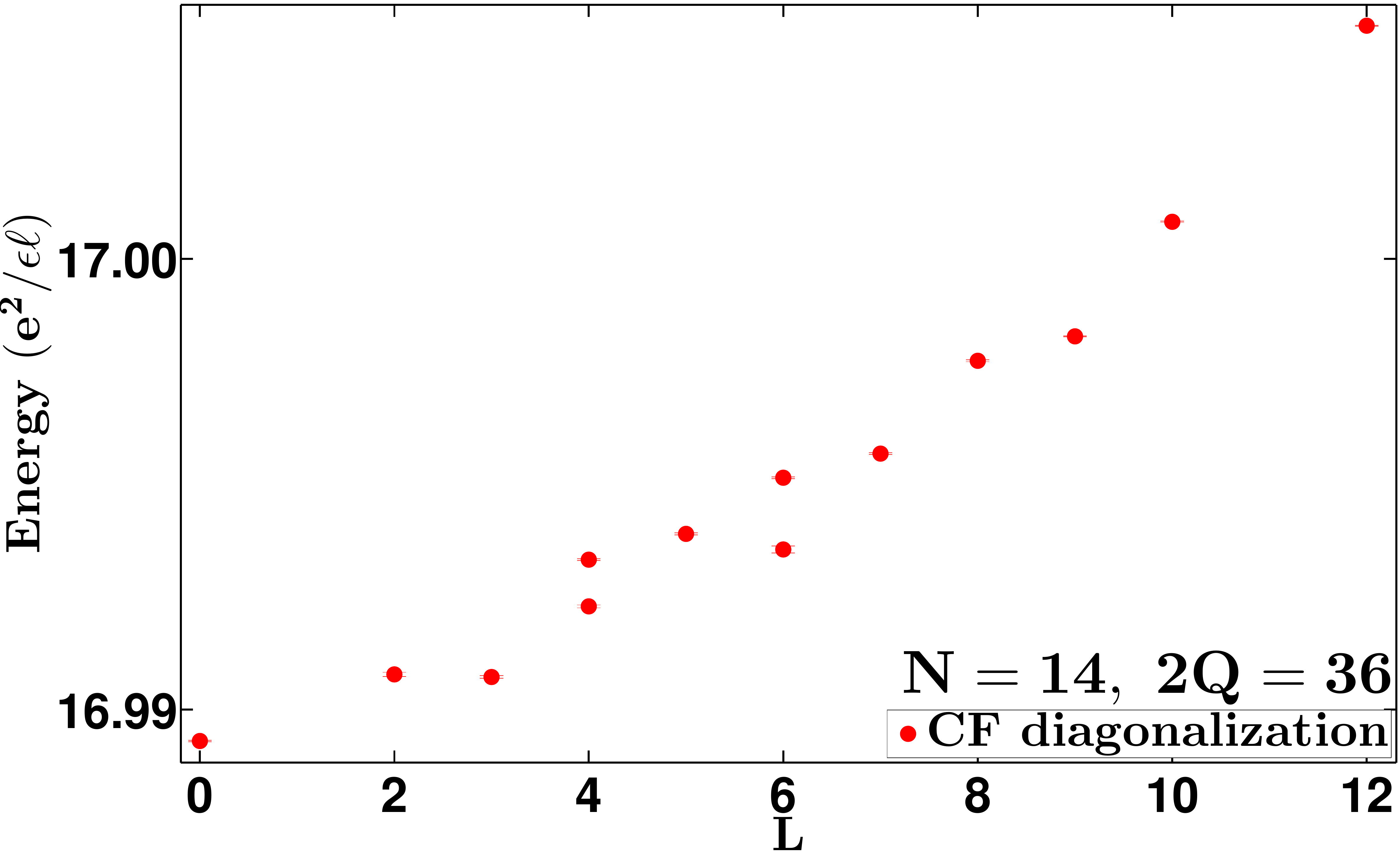}
\includegraphics[width=8cm,height=4.5cm]{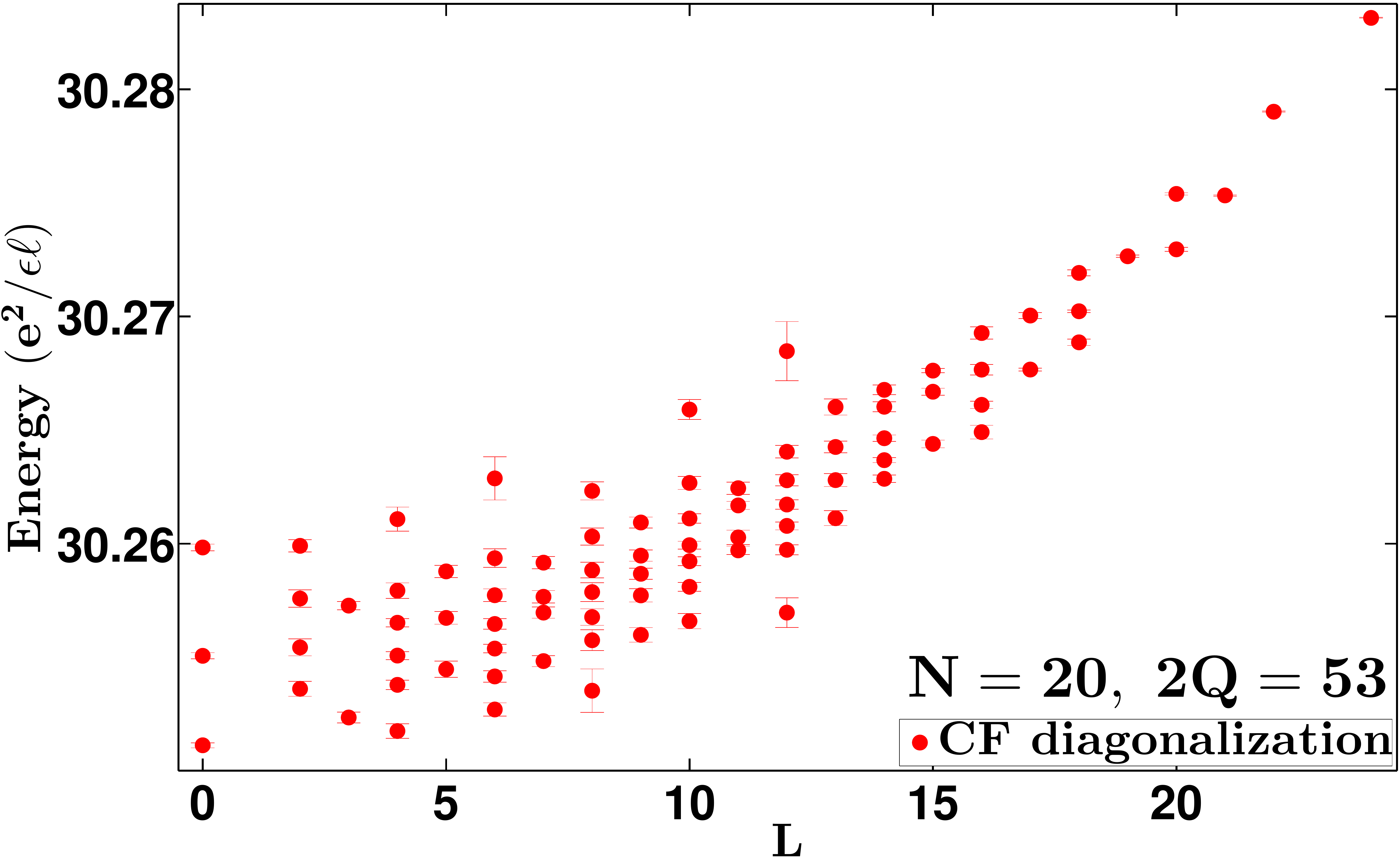}
\end{center}
\caption{\label{spectra_pp_6_17}Coulomb spectra in the spherical geometry for $N$ electrons at flux $2Q(hc/e)$ with total spin $S$ corresponding to the partially spin polarized state at $\nu=6/17$. The red dashes (dots) show the energies obtained from exact (CF) diagonalization.}
\end{figure}

The partially spin polarized 6/17 state 
$$[1,[1]_4]_{2} \leftrightarrow (5/17,1/17)~:\gamma=\frac{2}{3}$$
is obtained by composite-fermionizing the partially spin polarized 6/5 state: 
$$[1,[1]_4]\leftrightarrow(1,1/5)~:\gamma=\frac{2}{3}$$.
It corresponds to filling the L$\Lambda$L $\uparrow$ completely and forming a Lauglin 1/5 state in the L$\Lambda$L $\downarrow$. This state at $\nu=6/17$ occurs at $\mathcal{S}=11/3$ and has $S=(N-2)/3$. Fig. \ref{spectra_pp_6_17} shows the Coulomb spectra obtained from exact and CF diagonalization for this state.

\subsubsection{Spin singlet $6/17$}
\begin{figure}[htpb]
\begin{center}
\includegraphics[width=8cm,height=4.5cm]{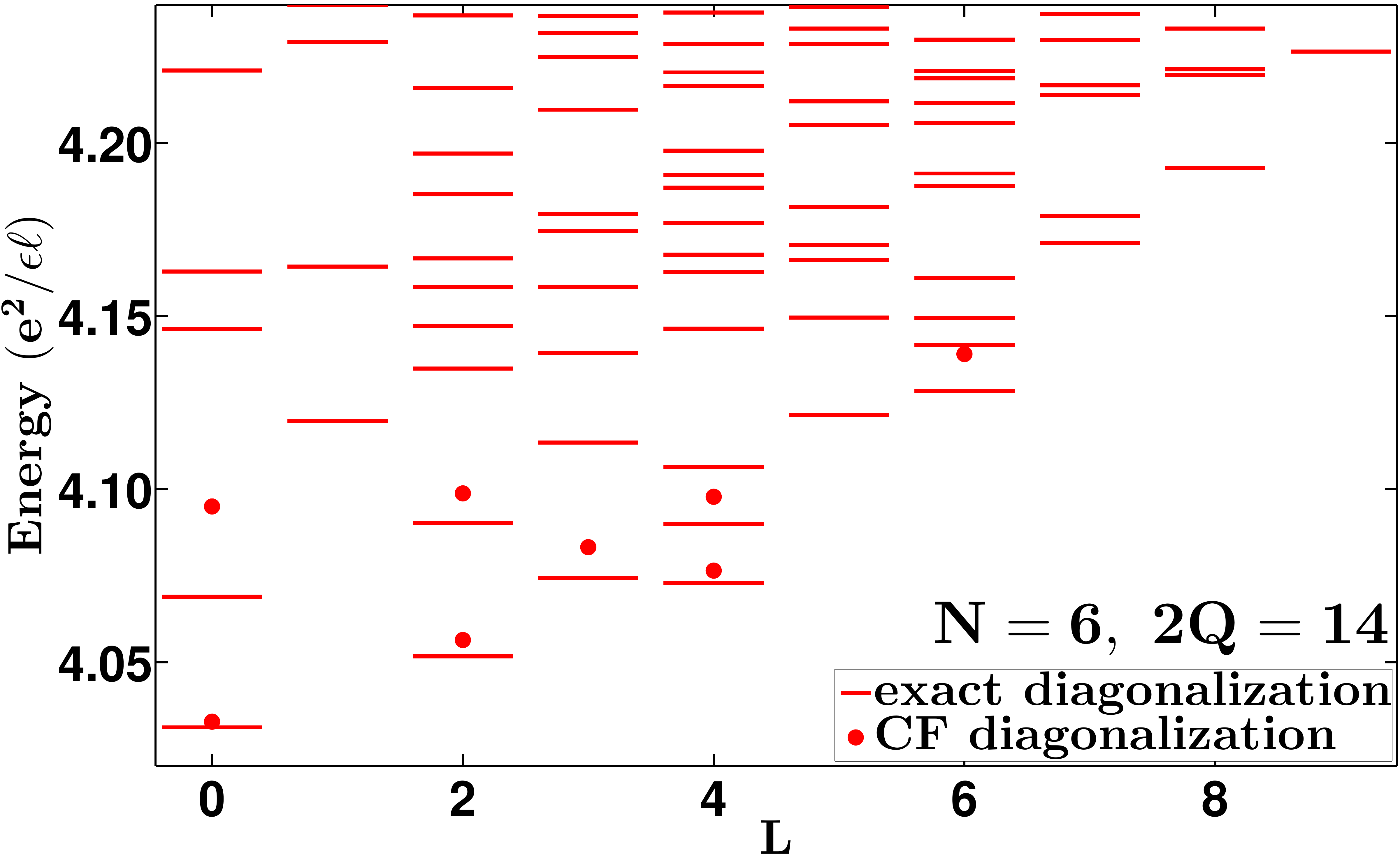}
\end{center}
\caption{\label{spectra_ss_6_17}Coulomb spectra in the spherical geometry for $N$ electrons with total spin $S=0$ at flux $2Q(hc/e)$ at $\nu=6/17$. The red dashes (dots) show the energies obtained from exact (CF) diagonalization.}
\end{figure}

The spin-singlet 6/17 state is
$$[\overline{[\overline{[1,1]_{-2}}]_{-2}}]_{2} \leftrightarrow (3/17,3/17)~:\gamma=0$$
is obtained from the 4/5 spin singlet state \cite{Balram15}
$$[\overline{[1,1]_{-2}}]_{-2} \leftrightarrow (2/5,2/5)~:\gamma=0$$
by taking its particle hole conjugate to produce a spin singlet state at 6/5 of the form $(3/5, 3/5)$, and then composite-fermionizing it. Note that $(3/5, 3/5)$ is {\em not} a direct product of two one-component 3/5 states in the two spin sectors. This state at $\nu=6/17$ occurs at $\mathcal{S}=3$ and has $S=0$. Figure \ref{spectra_ss_6_17} shows the Coulomb spectra for the smallest and only system amenable to exact and CF diagonalization. For the next system I have only been able to calculate the ground state energy using CFD in the $L=0$ and $S=0$ sector. \\

\begin{figure}[htpb]
\begin{center}
\includegraphics[width=8cm,height=4.5cm]{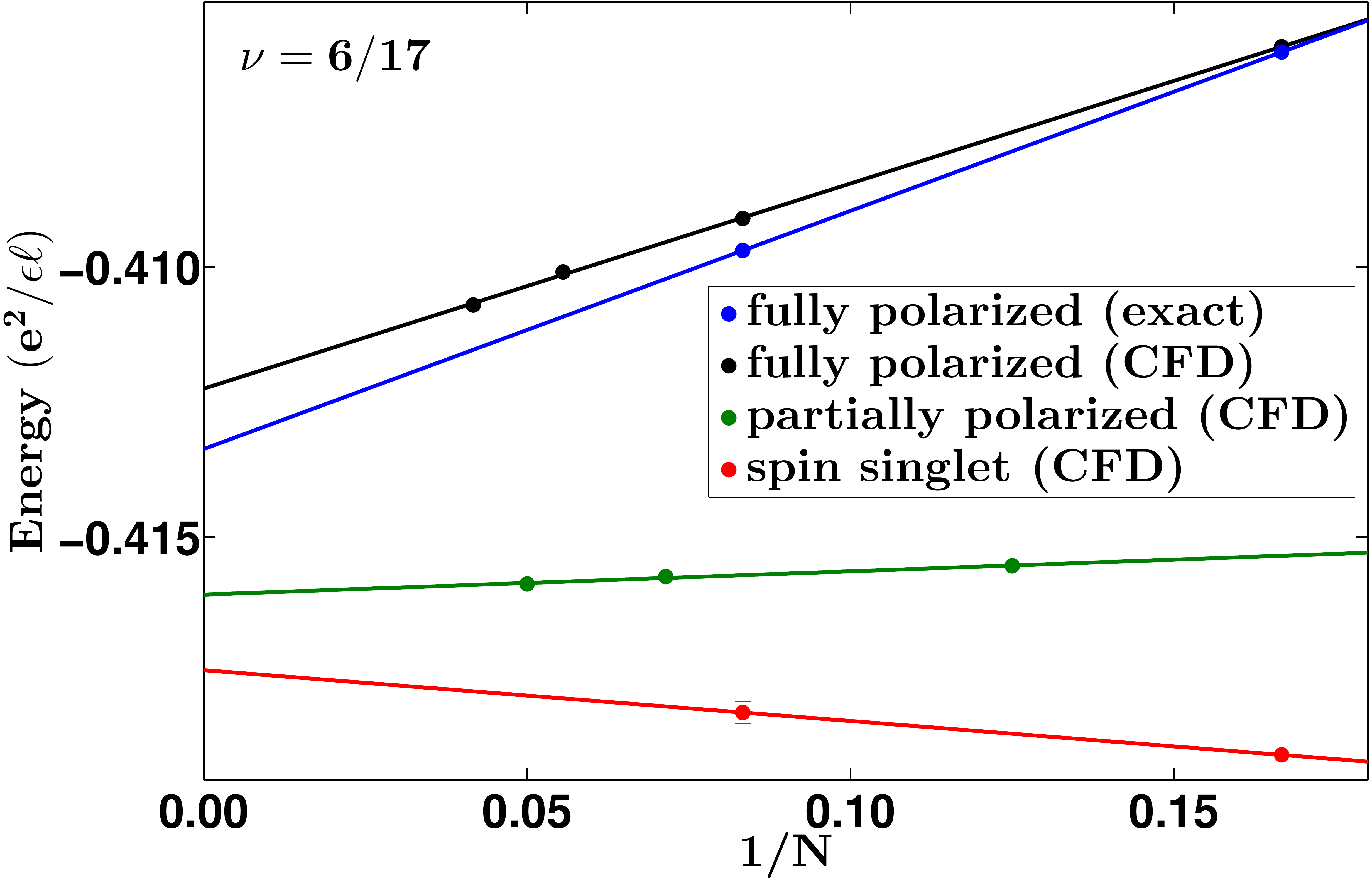}
\end{center}
\caption{\label{extrap_6_17}
Thermodynamic extrapolation of the lowest Landau level Coulomb ground state energy for different spin polarized states at $\nu=6/17$.}
\end{figure}

Figure \ref{extrap_6_17} shows the thermodynamic extrapolation of the LLL Coulomb ground state energies for these states. Table \ref{tab:6_5} shows the thermodynamic energies of these states as well as the critical Zeeman energies for the spin transitions among these states.

\begin{table*}[htpb]
\begin{center}
\begin{tabular}{|c|c|c|c|c|c|c|c|c|c|c|}
\hline
\multicolumn{1}{|c|}{$\nu$} & \multicolumn{2}{|c|}{fully spin polarized} & \multicolumn{2}{|c|}{partially spin polarized} & \multicolumn{2}{|c|}{spin singlet}& \multicolumn{2}{|c|}{$\kappa_{1}$}& \multicolumn{2}{|c|}{$\kappa_{2}$}		\\ \hline
			    & exact		    & CFD 	    & exact		     & 	CFD    		& 	exact		& CFD& 	exact		& CFD& 	exact		& CFD			\\ \hline
6/17 & \underbar{-0.41337(0)}& -0.41228(6)& -           	     &-0.41607(5)       & -	       		&\underbar{-0.41747(0)}&-&0.0228(7)&-&0.0042(1) \\ \hline
6/7  & -0.56827(26)& -	  & \underbar{-0.57502(0)} &-		        & \underbar{-0.58030(0)}&		       &0.0404(15)&-&0.0159(0)&- \\ \hline
\end{tabular}
\end{center}
\caption{The lowest Landau level Coulomb energies (in units of $e^{2}/\epsilon\ell$) in the thermodynamic limit for the fully spin polarized, partially spin polarized and spin-singlet states at $\nu=6/17$ and $\nu=6/7$ obtained using exact and composite fermion diagonalization (CFD). Also shown are the two critical Zeeman energies (in dimensionless units $\kappa=E_{Z}/(e^{2}/\epsilon\ell)$, where $E_{Z}$ is the Zeeman energy) for spin transitions between the fully spin polarized and partially spin polarized states ($\kappa_1$) and between the partially spin polarized and spin singlet states ($\kappa_2$). For $E_{Z}>\kappa~e^{2}/\epsilon\ell$, the state with higher spin polarization is favored over the one with lower spin polarization. An underbar indicates that the thermodynamic extrapolation was done using only two systems. The numbers in the parenthesis are the error bars in the intercept obtained from a linear extrapolation of the finite system results as a function of $1/N$. } 
\label{tab:6_5} 
\end{table*}

\subsection{$\nu=6/7$ (parent state $\nu^{*}=6/5$)}
The $\nu=6/7$ state is obtained from $\nu^{*}=6/5$ by reverse vortex attachment with two vortices. Since the composite fermion wave functions projected in the manner as stated in Sec. \ref{sec:sphere_CF} are not very accurate when doing reverse vortex attachment \cite{Balram15a}, I only consider results obtained from exact diagonalization in this subsection. For $6/7$ we have the following three states:
\subsubsection{Fully spin polarized $6/7$}
\begin{figure*}[htpb]
\centering
\subfigure[~Laughlin]{
\includegraphics[width=6cm,height=3.5cm]{N_6_2Q_7.pdf}
\includegraphics[width=6cm,height=3.5cm]{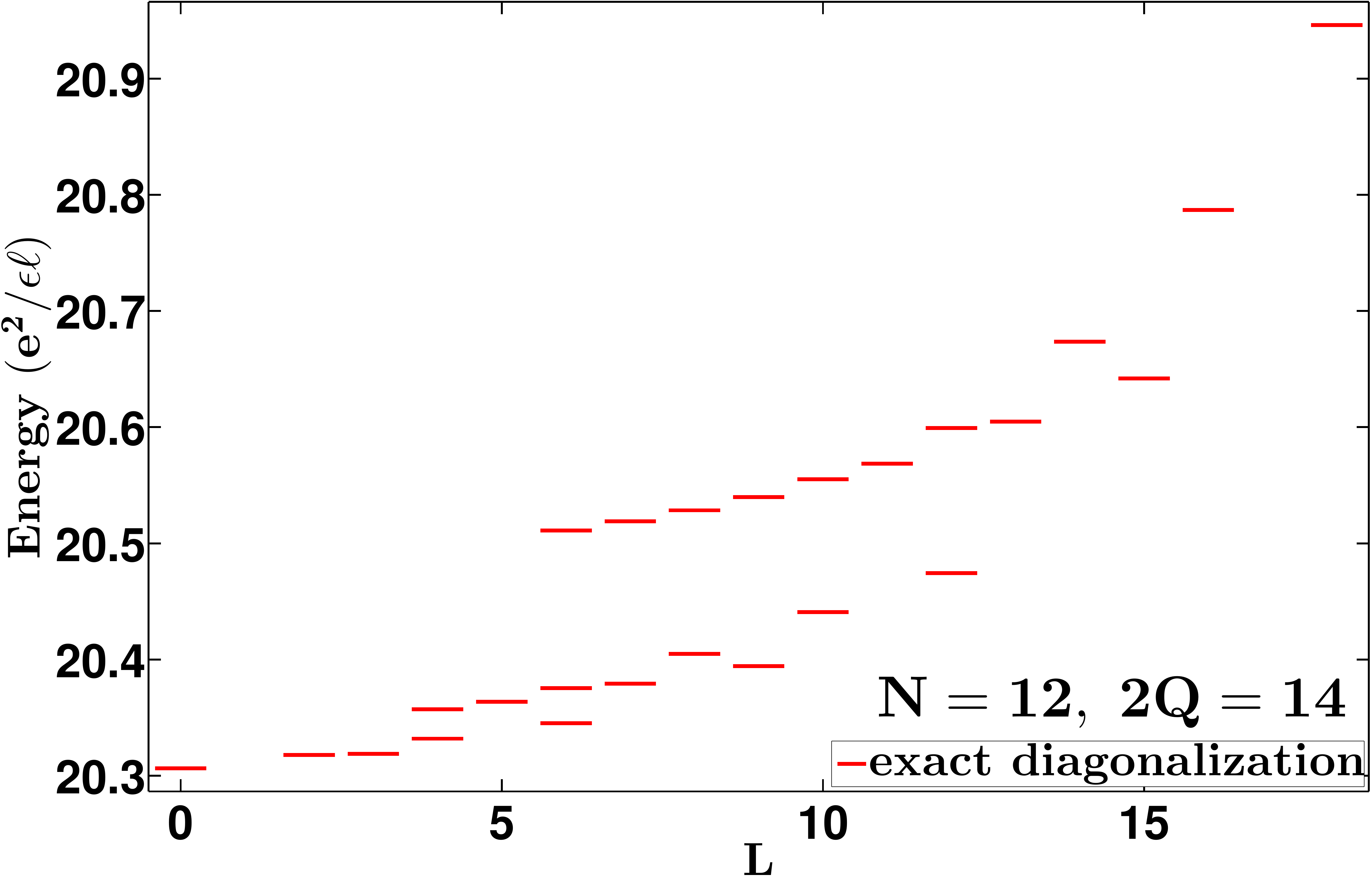}
\includegraphics[width=6cm,height=3.5cm]{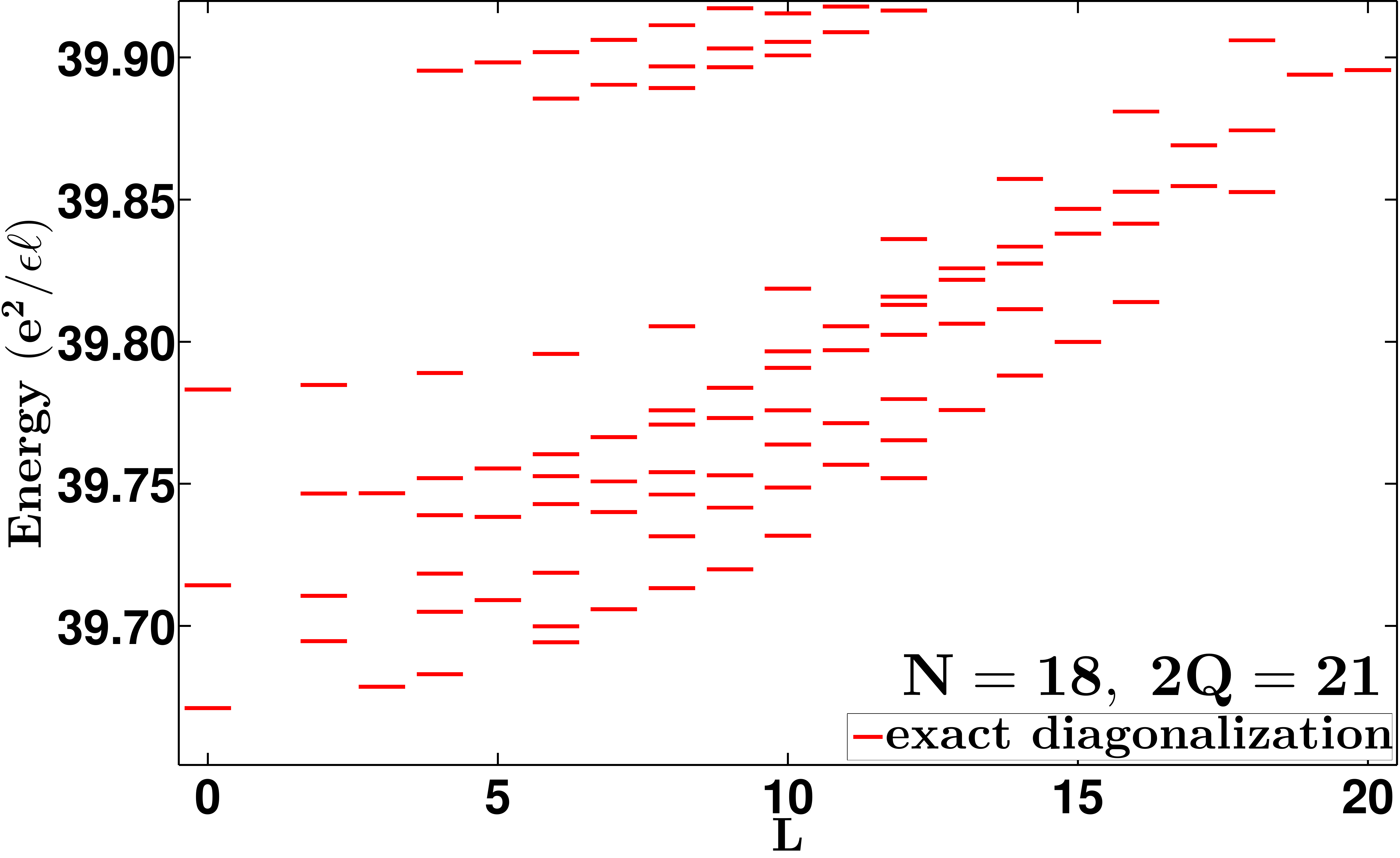}}
\subfigure[~Laughlin]{
\includegraphics[width=6cm,height=3.5cm]{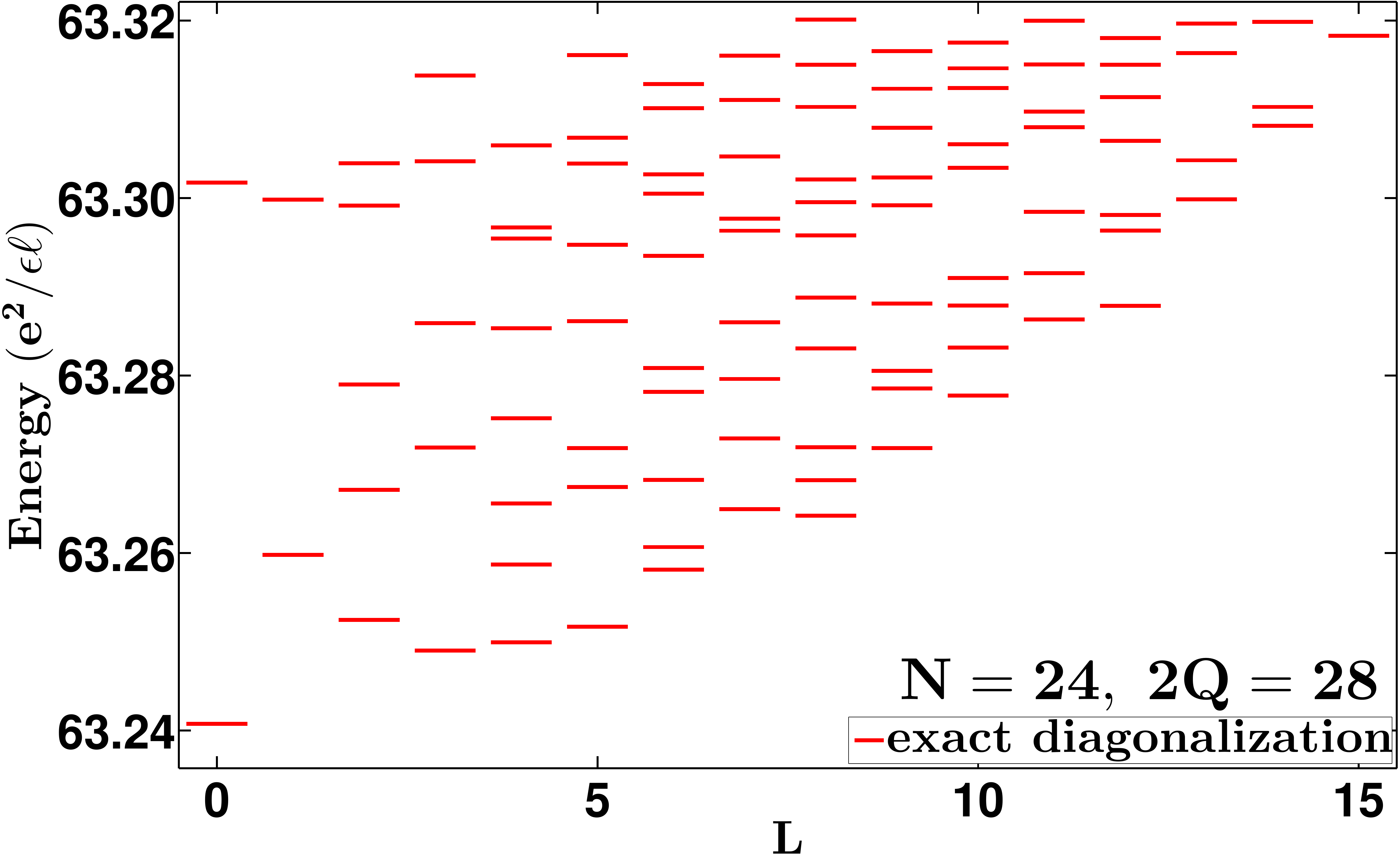}
\includegraphics[width=6cm,height=3.5cm]{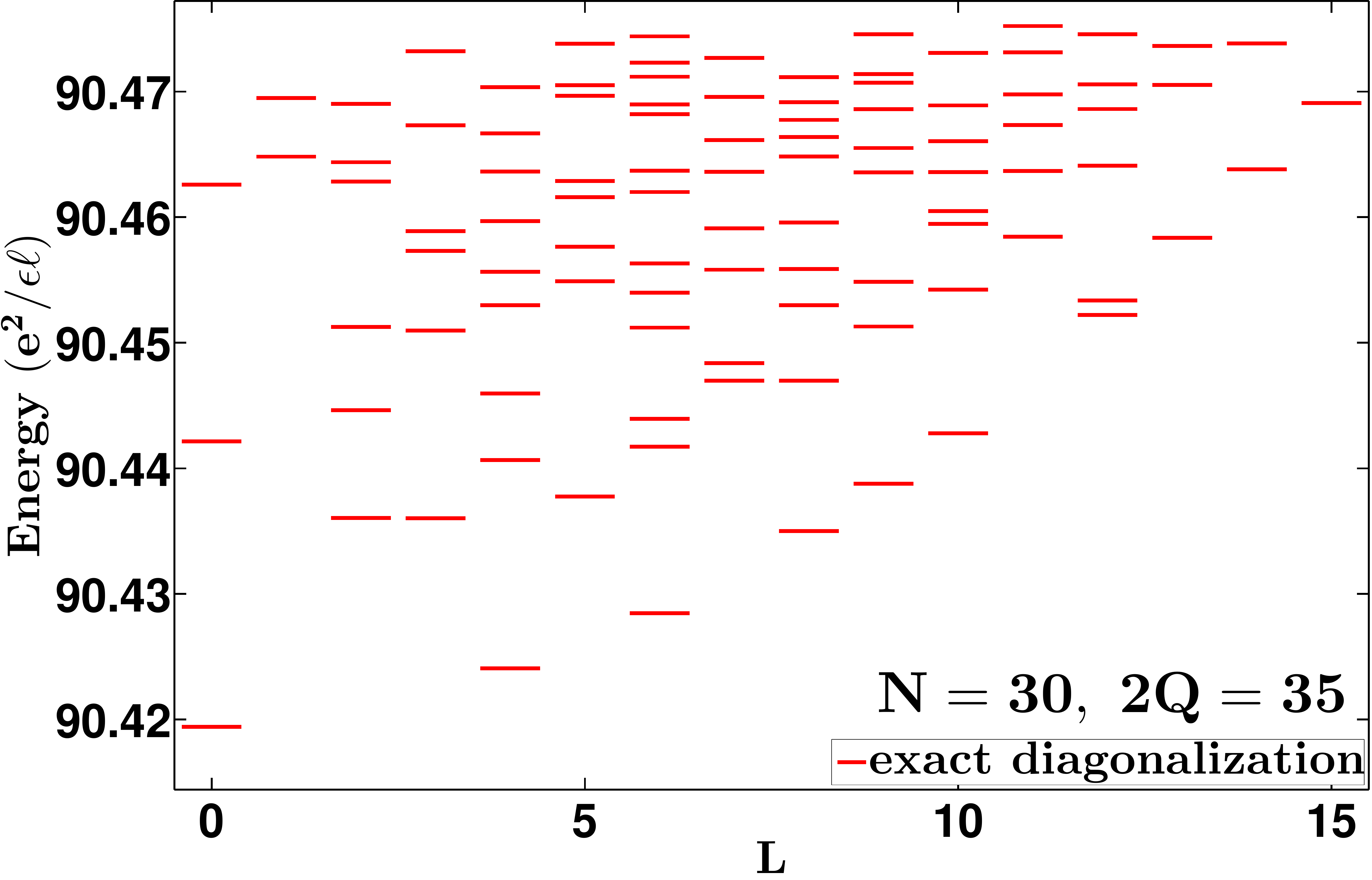}}
\quad
\subfigure[~WYQ]{
\includegraphics[width=6cm,height=3.5cm]{N_8_2Q_10.pdf}
\includegraphics[width=6cm,height=3.5cm]{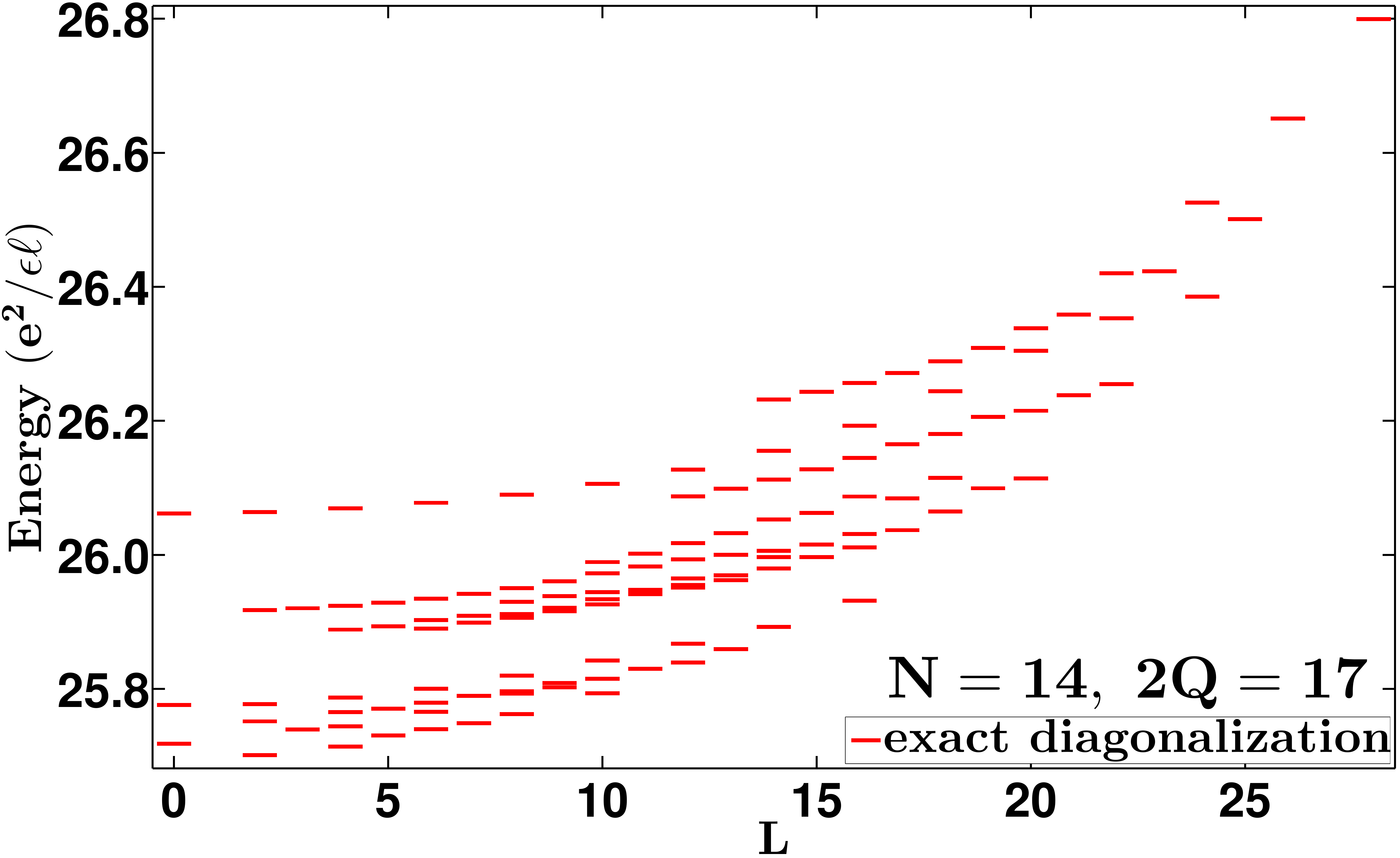}
\includegraphics[width=6cm,height=3.5cm]{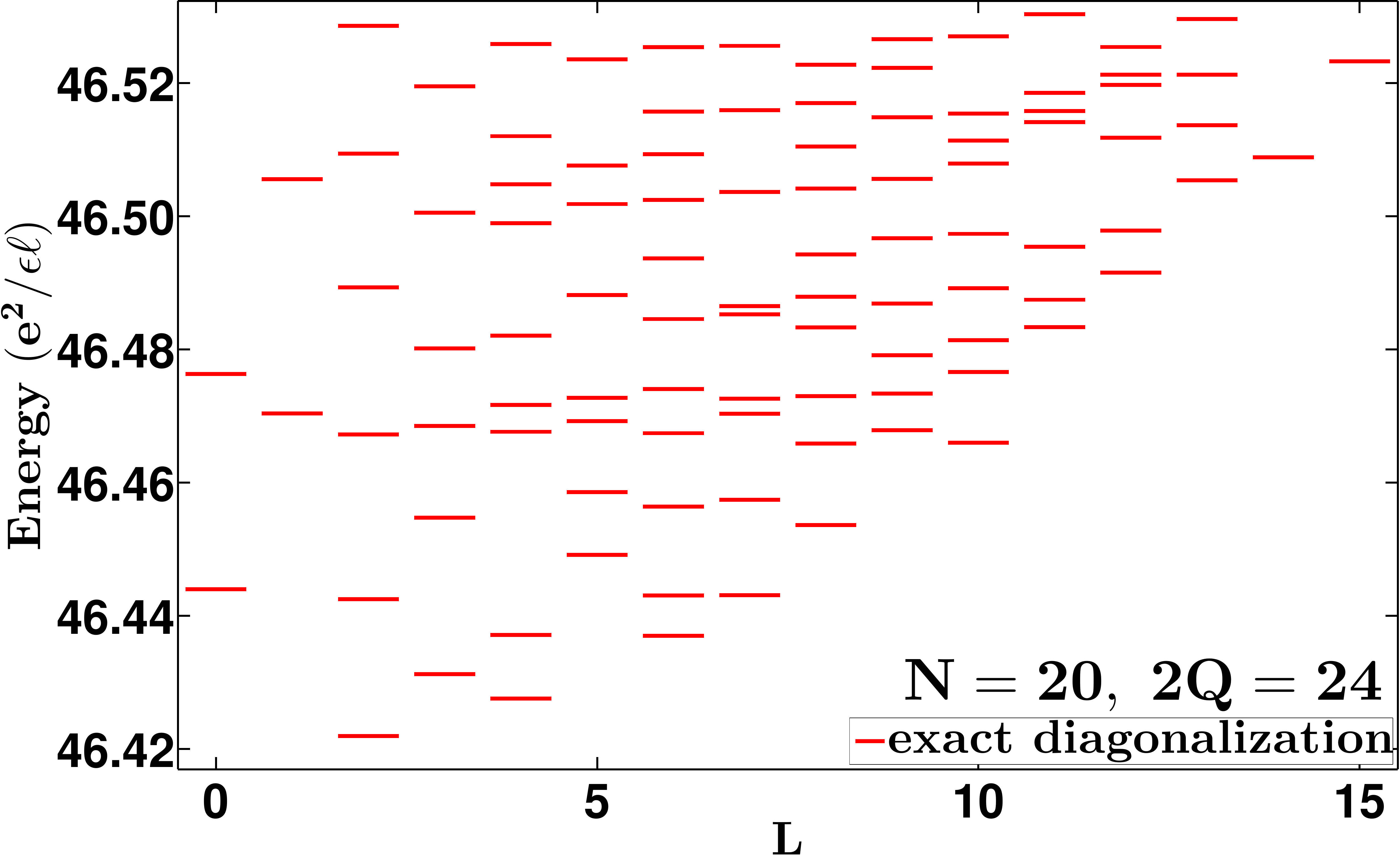}}
\subfigure[~WYQ]{
\includegraphics[width=6cm,height=3.5cm]{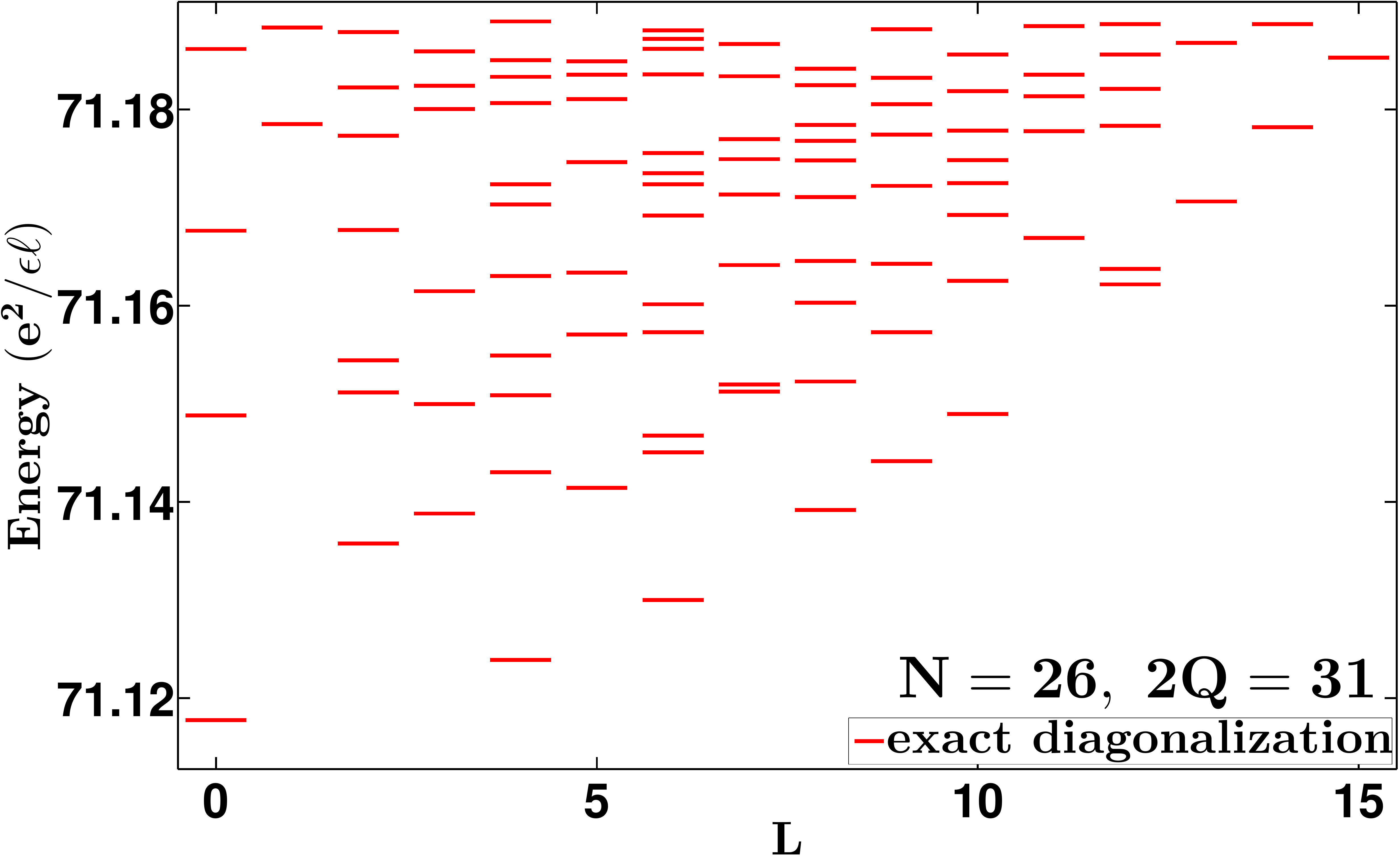}
\includegraphics[width=6cm,height=3.5cm]{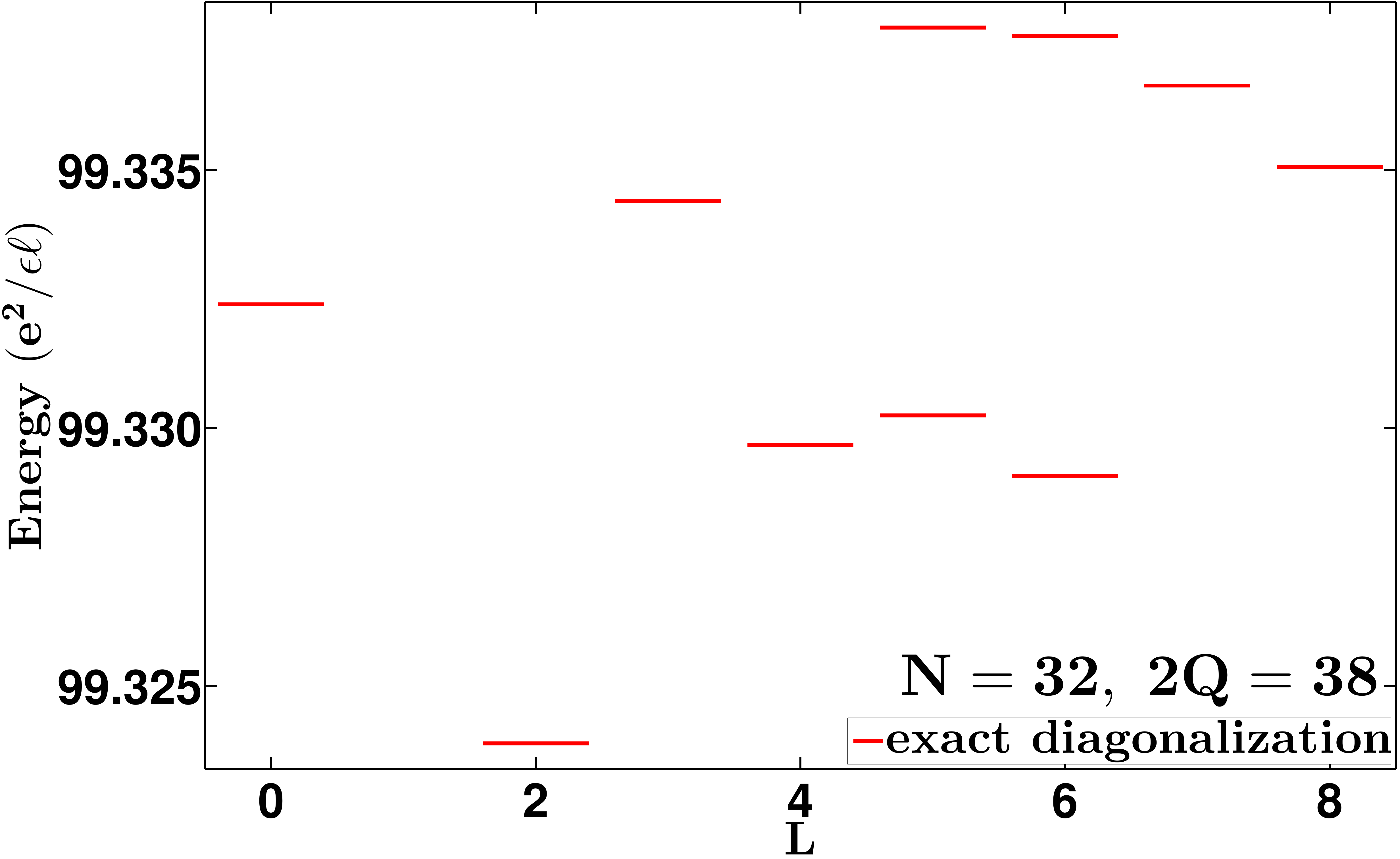}}
\caption{\label{spectra_fp_6_7}Coulomb spectra obtained from exact diagonalization in the spherical geometry for $N$ fully spin polarized electrons at $\nu=6/7$ at the total flux $2Q(hc/e)$  corresponding to the Laughlin (top two panels (a) and (b)) and WYQ state (bottom two panels (c) and (d)).}
\end{figure*}

For the fully spin polarized 6/7 the following two possibilities need to be considered:
\begin{itemize}
 \item The state:
 \begin{equation}
  [1+[1]_{4}]_{-2} \leftrightarrow(6/7)~:\gamma=1
  \label{eq_CF_6_5_m2}
 \end{equation}
corresponds to $\nu^*=6/5$, which is obtained by filling the L$\Lambda$L $\uparrow$ completely and forming a 1/5 Laughlin state in the S$\Lambda$L $\uparrow$. This state at $\nu=6/7$ occurs at $\mathcal{S}=0$ and has $S=N/2$. 
 \item The state:
 $$[1+1/5^{\rm WYQ}]_{-2} \leftrightarrow(6/7)~:\gamma=1$$
corresponds to $\nu^*=6/5$, which is obtained by filling the L$\Lambda$L $\uparrow$ completely and forming a 1/5 WYQ state in the S$\Lambda$L $\uparrow$. This state at $\nu=6/7$ occurs at $\mathcal{S}=-2/3$ and has $S=N/2$. 
\end{itemize}
The Coulomb spectra for both these cases obtained using exact diagonalization is shown in Fig. \ref{spectra_fp_6_7}. With a 1/5 Laughlin state in the S$\Lambda$L $\uparrow$ the ground state is incompressible for all system sizes while this is not the case for all systems supporting the 1/5 WYQ state in the S$\Lambda$L $\uparrow$. Thus the fully spin polarized 6/7 state is an FQHE state of composite fermions with a filled L$\Lambda$L $\uparrow$ and a conventional 1/5 Laughlin state in the S$\Lambda$L $\uparrow$. \\

Just as at the fully spin polarized $\nu=4/5$ and $\nu=5/7$, there exists another plausible candidate state for the fully spin polarized FQHE at $6/7$. This state is 
\begin{equation}
\overline{[1]_{6}} \leftrightarrow(6/7)~:\gamma=1
\label{eq_hole_1_7}
\end{equation}
which is the hole conjugate of the 1/7 Laughlin state. By now it should be clear that this difference is merely artificial and the wave functions for the above two states of Eq. \ref{eq_CF_6_5_m2} and Eq. \ref{eq_hole_1_7} are completely equivalent to each other. Following the analysis of the 4/5 state, one can show in a straightforward manner that the the maximum total orbital angular momentum of the exciton state here is: $L_{\rm max}=2Q+1-N$. As was emphasized in Ref. \cite{Balram15}, out of the two forms of Eq. \ref{eq_CF_6_5_m2} and Eq. \ref{eq_hole_1_7}, an understanding of the 6/7 as arising from the parent state of $\nu^*=6/5$ as given in Eq. \ref{eq_CF_6_5_m2} is more useful in bringing out the spin physics of the state which is discussed next. 

\subsubsection{Partially spin polarized $6/7$}
The partially spin polarized 6/7 state 
$$[1,[1]_{4}]_{-2} \leftrightarrow (5/7,1/7)~:\gamma=\frac{2}{3}$$
is obtained from the partially spin polarized 6/5 state
$$[1,[1]_4] \leftrightarrow(1,1/5)~:\gamma=\frac{2}{3}$$
which corresponds to filling the L$\Lambda$L $\uparrow$ of spin up completely and forming a 1/5 state in the L$\Lambda$L $\downarrow$. This state at $\nu=6/7$ occurs at $\mathcal{S}=1/3$ and has $S=(N-2)/3$. 

\subsubsection{Spin singlet $6/7$}
The state 
$$[\overline{[\overline{[1,1]_{-2}}]_{-2}}]_{-2} \leftrightarrow (3/7,3/7)~:\gamma=0$$
is obtained from the 4/5 spin singlet state \cite{Balram15}
$$[\overline{[1,1]_{-2}}]_{-2} \leftrightarrow (2/5,2/5)~:\gamma=0$$
by taking its particle hole conjugate to produce a singlet state at 6/5 and then composite-fermionizing it with reverse vortex attachment. This state at $\nu=6/7$ occurs at $\mathcal{S}=1$ and has $S=0$. \\

\begin{figure}[htbp]
\begin{center}
\includegraphics[width=8cm,height=4.5cm]{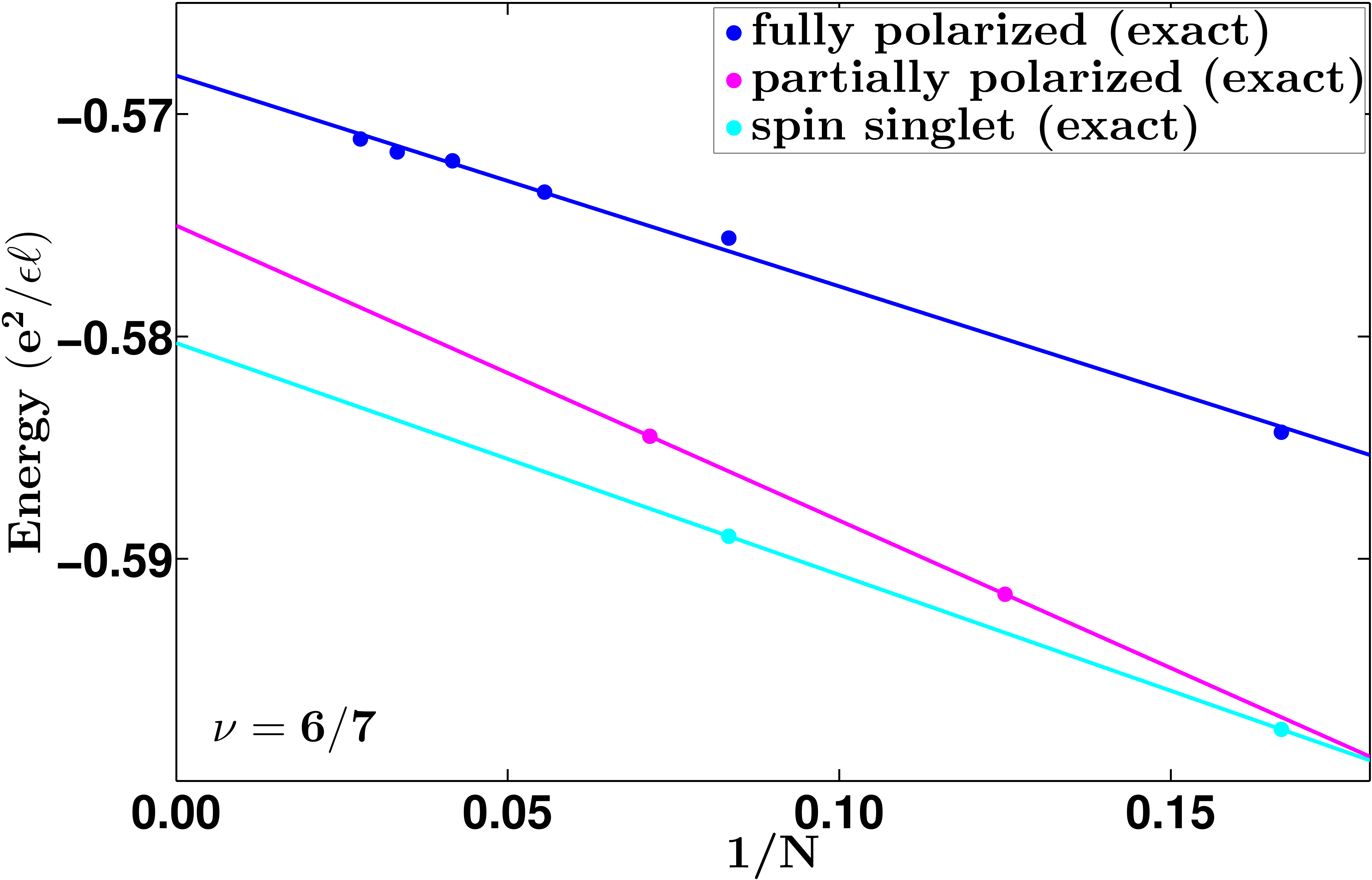}
\end{center}
\caption{\label{extrap_6_7}
Thermodynamic extrapolation of the lowest Landau level Coulomb ground state energy for different spin polarized states at $\nu=6/7$.}
\end{figure}

Fig. \ref{extrap_6_7} shows the thermodynamic extrapolation of the LLL Coulomb ground state energies for these states. Table \ref{tab:6_5} shows the thermodynamic energies of these states as well as the critical Zeeman energies for the transitions among these states. 

\section{Conclusions}
\label{sec:conclusions}
In this work I calculated in detail the interaction energy of two spinful composite fermions on top of a filled $\Lambda$L. I then use these CF pseudopotentials as a guide to narrow the search for FQHSs likely to support the unconventional WYQ type state of composite fermions (which is believed to be the underlying mechanism for FQHE at the fully polarized $\nu=4/11$, $4/13$, $5/13$ and $5/17$). Among the candidate states, I considered states at $\nu=4/5,~5/7,~6/7,$ and $6/17$. For the fully spin polarized version of these states, I find from detailed calculations that the state of composite fermions in the second $\Lambda$L is of the conventional Laughlin kind \cite{Laughlin83} and not of the unconventional WYQ type \cite{Wojs04}. \\

I calculated the ground state energies of the fully spin polarized, partially spin polarized and spin-singlet $6/17$ and $6/7$ states and obtained the critical Zeeman energies for the spin transitions between them. These can be measured in experiments by either tuning the density or tilting the magnetic field \cite{Yeh99,Liu14a,Pan15} or by doing resonant inelastic light scattering experiments and analyzing the excitation spectra \cite{Balram15d} as has been done at the nearby filling factors.\\

\begin{figure}[htbp]
\begin{center}
\includegraphics[width=8cm,height=4.5cm]{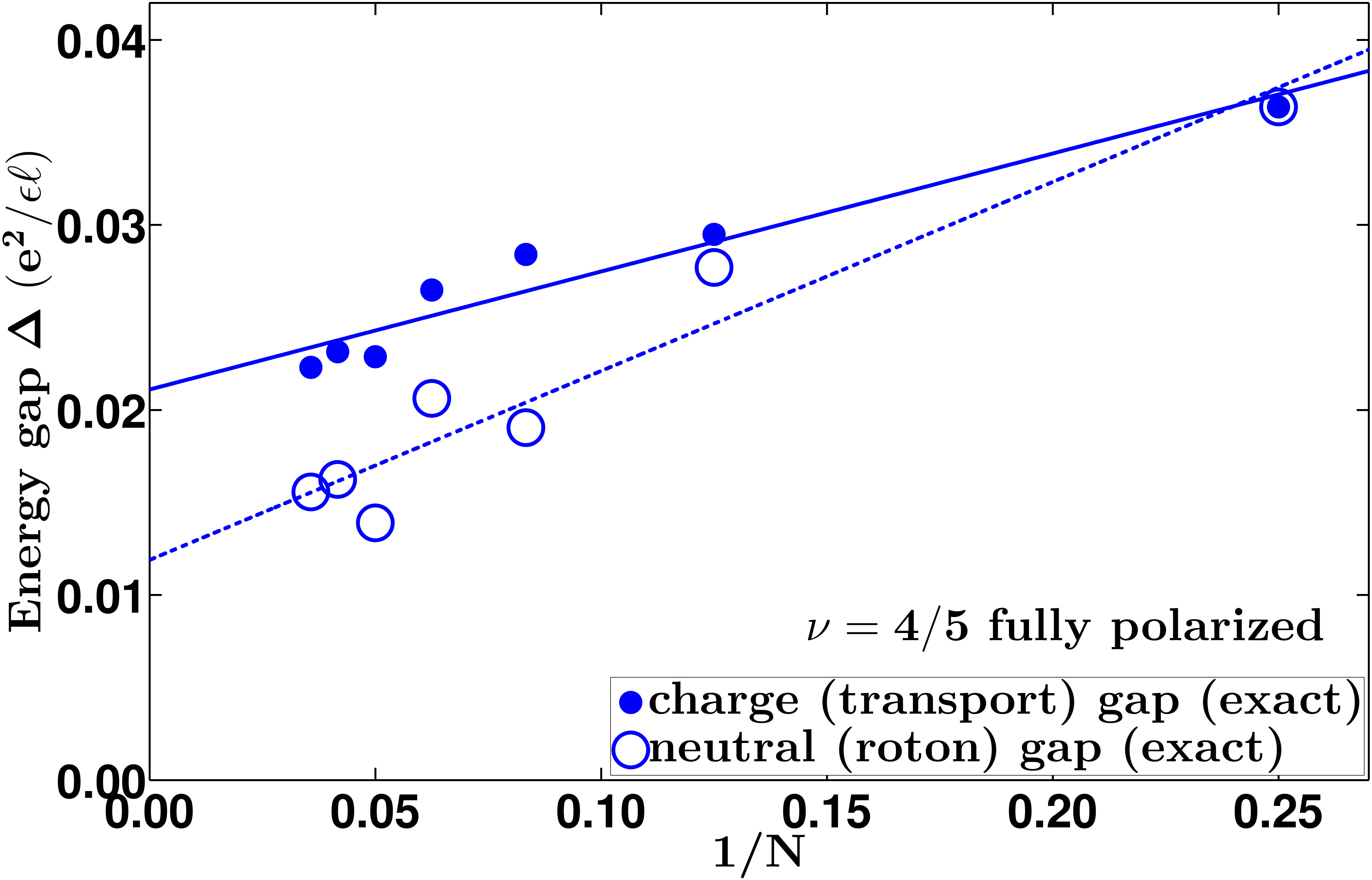}
\includegraphics[width=8cm,height=4.5cm]{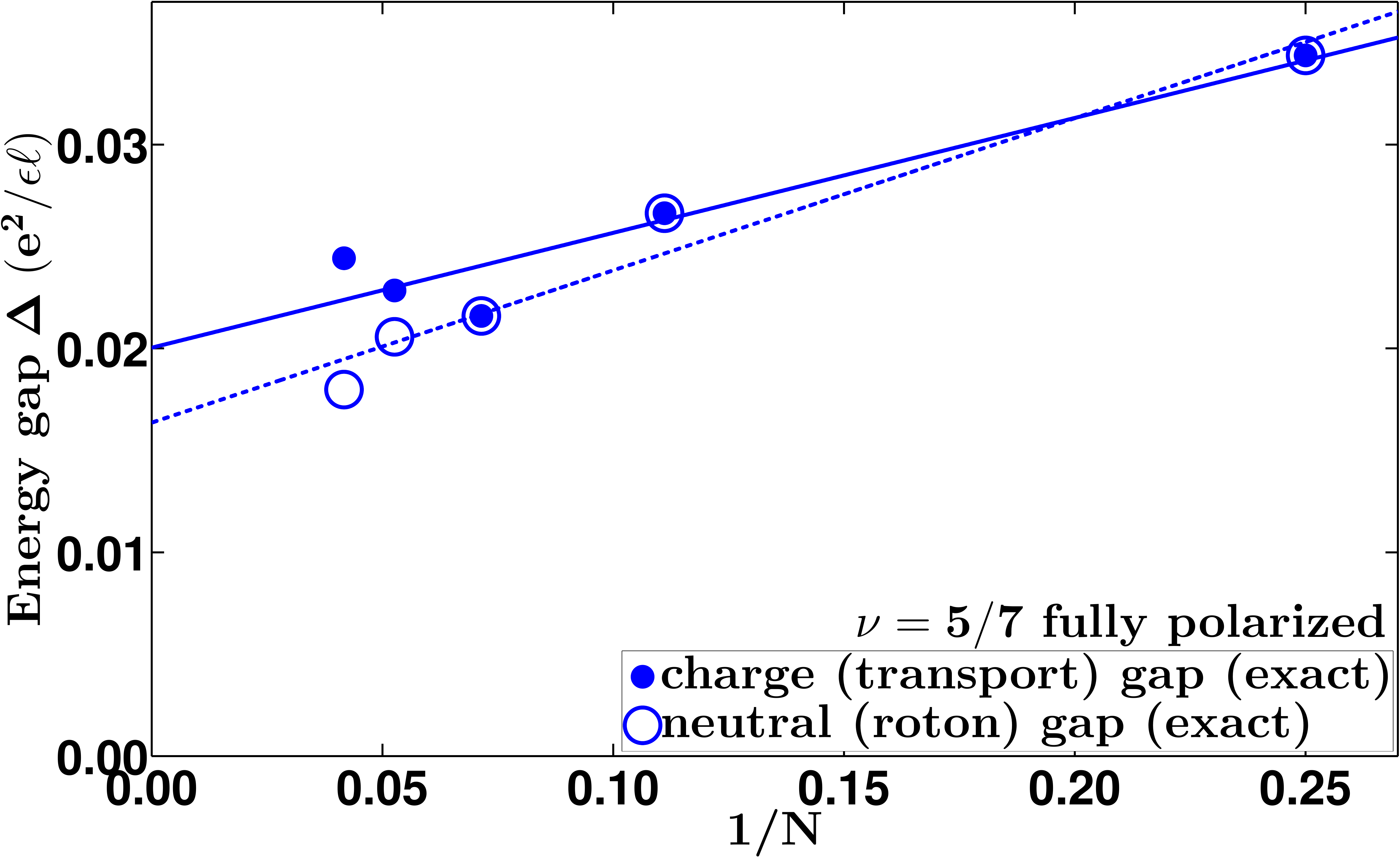}
\includegraphics[width=8cm,height=4.5cm]{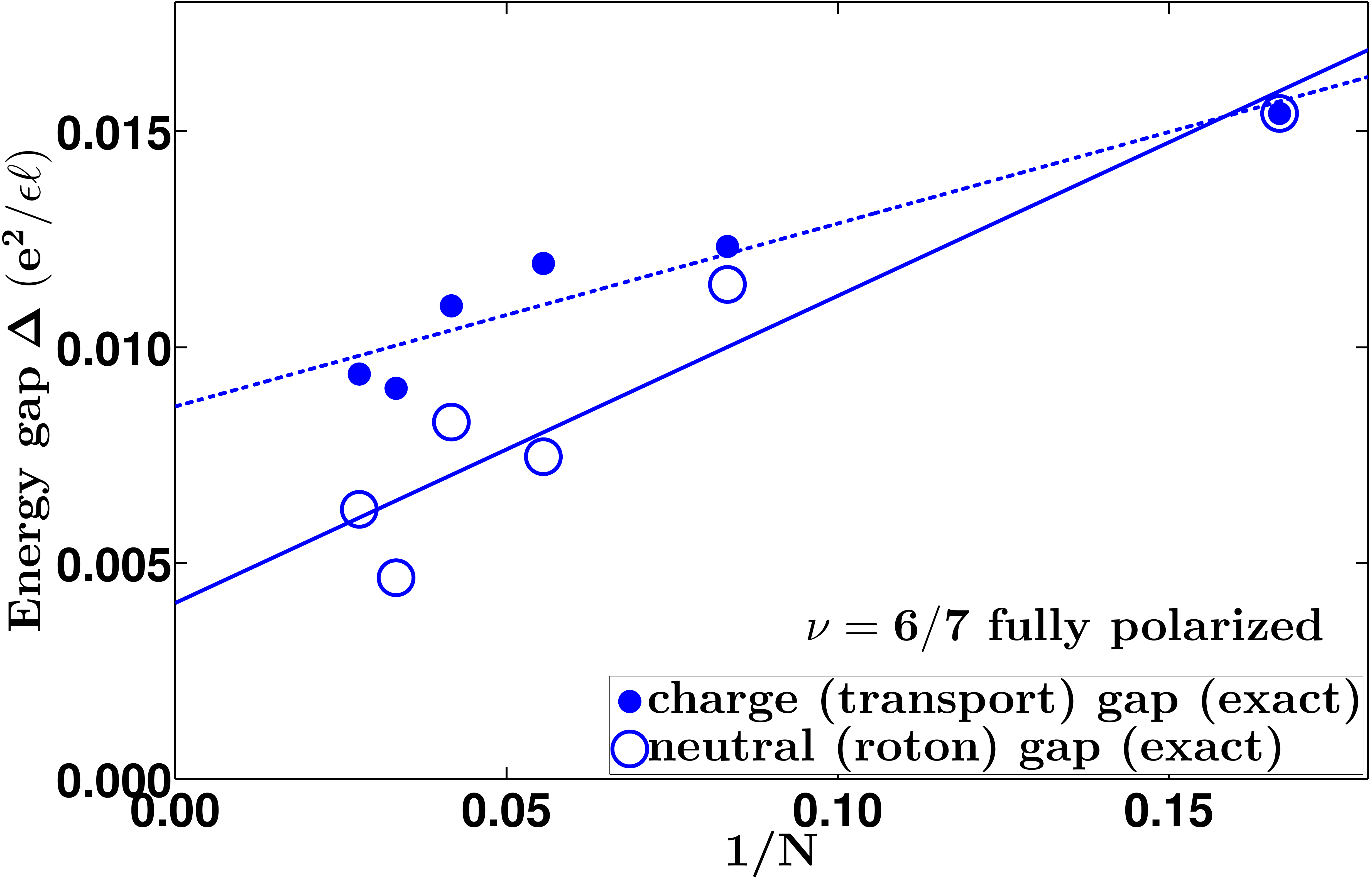}
\end{center}
\caption{\label{extrap_gaps}
Thermodynamic extrapolation of the neutral and charge Coulomb excitation gap at the fully spin polarized $\nu=4/5$ (top panel), $\nu=5/7$ (middle panel) and $\nu=6/7$ (bottom panel) state obtained from exact diagonalization in the spherical geometry.}
\end{figure}

Due to limitations arising from size of the Hilbert space dimension I have only been able to reliably determine the gaps for the fully spin polarized $\nu=4/5,~5/7,$ and $6/7$ states. The thermodynamic extrapolation of the gaps from finite system results for these three states is shown in Fig. \ref{extrap_gaps}. In the presence of particle-hole symmetry, the gap at the fully spin polarized $\nu=4/5$ and $\nu=6/7$ is the same as that at $\nu=1/5$ and $\nu=1/7$, respectively (density correction gives slightly different numbers) which has been evaluated previously. Reference \cite{Jain97} quoted neutral exciton gaps of $0.0009(5)~(0.0095(6))$ $e^{2}/\epsilon\ell$ and $0.0063~(0.017)$ $e^{2}/\epsilon\ell$ at $\nu=1/7$ ($\nu=1/5$) obtained from methods of composite fermion theory and single-mode approximation \cite{Girvin85,Girvin86} respectively. Our extrapolated neutral exciton gap lies between these two numbers. The charge gap at $4/5$ shown in Fig. \ref{extrap_gaps} is consistent with the known value of the $1/5$ charge gap: $0.025(3)$ $e^{2}/\epsilon\ell$ \cite{Jain97}. \\

From the largest system sizes considered in this work, I estimate the neutral exciton gap for the fully spin polarized and partially spin polarized $\nu=6/17$ state to be $\sim0.002$ and $\sim0.001$ $e^2/\epsilon\ell$ respectively. These numbers are similar to the numerical estimate of the energy gap of the fully (partially) polarized 4/11 state: $\sim0.002$ $(\sim0.001)$ $e^2/\epsilon\ell$ \cite{Mukherjee14,Chang03} for an ideal system (for the spin-singlet state at $\nu=6/17$ I have not been able to evaluate the Coulomb spectra for a reasonably large system to extract the gap). \\

The gap to excitations at $4/11$ and $4/5$ has been measured in transport experiments recently. Liu \emph{et al.} \cite{Liu14a} find a gap of $\sim 0.35$~K ($\approx0.003$ $e^{2}/\epsilon\ell$) at $\nu=4/5$ in a wide sample (quantum well width of $65$ nm) which is about an order of magnitude smaller than the aformentioned theoretical gap at $4/5$. Samkharadze \emph{et al.} \cite{Samkharadze15b} report an activation gap of $\sim 15$ mK at 4/11 and $\sim 3$ mK at 5/13 while Pan \emph{et al.} \cite{Pan15} report a gap of $\sim7$ mK at 4/11. For the parameters of Ref. \cite{Pan15} (Ref. \cite{Samkharadze15b}) the theoretical gap at $4/11$ \cite{Mukherjee14} $0.002~e^2/\epsilon\ell \approx 0.37$ K ($0.31$ K) is an order of magnitude larger than the experimental numbers. Similar discrepancies between experimental and theoretical estimates of the gap have been seen at $5/2$ \cite{Pan99}. The precise origin of this mismatch is not yet understood, though it is widely believed that disorder broadening plays a key role. Including corrections from disorder-broadening Samkharadze \emph{et al.} estimate the gap at 4/11 to be $0.0028~e^2/\epsilon\ell$ which is close to the numerical estimate for the fully spin polarized state \cite{Mukherjee14}. Pan \emph{et al.} include disorder corrections to the theoretical values and estimate that for their samples the fully (partially) polarized gap is 0.12 (0.05) K which is about a third of the theoretical estimate. Including further corrections from effects of LL mixing, which is known to reduce gaps \cite{Melik-Alaverdian97}, it may be possible to find an agreement between the theoretical and experimental numbers. The theoretical and experimental gaps at $\nu=4/5$ are larger than the corresponding numbers at $\nu=4/11$ which suggests that conventional Laughlin type correlations with reverse vortex attachment are more robust than their WYQ counterparts with parallel vortex attachment. \\ 

I end with a summary of the various mechanisms that have so far been identified for the observed FQHSs.  (i) The IQHE of composite fermions explains the Jain fractions $\nu=n/(2pn\pm 1)$ in the LLL, where $n$ and $p$ are positive integers, and the related fractions $1\pm n/(2pn\pm 1)$ and $2-n/(2pn\pm 1)$, and also the second LL fractions 10/3, 11/3, 16/5, 19/5, \cite{Eisenstein02}, 11/5, 14/5, 25/9 \cite{Kumar10}, 16/7, 19/7 \cite{Zhang12}. Although the $2+1/3$ state is generally believed to be described by the Laughlin wave function \cite{Laughlin83}, the situation is less convincing \cite{Ambrumenil88,Balram13b}, and other proposals have been investigated \cite{Jeong16}. (ii) The states at 4/11, 4/13, 5/13 and 5/17 \cite{Pan03} are likely described as W\'ojs-Yi-Quinn \cite{Wojs04,Mukherjee14,Mukherjee15b} FQHE of composite fermions, but no satisfactory trial wave functions yet exist (although some have been proposed \cite{Bergholtz07,Bergholtz08,Hansson09}). Signatures of FQHE have also been observed at $7/11$ \cite{Pan03,Goldman90a} and $9/13$ \cite{Goldman90a}. These would be described as FQHE of composite fermions carrying two vortices in the direction opposite to the external magnetic field at $\nu^{*}=7/3~(\nu=7/11)$ and $\nu^{*}=9/5~(\nu=9/13)$ respectively, but their precise mechanism is not yet known (the fully spin polarized $\nu=7/11$ and $\nu=9/13$ states are the particle-hole conjugates of the fully spin polarized $4/11$ and $4/13$ states). (iii) The 5/2 FQHE \cite{Willett87} (as well as the closely related 7/2 FQHE \cite{Eisenstein02}) is believed to arise from a pairing of composite fermions \cite{Moore91,Read00}. (iv) Numerical diagonalization studies \cite{Bonderson12,Sreejith13} make a clear case that the FQHE at $2+2/5$ \cite{Xia04,Pan08,Kumar10} is either a Bonderson-Slingerland state \cite{Bonderson08,Bonderson12} or the particle hole conjugate of the $k=3$ Read-Rezayi state \cite{Read99,Rezayi09}, with the more recent calculations favoring the latter \cite{Pakrouski16,Mong15,Zhu15}. (v) Mukherjee {\em et al.} \cite{Mukherjee12,Mukherjee14c} showed that pairing is plausible for composite fermions in the second $\Lambda$L and spin down lowest $\Lambda$L which leads to the fully and partially spin polarized 3/8 FQHE \cite{Pan03,Bellani10} respectively. (vi) Mukherjee and Mandal \cite{Mukherjee15b} have shown that an incompressible fully spin polarized FQHE state at $\nu=3/10$ occurs due to pairing of composite fermions carrying four vortices in the second $\Lambda$L. (vii) Hutasoit \emph{et al.} \cite{Hutasoit16} have made a case that the most plausible explanation of the 2+3/8 \cite{Xia04,Pan08,Kumar10} FQHE state is in terms of a Bonderson-Slingerland state. (viii) In this work I have put forth the case that the FQHE at 6/17 \cite{Pan03} and 6/7 arises from a conventional FQHE of composite fermions at $\nu^{*}=6/5$. (ix) Kumar {\em et al.} \cite{Kumar10} have observed FQHE at $\nu=2+6/13$, for which no proposal yet exists. \\

{\bf Acknowledgment}
I thank W. Pan for sharing his data. I am grateful to J. K. Jain for useful discussions and his comments on the manuscript. This work was supported by the U. S. National Science Foundation Grant no. DMR-1401636 and the Villum Foundation. I acknowledge the Research Computing and Cyberinfrastructure at Pennsylvania State University which is in part funded by the National Science Foundation Grant No. OCI-0821527. Some of the numerical calculations were performed using the DiagHam package, for which I am grateful to its authors.\\

\emph{Note added}: During the completion of this work, I became aware of a recent calculation \cite{Regnault16} on the torus geometry which proposed an alternate possibility that the WYQ model $V_{m}=\delta_{m,3}$ leads to a compressible stripe phase at $\nu=1/3$.

\bibliography{../../../Latex-Revtex-etc./biblio_fqhe}

\begin{thebibliography}{90}
\expandafter\ifx\csname natexlab\endcsname\relax\def\natexlab#1{#1}\fi
\expandafter\ifx\csname bibnamefont\endcsname\relax
  \def\bibnamefont#1{#1}\fi
\expandafter\ifx\csname bibfnamefont\endcsname\relax
  \def\bibfnamefont#1{#1}\fi
\expandafter\ifx\csname citenamefont\endcsname\relax
  \def\citenamefont#1{#1}\fi
\expandafter\ifx\csname url\endcsname\relax
  \def\url#1{\texttt{#1}}\fi
\expandafter\ifx\csname urlprefix\endcsname\relax\def\urlprefix{URL }\fi
\providecommand{\bibinfo}[2]{#2}
\providecommand{\eprint}[2][]{\url{#2}}

\bibitem[{\citenamefont{Tsui et~al.}(1982)\citenamefont{Tsui, Stormer, and
  Gossard}}]{Tsui82}
\bibinfo{author}{\bibfnamefont{D.~C.} \bibnamefont{Tsui}},
  \bibinfo{author}{\bibfnamefont{H.~L.} \bibnamefont{Stormer}},
  \bibnamefont{and} \bibinfo{author}{\bibfnamefont{A.~C.}
  \bibnamefont{Gossard}}, \bibinfo{journal}{Phys. Rev. Lett.}
  \textbf{\bibinfo{volume}{48}}, \bibinfo{pages}{1559} (\bibinfo{year}{1982}),
  \urlprefix\url{http://link.aps.org/doi/10.1103/PhysRevLett.48.1559}.

\bibitem[{\citenamefont{Jain}(1989)}]{Jain89}
\bibinfo{author}{\bibfnamefont{J.~K.} \bibnamefont{Jain}},
  \bibinfo{journal}{Phys. Rev. Lett.} \textbf{\bibinfo{volume}{63}},
  \bibinfo{pages}{199} (\bibinfo{year}{1989}),
  \urlprefix\url{http://link.aps.org/doi/10.1103/PhysRevLett.63.199}.

\bibitem[{\citenamefont{Jain}(2007)}]{Jain07}
\bibinfo{author}{\bibfnamefont{J.~K.} \bibnamefont{Jain}},
  \emph{\bibinfo{title}{Composite Fermions}} (\bibinfo{publisher}{Cambridge
  University Press, New York, US (Cambridge Books Online)},
  \bibinfo{year}{2007}).

\bibitem[{\citenamefont{Wu et~al.}(1993)\citenamefont{Wu, Dev, and
  Jain}}]{Wu93}
\bibinfo{author}{\bibfnamefont{X.~G.} \bibnamefont{Wu}},
  \bibinfo{author}{\bibfnamefont{G.}~\bibnamefont{Dev}}, \bibnamefont{and}
  \bibinfo{author}{\bibfnamefont{J.~K.} \bibnamefont{Jain}},
  \bibinfo{journal}{Phys. Rev. Lett.} \textbf{\bibinfo{volume}{71}},
  \bibinfo{pages}{153} (\bibinfo{year}{1993}),
  \urlprefix\url{http://link.aps.org/doi/10.1103/PhysRevLett.71.153}.

\bibitem[{\citenamefont{Park and Jain}(1998)}]{Park98}
\bibinfo{author}{\bibfnamefont{K.}~\bibnamefont{Park}} \bibnamefont{and}
  \bibinfo{author}{\bibfnamefont{J.~K.} \bibnamefont{Jain}},
  \bibinfo{journal}{Phys. Rev. Lett.} \textbf{\bibinfo{volume}{80}},
  \bibinfo{pages}{4237} (\bibinfo{year}{1998}),
  \urlprefix\url{http://link.aps.org/doi/10.1103/PhysRevLett.80.4237}.

\bibitem[{\citenamefont{Park et~al.}(1998)\citenamefont{Park, Melik-Alaverdian,
  Bonesteel, and Jain}}]{Park98b}
\bibinfo{author}{\bibfnamefont{K.}~\bibnamefont{Park}},
  \bibinfo{author}{\bibfnamefont{V.}~\bibnamefont{Melik-Alaverdian}},
  \bibinfo{author}{\bibfnamefont{N.~E.} \bibnamefont{Bonesteel}},
  \bibnamefont{and} \bibinfo{author}{\bibfnamefont{J.~K.} \bibnamefont{Jain}},
  \bibinfo{journal}{Phys. Rev. B} \textbf{\bibinfo{volume}{58}},
  \bibinfo{pages}{R10167} (\bibinfo{year}{1998}),
  \urlprefix\url{http://link.aps.org/doi/10.1103/PhysRevB.58.R10167}.

\bibitem[{\citenamefont{Park and Jain}(2001)}]{Park01}
\bibinfo{author}{\bibfnamefont{K.}~\bibnamefont{Park}} \bibnamefont{and}
  \bibinfo{author}{\bibfnamefont{J.~K.} \bibnamefont{Jain}},
  \bibinfo{journal}{Solid State Commun.} \textbf{\bibinfo{volume}{119}},
  \bibinfo{pages}{291} (\bibinfo{year}{2001}).

\bibitem[{\citenamefont{Balram et~al.}(2015{\natexlab{a}})\citenamefont{Balram,
  T\"oke, W\'ojs, and Jain}}]{Balram15a}
\bibinfo{author}{\bibfnamefont{A.~C.} \bibnamefont{Balram}},
  \bibinfo{author}{\bibfnamefont{C.}~\bibnamefont{T\"oke}},
  \bibinfo{author}{\bibfnamefont{A.}~\bibnamefont{W\'ojs}}, \bibnamefont{and}
  \bibinfo{author}{\bibfnamefont{J.~K.} \bibnamefont{Jain}},
  \bibinfo{journal}{Phys. Rev. B} \textbf{\bibinfo{volume}{92}},
  \bibinfo{pages}{075410} (\bibinfo{year}{2015}{\natexlab{a}}),
  \urlprefix\url{http://link.aps.org/doi/10.1103/PhysRevB.92.075410}.

\bibitem[{\citenamefont{Eisenstein et~al.}(1989)\citenamefont{Eisenstein,
  Stormer, Pfeiffer, and West}}]{Eisenstein89}
\bibinfo{author}{\bibfnamefont{J.~P.} \bibnamefont{Eisenstein}},
  \bibinfo{author}{\bibfnamefont{H.~L.} \bibnamefont{Stormer}},
  \bibinfo{author}{\bibfnamefont{L.}~\bibnamefont{Pfeiffer}}, \bibnamefont{and}
  \bibinfo{author}{\bibfnamefont{K.~W.} \bibnamefont{West}},
  \bibinfo{journal}{Phys. Rev. Lett.} \textbf{\bibinfo{volume}{62}},
  \bibinfo{pages}{1540} (\bibinfo{year}{1989}),
  \urlprefix\url{http://link.aps.org/doi/10.1103/PhysRevLett.62.1540}.

\bibitem[{\citenamefont{Engel et~al.}(1992)\citenamefont{Engel, Hwang, Sajoto,
  Tsui, and Shayegan}}]{Engel92}
\bibinfo{author}{\bibfnamefont{L.~W.} \bibnamefont{Engel}},
  \bibinfo{author}{\bibfnamefont{S.~W.} \bibnamefont{Hwang}},
  \bibinfo{author}{\bibfnamefont{T.}~\bibnamefont{Sajoto}},
  \bibinfo{author}{\bibfnamefont{D.~C.} \bibnamefont{Tsui}}, \bibnamefont{and}
  \bibinfo{author}{\bibfnamefont{M.}~\bibnamefont{Shayegan}},
  \bibinfo{journal}{Phys. Rev. B} \textbf{\bibinfo{volume}{45}},
  \bibinfo{pages}{3418} (\bibinfo{year}{1992}),
  \urlprefix\url{http://link.aps.org/doi/10.1103/PhysRevB.45.3418}.

\bibitem[{\citenamefont{Du et~al.}(1995)\citenamefont{Du, Yeh, Stormer, Tsui,
  Pfeiffer, and West}}]{Du95}
\bibinfo{author}{\bibfnamefont{R.~R.} \bibnamefont{Du}},
  \bibinfo{author}{\bibfnamefont{A.~S.} \bibnamefont{Yeh}},
  \bibinfo{author}{\bibfnamefont{H.~L.} \bibnamefont{Stormer}},
  \bibinfo{author}{\bibfnamefont{D.~C.} \bibnamefont{Tsui}},
  \bibinfo{author}{\bibfnamefont{L.~N.} \bibnamefont{Pfeiffer}},
  \bibnamefont{and} \bibinfo{author}{\bibfnamefont{K.~W.} \bibnamefont{West}},
  \bibinfo{journal}{Phys. Rev. Lett.} \textbf{\bibinfo{volume}{75}},
  \bibinfo{pages}{3926} (\bibinfo{year}{1995}),
  \urlprefix\url{http://link.aps.org/doi/10.1103/PhysRevLett.75.3926}.

\bibitem[{\citenamefont{Kang et~al.}(1997)\citenamefont{Kang, Young, Hannahs,
  Palm, Campman, and Gossard}}]{Kang97}
\bibinfo{author}{\bibfnamefont{W.}~\bibnamefont{Kang}},
  \bibinfo{author}{\bibfnamefont{J.~B.} \bibnamefont{Young}},
  \bibinfo{author}{\bibfnamefont{S.~T.} \bibnamefont{Hannahs}},
  \bibinfo{author}{\bibfnamefont{E.}~\bibnamefont{Palm}},
  \bibinfo{author}{\bibfnamefont{K.~L.} \bibnamefont{Campman}},
  \bibnamefont{and} \bibinfo{author}{\bibfnamefont{A.~C.}
  \bibnamefont{Gossard}}, \bibinfo{journal}{Phys. Rev. B}
  \textbf{\bibinfo{volume}{56}}, \bibinfo{pages}{R12776}
  (\bibinfo{year}{1997}),
  \urlprefix\url{http://link.aps.org/doi/10.1103/PhysRevB.56.R12776}.

\bibitem[{\citenamefont{Yeh et~al.}(1999)\citenamefont{Yeh, Stormer, Tsui,
  Pfeiffer, Baldwin, and West}}]{Yeh99}
\bibinfo{author}{\bibfnamefont{A.~S.} \bibnamefont{Yeh}},
  \bibinfo{author}{\bibfnamefont{H.~L.} \bibnamefont{Stormer}},
  \bibinfo{author}{\bibfnamefont{D.~C.} \bibnamefont{Tsui}},
  \bibinfo{author}{\bibfnamefont{L.~N.} \bibnamefont{Pfeiffer}},
  \bibinfo{author}{\bibfnamefont{K.~W.} \bibnamefont{Baldwin}},
  \bibnamefont{and} \bibinfo{author}{\bibfnamefont{K.~W.} \bibnamefont{West}},
  \bibinfo{journal}{Phys. Rev. Lett.} \textbf{\bibinfo{volume}{82}},
  \bibinfo{pages}{592} (\bibinfo{year}{1999}),
  \urlprefix\url{http://link.aps.org/doi/10.1103/PhysRevLett.82.592}.

\bibitem[{\citenamefont{Kukushkin et~al.}(1999)\citenamefont{Kukushkin,
  v.~Klitzing, and Eberl}}]{Kukushkin99}
\bibinfo{author}{\bibfnamefont{I.~V.} \bibnamefont{Kukushkin}},
  \bibinfo{author}{\bibfnamefont{K.}~\bibnamefont{v.~Klitzing}},
  \bibnamefont{and} \bibinfo{author}{\bibfnamefont{K.}~\bibnamefont{Eberl}},
  \bibinfo{journal}{Phys. Rev. Lett.} \textbf{\bibinfo{volume}{82}},
  \bibinfo{pages}{3665} (\bibinfo{year}{1999}),
  \urlprefix\url{http://link.aps.org/doi/10.1103/PhysRevLett.82.3665}.

\bibitem[{\citenamefont{Kukushkin et~al.}(2000)\citenamefont{Kukushkin, Smet,
  von Klitzing, and Eberl}}]{Kukushkin00}
\bibinfo{author}{\bibfnamefont{I.~V.} \bibnamefont{Kukushkin}},
  \bibinfo{author}{\bibfnamefont{J.~H.} \bibnamefont{Smet}},
  \bibinfo{author}{\bibfnamefont{K.}~\bibnamefont{von Klitzing}},
  \bibnamefont{and} \bibinfo{author}{\bibfnamefont{K.}~\bibnamefont{Eberl}},
  \bibinfo{journal}{Phys. Rev. Lett.} \textbf{\bibinfo{volume}{85}},
  \bibinfo{pages}{3688} (\bibinfo{year}{2000}),
  \urlprefix\url{http://link.aps.org/doi/10.1103/PhysRevLett.85.3688}.

\bibitem[{\citenamefont{Melinte et~al.}(2000)\citenamefont{Melinte, Freytag,
  Horvatic, Berthier, L\'evy, Bayot, and Shayegan}}]{Melinte00}
\bibinfo{author}{\bibfnamefont{S.}~\bibnamefont{Melinte}},
  \bibinfo{author}{\bibfnamefont{N.}~\bibnamefont{Freytag}},
  \bibinfo{author}{\bibfnamefont{M.}~\bibnamefont{Horvatic}},
  \bibinfo{author}{\bibfnamefont{C.}~\bibnamefont{Berthier}},
  \bibinfo{author}{\bibfnamefont{L.~P.} \bibnamefont{L\'evy}},
  \bibinfo{author}{\bibfnamefont{V.}~\bibnamefont{Bayot}}, \bibnamefont{and}
  \bibinfo{author}{\bibfnamefont{M.}~\bibnamefont{Shayegan}},
  \bibinfo{journal}{Phys. Rev. Lett.} \textbf{\bibinfo{volume}{84}},
  \bibinfo{pages}{354} (\bibinfo{year}{2000}),
  \urlprefix\url{http://link.aps.org/doi/10.1103/PhysRevLett.84.354}.

\bibitem[{\citenamefont{Freytag et~al.}(2001)\citenamefont{Freytag, Tokunaga,
  Horvati\'{c}, Berthier, Shayegan, and L\'evy}}]{Freytag01}
\bibinfo{author}{\bibfnamefont{N.}~\bibnamefont{Freytag}},
  \bibinfo{author}{\bibfnamefont{Y.}~\bibnamefont{Tokunaga}},
  \bibinfo{author}{\bibfnamefont{M.}~\bibnamefont{Horvati\'{c}}},
  \bibinfo{author}{\bibfnamefont{C.}~\bibnamefont{Berthier}},
  \bibinfo{author}{\bibfnamefont{M.}~\bibnamefont{Shayegan}}, \bibnamefont{and}
  \bibinfo{author}{\bibfnamefont{L.~P.} \bibnamefont{L\'evy}},
  \bibinfo{journal}{Phys. Rev. Lett.} \textbf{\bibinfo{volume}{87}},
  \bibinfo{pages}{136801} (\bibinfo{year}{2001}),
  \urlprefix\url{http://link.aps.org/doi/10.1103/PhysRevLett.87.136801}.

\bibitem[{\citenamefont{Tiemann et~al.}(2012)\citenamefont{Tiemann, Gamez,
  Kumada, and Muraki}}]{Tiemann12}
\bibinfo{author}{\bibfnamefont{L.}~\bibnamefont{Tiemann}},
  \bibinfo{author}{\bibfnamefont{G.}~\bibnamefont{Gamez}},
  \bibinfo{author}{\bibfnamefont{N.}~\bibnamefont{Kumada}}, \bibnamefont{and}
  \bibinfo{author}{\bibfnamefont{K.}~\bibnamefont{Muraki}},
  \bibinfo{journal}{Science} \textbf{\bibinfo{volume}{335}},
  \bibinfo{pages}{828} (\bibinfo{year}{2012}),
  \eprint{http://www.sciencemag.org/content/335/6070/828.full.pdf},
  \urlprefix\url{http://www.sciencemag.org/content/335/6070/828.abstract}.

\bibitem[{\citenamefont{Feldman et~al.}(2013)\citenamefont{Feldman, Levin,
  Krauss, Abanin, Halperin, Smet, and Yacoby}}]{Feldman13}
\bibinfo{author}{\bibfnamefont{B.~E.} \bibnamefont{Feldman}},
  \bibinfo{author}{\bibfnamefont{A.~J.} \bibnamefont{Levin}},
  \bibinfo{author}{\bibfnamefont{B.}~\bibnamefont{Krauss}},
  \bibinfo{author}{\bibfnamefont{D.~A.} \bibnamefont{Abanin}},
  \bibinfo{author}{\bibfnamefont{B.~I.} \bibnamefont{Halperin}},
  \bibinfo{author}{\bibfnamefont{J.~H.} \bibnamefont{Smet}}, \bibnamefont{and}
  \bibinfo{author}{\bibfnamefont{A.}~\bibnamefont{Yacoby}},
  \bibinfo{journal}{Phys. Rev. Lett.} \textbf{\bibinfo{volume}{111}},
  \bibinfo{pages}{076802} (\bibinfo{year}{2013}),
  \urlprefix\url{http://link.aps.org/doi/10.1103/PhysRevLett.111.076802}.

\bibitem[{\citenamefont{Liu et~al.}(2014{\natexlab{a}})\citenamefont{Liu,
  Hasdemir, W\'ojs, Jain, Pfeiffer, West, Baldwin, and Shayegan}}]{Liu14}
\bibinfo{author}{\bibfnamefont{Y.}~\bibnamefont{Liu}},
  \bibinfo{author}{\bibfnamefont{S.}~\bibnamefont{Hasdemir}},
  \bibinfo{author}{\bibfnamefont{A.}~\bibnamefont{W\'ojs}},
  \bibinfo{author}{\bibfnamefont{J.~K.} \bibnamefont{Jain}},
  \bibinfo{author}{\bibfnamefont{L.~N.} \bibnamefont{Pfeiffer}},
  \bibinfo{author}{\bibfnamefont{K.~W.} \bibnamefont{West}},
  \bibinfo{author}{\bibfnamefont{K.~W.} \bibnamefont{Baldwin}},
  \bibnamefont{and} \bibinfo{author}{\bibfnamefont{M.}~\bibnamefont{Shayegan}},
  \bibinfo{journal}{Phys. Rev. B} \textbf{\bibinfo{volume}{90}},
  \bibinfo{pages}{085301} (\bibinfo{year}{2014}{\natexlab{a}}),
  \urlprefix\url{http://link.aps.org/doi/10.1103/PhysRevB.90.085301}.

\bibitem[{\citenamefont{Moore and Read}(1991)}]{Moore91}
\bibinfo{author}{\bibfnamefont{G.}~\bibnamefont{Moore}} \bibnamefont{and}
  \bibinfo{author}{\bibfnamefont{N.}~\bibnamefont{Read}},
  \bibinfo{journal}{Nucl. Phys. B} \textbf{\bibinfo{volume}{360}},
  \bibinfo{pages}{362 } (\bibinfo{year}{1991}), ISSN \bibinfo{issn}{0550-3213},
  \urlprefix\url{http://www.sciencedirect.com/science/article/pii/055032139190407O}.

\bibitem[{\citenamefont{Read and Green}(2000)}]{Read00}
\bibinfo{author}{\bibfnamefont{N.}~\bibnamefont{Read}} \bibnamefont{and}
  \bibinfo{author}{\bibfnamefont{D.}~\bibnamefont{Green}},
  \bibinfo{journal}{Phys. Rev. B} \textbf{\bibinfo{volume}{61}},
  \bibinfo{pages}{10267} (\bibinfo{year}{2000}),
  \urlprefix\url{http://link.aps.org/doi/10.1103/PhysRevB.61.10267}.

\bibitem[{\citenamefont{Willett et~al.}(1987)\citenamefont{Willett, Eisenstein,
  St\"ormer, Tsui, Gossard, and English}}]{Willett87}
\bibinfo{author}{\bibfnamefont{R.}~\bibnamefont{Willett}},
  \bibinfo{author}{\bibfnamefont{J.~P.} \bibnamefont{Eisenstein}},
  \bibinfo{author}{\bibfnamefont{H.~L.} \bibnamefont{St\"ormer}},
  \bibinfo{author}{\bibfnamefont{D.~C.} \bibnamefont{Tsui}},
  \bibinfo{author}{\bibfnamefont{A.~C.} \bibnamefont{Gossard}},
  \bibnamefont{and} \bibinfo{author}{\bibfnamefont{J.~H.}
  \bibnamefont{English}}, \bibinfo{journal}{Phys. Rev. Lett.}
  \textbf{\bibinfo{volume}{59}}, \bibinfo{pages}{1776} (\bibinfo{year}{1987}),
  \urlprefix\url{http://link.aps.org/doi/10.1103/PhysRevLett.59.1776}.

\bibitem[{\citenamefont{Mukherjee et~al.}(2012)\citenamefont{Mukherjee, Mandal,
  W\'ojs, and Jain}}]{Mukherjee12}
\bibinfo{author}{\bibfnamefont{S.}~\bibnamefont{Mukherjee}},
  \bibinfo{author}{\bibfnamefont{S.~S.} \bibnamefont{Mandal}},
  \bibinfo{author}{\bibfnamefont{A.}~\bibnamefont{W\'ojs}}, \bibnamefont{and}
  \bibinfo{author}{\bibfnamefont{J.~K.} \bibnamefont{Jain}},
  \bibinfo{journal}{Phys. Rev. Lett.} \textbf{\bibinfo{volume}{109}},
  \bibinfo{pages}{256801} (\bibinfo{year}{2012}),
  \urlprefix\url{http://link.aps.org/doi/10.1103/PhysRevLett.109.256801}.

\bibitem[{\citenamefont{Mukherjee
  et~al.}(2014{\natexlab{a}})\citenamefont{Mukherjee, Jain, and
  Mandal}}]{Mukherjee14c}
\bibinfo{author}{\bibfnamefont{S.}~\bibnamefont{Mukherjee}},
  \bibinfo{author}{\bibfnamefont{J.~K.} \bibnamefont{Jain}}, \bibnamefont{and}
  \bibinfo{author}{\bibfnamefont{S.~S.} \bibnamefont{Mandal}},
  \bibinfo{journal}{Phys. Rev. B} \textbf{\bibinfo{volume}{90}},
  \bibinfo{pages}{121305} (\bibinfo{year}{2014}{\natexlab{a}}),
  \urlprefix\url{http://link.aps.org/doi/10.1103/PhysRevB.90.121305}.

\bibitem[{\citenamefont{Mukherjee and Mandal}(2015)}]{Mukherjee15b}
\bibinfo{author}{\bibfnamefont{S.}~\bibnamefont{Mukherjee}} \bibnamefont{and}
  \bibinfo{author}{\bibfnamefont{S.~S.} \bibnamefont{Mandal}},
  \bibinfo{journal}{Phys. Rev. B} \textbf{\bibinfo{volume}{92}},
  \bibinfo{pages}{235302} (\bibinfo{year}{2015}),
  \urlprefix\url{http://link.aps.org/doi/10.1103/PhysRevB.92.235302}.

\bibitem[{\citenamefont{Pan et~al.}(2003)\citenamefont{Pan, Stormer, Tsui,
  Pfeiffer, Baldwin, and West}}]{Pan03}
\bibinfo{author}{\bibfnamefont{W.}~\bibnamefont{Pan}},
  \bibinfo{author}{\bibfnamefont{H.~L.} \bibnamefont{Stormer}},
  \bibinfo{author}{\bibfnamefont{D.~C.} \bibnamefont{Tsui}},
  \bibinfo{author}{\bibfnamefont{L.~N.} \bibnamefont{Pfeiffer}},
  \bibinfo{author}{\bibfnamefont{K.~W.} \bibnamefont{Baldwin}},
  \bibnamefont{and} \bibinfo{author}{\bibfnamefont{K.~W.} \bibnamefont{West}},
  \bibinfo{journal}{Phys. Rev. Lett.} \textbf{\bibinfo{volume}{90}},
  \bibinfo{pages}{016801} (\bibinfo{year}{2003}),
  \urlprefix\url{http://link.aps.org/doi/10.1103/PhysRevLett.90.016801}.

\bibitem[{\citenamefont{Bellani et~al.}(2010)\citenamefont{Bellani, Dionigi,
  Rossella, Amado, Diez, Biasiol, and Sorba}}]{Bellani10}
\bibinfo{author}{\bibfnamefont{V.}~\bibnamefont{Bellani}},
  \bibinfo{author}{\bibfnamefont{F.}~\bibnamefont{Dionigi}},
  \bibinfo{author}{\bibfnamefont{F.}~\bibnamefont{Rossella}},
  \bibinfo{author}{\bibfnamefont{M.}~\bibnamefont{Amado}},
  \bibinfo{author}{\bibfnamefont{E.}~\bibnamefont{Diez}},
  \bibinfo{author}{\bibfnamefont{G.}~\bibnamefont{Biasiol}}, \bibnamefont{and}
  \bibinfo{author}{\bibfnamefont{L.}~\bibnamefont{Sorba}},
  \bibinfo{journal}{Phys. Rev. B} \textbf{\bibinfo{volume}{81}},
  \bibinfo{pages}{155316} (\bibinfo{year}{2010}),
  \urlprefix\url{http://link.aps.org/doi/10.1103/PhysRevB.81.155316}.

\bibitem[{\citenamefont{Pan et~al.}(2015)\citenamefont{Pan, Baldwin, West,
  Pfeiffer, and Tsui}}]{Pan15}
\bibinfo{author}{\bibfnamefont{W.}~\bibnamefont{Pan}},
  \bibinfo{author}{\bibfnamefont{K.~W.} \bibnamefont{Baldwin}},
  \bibinfo{author}{\bibfnamefont{K.~W.} \bibnamefont{West}},
  \bibinfo{author}{\bibfnamefont{L.~N.} \bibnamefont{Pfeiffer}},
  \bibnamefont{and} \bibinfo{author}{\bibfnamefont{D.~C.} \bibnamefont{Tsui}},
  \bibinfo{journal}{Phys. Rev. B} \textbf{\bibinfo{volume}{91}},
  \bibinfo{pages}{041301} (\bibinfo{year}{2015}),
  \urlprefix\url{http://link.aps.org/doi/10.1103/PhysRevB.91.041301}.

\bibitem[{\citenamefont{Samkharadze et~al.}(2015)\citenamefont{Samkharadze,
  Arnold, Pfeiffer, West, and Cs\'athy}}]{Samkharadze15b}
\bibinfo{author}{\bibfnamefont{N.}~\bibnamefont{Samkharadze}},
  \bibinfo{author}{\bibfnamefont{I.}~\bibnamefont{Arnold}},
  \bibinfo{author}{\bibfnamefont{L.~N.} \bibnamefont{Pfeiffer}},
  \bibinfo{author}{\bibfnamefont{K.~W.} \bibnamefont{West}}, \bibnamefont{and}
  \bibinfo{author}{\bibfnamefont{G.~A.} \bibnamefont{Cs\'athy}},
  \bibinfo{journal}{Phys. Rev. B} \textbf{\bibinfo{volume}{91}},
  \bibinfo{pages}{081109} (\bibinfo{year}{2015}),
  \urlprefix\url{http://link.aps.org/doi/10.1103/PhysRevB.91.081109}.

\bibitem[{\citenamefont{Liu et~al.}(2014{\natexlab{b}})\citenamefont{Liu,
  Kamburov, Hasdemir, Shayegan, Pfeiffer, West, and Baldwin}}]{Liu14a}
\bibinfo{author}{\bibfnamefont{Y.}~\bibnamefont{Liu}},
  \bibinfo{author}{\bibfnamefont{D.}~\bibnamefont{Kamburov}},
  \bibinfo{author}{\bibfnamefont{S.}~\bibnamefont{Hasdemir}},
  \bibinfo{author}{\bibfnamefont{M.}~\bibnamefont{Shayegan}},
  \bibinfo{author}{\bibfnamefont{L.~N.} \bibnamefont{Pfeiffer}},
  \bibinfo{author}{\bibfnamefont{K.~W.} \bibnamefont{West}}, \bibnamefont{and}
  \bibinfo{author}{\bibfnamefont{K.~W.} \bibnamefont{Baldwin}},
  \bibinfo{journal}{Phys. Rev. Lett.} \textbf{\bibinfo{volume}{113}},
  \bibinfo{pages}{246803} (\bibinfo{year}{2014}{\natexlab{b}}),
  \urlprefix\url{http://link.aps.org/doi/10.1103/PhysRevLett.113.246803}.

\bibitem[{\citenamefont{W\'ojs and Quinn}(2000)}]{Wojs00}
\bibinfo{author}{\bibfnamefont{A.}~\bibnamefont{W\'ojs}} \bibnamefont{and}
  \bibinfo{author}{\bibfnamefont{J.~J.} \bibnamefont{Quinn}},
  \bibinfo{journal}{Phys. Rev. B} \textbf{\bibinfo{volume}{61}},
  \bibinfo{pages}{2846} (\bibinfo{year}{2000}),
  \urlprefix\url{http://link.aps.org/doi/10.1103/PhysRevB.61.2846}.

\bibitem[{\citenamefont{W\'ojs et~al.}(2004)\citenamefont{W\'ojs, Yi, and
  Quinn}}]{Wojs04}
\bibinfo{author}{\bibfnamefont{A.}~\bibnamefont{W\'ojs}},
  \bibinfo{author}{\bibfnamefont{K.-S.} \bibnamefont{Yi}}, \bibnamefont{and}
  \bibinfo{author}{\bibfnamefont{J.~J.} \bibnamefont{Quinn}},
  \bibinfo{journal}{Phys. Rev. B} \textbf{\bibinfo{volume}{69}},
  \bibinfo{pages}{205322} (\bibinfo{year}{2004}),
  \urlprefix\url{http://link.aps.org/doi/10.1103/PhysRevB.69.205322}.

\bibitem[{\citenamefont{Mukherjee
  et~al.}(2014{\natexlab{b}})\citenamefont{Mukherjee, Mandal, Wu, W\'ojs, and
  Jain}}]{Mukherjee14}
\bibinfo{author}{\bibfnamefont{S.}~\bibnamefont{Mukherjee}},
  \bibinfo{author}{\bibfnamefont{S.~S.} \bibnamefont{Mandal}},
  \bibinfo{author}{\bibfnamefont{Y.-H.} \bibnamefont{Wu}},
  \bibinfo{author}{\bibfnamefont{A.}~\bibnamefont{W\'ojs}}, \bibnamefont{and}
  \bibinfo{author}{\bibfnamefont{J.~K.} \bibnamefont{Jain}},
  \bibinfo{journal}{Phys. Rev. Lett.} \textbf{\bibinfo{volume}{112}},
  \bibinfo{pages}{016801} (\bibinfo{year}{2014}{\natexlab{b}}),
  \urlprefix\url{http://link.aps.org/doi/10.1103/PhysRevLett.112.016801}.

\bibitem[{\citenamefont{Laughlin}(1983)}]{Laughlin83}
\bibinfo{author}{\bibfnamefont{R.~B.} \bibnamefont{Laughlin}},
  \bibinfo{journal}{Phys. Rev. Lett.} \textbf{\bibinfo{volume}{50}},
  \bibinfo{pages}{1395} (\bibinfo{year}{1983}),
  \urlprefix\url{http://link.aps.org/doi/10.1103/PhysRevLett.50.1395}.

\bibitem[{\citenamefont{Balram et~al.}(2015{\natexlab{b}})\citenamefont{Balram,
  T\"oke, W\'ojs, and Jain}}]{Balram15}
\bibinfo{author}{\bibfnamefont{A.~C.} \bibnamefont{Balram}},
  \bibinfo{author}{\bibfnamefont{C.}~\bibnamefont{T\"oke}},
  \bibinfo{author}{\bibfnamefont{A.}~\bibnamefont{W\'ojs}}, \bibnamefont{and}
  \bibinfo{author}{\bibfnamefont{J.~K.} \bibnamefont{Jain}},
  \bibinfo{journal}{Phys. Rev. B} \textbf{\bibinfo{volume}{91}},
  \bibinfo{pages}{045109} (\bibinfo{year}{2015}{\natexlab{b}}),
  \urlprefix\url{http://link.aps.org/doi/10.1103/PhysRevB.91.045109}.

\bibitem[{\citenamefont{Sitko et~al.}(1996)\citenamefont{Sitko, Yi, Yi, and
  Quinn}}]{Sitko96}
\bibinfo{author}{\bibfnamefont{P.}~\bibnamefont{Sitko}},
  \bibinfo{author}{\bibfnamefont{S.~N.} \bibnamefont{Yi}},
  \bibinfo{author}{\bibfnamefont{K.~S.} \bibnamefont{Yi}}, \bibnamefont{and}
  \bibinfo{author}{\bibfnamefont{J.~J.} \bibnamefont{Quinn}},
  \bibinfo{journal}{Phys. Rev. Lett.} \textbf{\bibinfo{volume}{76}},
  \bibinfo{pages}{3396} (\bibinfo{year}{1996}),
  \urlprefix\url{http://link.aps.org/doi/10.1103/PhysRevLett.76.3396}.

\bibitem[{\citenamefont{Lee et~al.}(2001)\citenamefont{Lee, Scarola, and
  Jain}}]{Lee01}
\bibinfo{author}{\bibfnamefont{S.-Y.} \bibnamefont{Lee}},
  \bibinfo{author}{\bibfnamefont{V.~W.} \bibnamefont{Scarola}},
  \bibnamefont{and} \bibinfo{author}{\bibfnamefont{J.~K.} \bibnamefont{Jain}},
  \bibinfo{journal}{Phys. Rev. Lett.} \textbf{\bibinfo{volume}{87}},
  \bibinfo{pages}{256803} (\bibinfo{year}{2001}),
  \urlprefix\url{http://link.aps.org/doi/10.1103/PhysRevLett.87.256803}.

\bibitem[{\citenamefont{Lee et~al.}(2002)\citenamefont{Lee, Scarola, and
  Jain}}]{Lee02}
\bibinfo{author}{\bibfnamefont{S.-Y.} \bibnamefont{Lee}},
  \bibinfo{author}{\bibfnamefont{V.~W.} \bibnamefont{Scarola}},
  \bibnamefont{and} \bibinfo{author}{\bibfnamefont{J.~K.} \bibnamefont{Jain}},
  \bibinfo{journal}{Phys. Rev. B} \textbf{\bibinfo{volume}{66}},
  \bibinfo{pages}{085336} (\bibinfo{year}{2002}),
  \urlprefix\url{http://link.aps.org/doi/10.1103/PhysRevB.66.085336}.

\bibitem[{\citenamefont{Haldane}(1983)}]{Haldane83}
\bibinfo{author}{\bibfnamefont{F.~D.~M.} \bibnamefont{Haldane}},
  \bibinfo{journal}{Phys. Rev. Lett.} \textbf{\bibinfo{volume}{51}},
  \bibinfo{pages}{605} (\bibinfo{year}{1983}),
  \urlprefix\url{http://link.aps.org/doi/10.1103/PhysRevLett.51.605}.

\bibitem[{\citenamefont{Greiter}(2011)}]{Greiter11}
\bibinfo{author}{\bibfnamefont{M.}~\bibnamefont{Greiter}},
  \bibinfo{journal}{Phys. Rev. B} \textbf{\bibinfo{volume}{83}},
  \bibinfo{pages}{115129} (\bibinfo{year}{2011}),
  \urlprefix\url{http://link.aps.org/doi/10.1103/PhysRevB.83.115129}.

\bibitem[{\citenamefont{Wen and Zee}(1992)}]{Wen92}
\bibinfo{author}{\bibfnamefont{X.~G.} \bibnamefont{Wen}} \bibnamefont{and}
  \bibinfo{author}{\bibfnamefont{A.}~\bibnamefont{Zee}},
  \bibinfo{journal}{Phys. Rev. Lett.} \textbf{\bibinfo{volume}{69}},
  \bibinfo{pages}{953} (\bibinfo{year}{1992}),
  \urlprefix\url{http://link.aps.org/doi/10.1103/PhysRevLett.69.953}.

\bibitem[{\citenamefont{Haldane and Rezayi}(1985{\natexlab{a}})}]{Haldane85}
\bibinfo{author}{\bibfnamefont{F.~D.~M.} \bibnamefont{Haldane}}
  \bibnamefont{and} \bibinfo{author}{\bibfnamefont{E.~H.}
  \bibnamefont{Rezayi}}, \bibinfo{journal}{Phys. Rev. B}
  \textbf{\bibinfo{volume}{31}}, \bibinfo{pages}{2529}
  (\bibinfo{year}{1985}{\natexlab{a}}),
  \urlprefix\url{http://link.aps.org/doi/10.1103/PhysRevB.31.2529}.

\bibitem[{\citenamefont{Haldane}(1985)}]{Haldane85b}
\bibinfo{author}{\bibfnamefont{F.~D.~M.} \bibnamefont{Haldane}},
  \bibinfo{journal}{Phys. Rev. Lett.} \textbf{\bibinfo{volume}{55}},
  \bibinfo{pages}{2095} (\bibinfo{year}{1985}),
  \urlprefix\url{http://link.aps.org/doi/10.1103/PhysRevLett.55.2095}.

\bibitem[{\citenamefont{Hermanns}(2013)}]{Hermanns13}
\bibinfo{author}{\bibfnamefont{M.}~\bibnamefont{Hermanns}},
  \bibinfo{journal}{Phys. Rev. B} \textbf{\bibinfo{volume}{87}},
  \bibinfo{pages}{235128} (\bibinfo{year}{2013}),
  \urlprefix\url{http://link.aps.org/doi/10.1103/PhysRevB.87.235128}.

\bibitem[{\citenamefont{Fremling}(2013)}]{Fremling13}
\bibinfo{author}{\bibfnamefont{M.}~\bibnamefont{Fremling}},
  \bibinfo{journal}{Journal of Physics A: Mathematical and Theoretical}
  \textbf{\bibinfo{volume}{46}}, \bibinfo{pages}{275302}
  (\bibinfo{year}{2013}),
  \urlprefix\url{http://stacks.iop.org/1751-8121/46/i=27/a=275302}.

\bibitem[{\citenamefont{{Fremling}}(2014)}]{Fremling14}
\bibinfo{author}{\bibfnamefont{M.}~\bibnamefont{{Fremling}}},
  \bibinfo{journal}{ArXiv e-prints}  (\bibinfo{year}{2014}),
  \eprint{1401.6834}.

\bibitem[{\citenamefont{Jain and Kamilla}(1997{\natexlab{a}})}]{Jain97}
\bibinfo{author}{\bibfnamefont{J.~K.} \bibnamefont{Jain}} \bibnamefont{and}
  \bibinfo{author}{\bibfnamefont{R.~K.} \bibnamefont{Kamilla}},
  \bibinfo{journal}{Int. J. Mod. Phys. B} \textbf{\bibinfo{volume}{11}},
  \bibinfo{pages}{2621} (\bibinfo{year}{1997}{\natexlab{a}}).

\bibitem[{\citenamefont{Jain and Kamilla}(1997{\natexlab{b}})}]{Jain97b}
\bibinfo{author}{\bibfnamefont{J.~K.} \bibnamefont{Jain}} \bibnamefont{and}
  \bibinfo{author}{\bibfnamefont{R.~K.} \bibnamefont{Kamilla}},
  \bibinfo{journal}{Phys. Rev. B} \textbf{\bibinfo{volume}{55}},
  \bibinfo{pages}{R4895} (\bibinfo{year}{1997}{\natexlab{b}}),
  \urlprefix\url{http://link.aps.org/doi/10.1103/PhysRevB.55.R4895}.

\bibitem[{\citenamefont{M\"oller and Simon}(2005)}]{Moller05}
\bibinfo{author}{\bibfnamefont{G.}~\bibnamefont{M\"oller}} \bibnamefont{and}
  \bibinfo{author}{\bibfnamefont{S.~H.} \bibnamefont{Simon}},
  \bibinfo{journal}{Phys. Rev. B} \textbf{\bibinfo{volume}{72}},
  \bibinfo{pages}{045344} (\bibinfo{year}{2005}),
  \urlprefix\url{http://link.aps.org/doi/10.1103/PhysRevB.72.045344}.

\bibitem[{\citenamefont{Davenport and Simon}(2012)}]{Davenport12}
\bibinfo{author}{\bibfnamefont{S.~C.} \bibnamefont{Davenport}}
  \bibnamefont{and} \bibinfo{author}{\bibfnamefont{S.~H.} \bibnamefont{Simon}},
  \bibinfo{journal}{Phys. Rev. B} \textbf{\bibinfo{volume}{85}},
  \bibinfo{pages}{245303} (\bibinfo{year}{2012}),
  \urlprefix\url{http://link.aps.org/doi/10.1103/PhysRevB.85.245303}.

\bibitem[{\citenamefont{Mandal and Jain}(2002)}]{Mandal02}
\bibinfo{author}{\bibfnamefont{S.~S.} \bibnamefont{Mandal}} \bibnamefont{and}
  \bibinfo{author}{\bibfnamefont{J.~K.} \bibnamefont{Jain}},
  \bibinfo{journal}{Phys. Rev. B} \textbf{\bibinfo{volume}{66}},
  \bibinfo{pages}{155302} (\bibinfo{year}{2002}),
  \urlprefix\url{http://link.aps.org/doi/10.1103/PhysRevB.66.155302}.

\bibitem[{\citenamefont{Hamermesh}(1962)}]{Hamermesh62}
\bibinfo{author}{\bibfnamefont{M.}~\bibnamefont{Hamermesh}},
  \emph{\bibinfo{title}{Group Theory and Its Application to Physical Problems}}
  (\bibinfo{publisher}{New York: Dover, US}, \bibinfo{year}{1962}).

\bibitem[{\citenamefont{Pinczuk et~al.}(1993)\citenamefont{Pinczuk, Dennis,
  Pfeiffer, and West}}]{Pinczuk93}
\bibinfo{author}{\bibfnamefont{A.}~\bibnamefont{Pinczuk}},
  \bibinfo{author}{\bibfnamefont{B.~S.} \bibnamefont{Dennis}},
  \bibinfo{author}{\bibfnamefont{L.~N.} \bibnamefont{Pfeiffer}},
  \bibnamefont{and} \bibinfo{author}{\bibfnamefont{K.}~\bibnamefont{West}},
  \bibinfo{journal}{Phys. Rev. Lett.} \textbf{\bibinfo{volume}{70}},
  \bibinfo{pages}{3983} (\bibinfo{year}{1993}),
  \urlprefix\url{http://link.aps.org/doi/10.1103/PhysRevLett.70.3983}.

\bibitem[{\citenamefont{Mellor et~al.}(1995)\citenamefont{Mellor, Eyles, Digby,
  Kent, Benedict, Challis, Henini, Foxon, and Harris}}]{Mellor95}
\bibinfo{author}{\bibfnamefont{C.~J.} \bibnamefont{Mellor}},
  \bibinfo{author}{\bibfnamefont{R.~H.} \bibnamefont{Eyles}},
  \bibinfo{author}{\bibfnamefont{J.~E.} \bibnamefont{Digby}},
  \bibinfo{author}{\bibfnamefont{A.~J.} \bibnamefont{Kent}},
  \bibinfo{author}{\bibfnamefont{K.~A.} \bibnamefont{Benedict}},
  \bibinfo{author}{\bibfnamefont{L.~J.} \bibnamefont{Challis}},
  \bibinfo{author}{\bibfnamefont{M.}~\bibnamefont{Henini}},
  \bibinfo{author}{\bibfnamefont{C.~T.} \bibnamefont{Foxon}}, \bibnamefont{and}
  \bibinfo{author}{\bibfnamefont{J.~J.} \bibnamefont{Harris}},
  \bibinfo{journal}{Phys. Rev. Lett.} \textbf{\bibinfo{volume}{74}},
  \bibinfo{pages}{2339} (\bibinfo{year}{1995}),
  \urlprefix\url{http://link.aps.org/doi/10.1103/PhysRevLett.74.2339}.

\bibitem[{\citenamefont{Morf et~al.}(1986)\citenamefont{Morf, d'Ambrumenil, and
  Halperin}}]{Morf86}
\bibinfo{author}{\bibfnamefont{R.}~\bibnamefont{Morf}},
  \bibinfo{author}{\bibfnamefont{N.}~\bibnamefont{d'Ambrumenil}},
  \bibnamefont{and} \bibinfo{author}{\bibfnamefont{B.~I.}
  \bibnamefont{Halperin}}, \bibinfo{journal}{Phys. Rev. B}
  \textbf{\bibinfo{volume}{34}}, \bibinfo{pages}{3037} (\bibinfo{year}{1986}),
  \urlprefix\url{http://link.aps.org/doi/10.1103/PhysRevB.34.3037}.

\bibitem[{\citenamefont{Haldane and Rezayi}(1985{\natexlab{b}})}]{Haldane85a}
\bibinfo{author}{\bibfnamefont{F.~D.~M.} \bibnamefont{Haldane}}
  \bibnamefont{and} \bibinfo{author}{\bibfnamefont{E.~H.}
  \bibnamefont{Rezayi}}, \bibinfo{journal}{Phys. Rev. Lett.}
  \textbf{\bibinfo{volume}{54}}, \bibinfo{pages}{237}
  (\bibinfo{year}{1985}{\natexlab{b}}),
  \urlprefix\url{http://link.aps.org/doi/10.1103/PhysRevLett.54.237}.

\bibitem[{\citenamefont{Fano et~al.}(1986)\citenamefont{Fano, Ortolani, and
  Colombo}}]{Fano86}
\bibinfo{author}{\bibfnamefont{G.}~\bibnamefont{Fano}},
  \bibinfo{author}{\bibfnamefont{F.}~\bibnamefont{Ortolani}}, \bibnamefont{and}
  \bibinfo{author}{\bibfnamefont{E.}~\bibnamefont{Colombo}},
  \bibinfo{journal}{Phys. Rev. B} \textbf{\bibinfo{volume}{34}},
  \bibinfo{pages}{2670} (\bibinfo{year}{1986}),
  \urlprefix\url{http://link.aps.org/doi/10.1103/PhysRevB.34.2670}.

\bibitem[{\citenamefont{Balram et~al.}(2015{\natexlab{c}})\citenamefont{Balram,
  T\H{o}ke, and Jain}}]{Balram15b}
\bibinfo{author}{\bibfnamefont{A.~C.} \bibnamefont{Balram}},
  \bibinfo{author}{\bibfnamefont{C.}~\bibnamefont{T\H{o}ke}}, \bibnamefont{and}
  \bibinfo{author}{\bibfnamefont{J.~K.} \bibnamefont{Jain}},
  \bibinfo{journal}{Phys. Rev. Lett.} \textbf{\bibinfo{volume}{115}},
  \bibinfo{pages}{186805} (\bibinfo{year}{2015}{\natexlab{c}}),
  \urlprefix\url{http://link.aps.org/doi/10.1103/PhysRevLett.115.186805}.

\bibitem[{\citenamefont{Jolicoeur}(2007)}]{Jolicoeur07}
\bibinfo{author}{\bibfnamefont{T.}~\bibnamefont{Jolicoeur}},
  \bibinfo{journal}{Phys. Rev. Lett.} \textbf{\bibinfo{volume}{99}},
  \bibinfo{pages}{036805} (\bibinfo{year}{2007}),
  \urlprefix\url{http://link.aps.org/doi/10.1103/PhysRevLett.99.036805}.

\bibitem[{\citenamefont{Read and Rezayi}(1999)}]{Read99}
\bibinfo{author}{\bibfnamefont{N.}~\bibnamefont{Read}} \bibnamefont{and}
  \bibinfo{author}{\bibfnamefont{E.}~\bibnamefont{Rezayi}},
  \bibinfo{journal}{Phys. Rev. B} \textbf{\bibinfo{volume}{59}},
  \bibinfo{pages}{8084} (\bibinfo{year}{1999}),
  \urlprefix\url{http://link.aps.org/doi/10.1103/PhysRevB.59.8084}.

\bibitem[{\citenamefont{Rezayi and Read}(2009)}]{Rezayi09}
\bibinfo{author}{\bibfnamefont{E.~H.} \bibnamefont{Rezayi}} \bibnamefont{and}
  \bibinfo{author}{\bibfnamefont{N.}~\bibnamefont{Read}},
  \bibinfo{journal}{Phys. Rev. B} \textbf{\bibinfo{volume}{79}},
  \bibinfo{pages}{075306} (\bibinfo{year}{2009}),
  \urlprefix\url{http://link.aps.org/doi/10.1103/PhysRevB.79.075306}.

\bibitem[{\citenamefont{Girvin and Jach}(1984)}]{Girvin84b}
\bibinfo{author}{\bibfnamefont{S.~M.} \bibnamefont{Girvin}} \bibnamefont{and}
  \bibinfo{author}{\bibfnamefont{T.}~\bibnamefont{Jach}},
  \bibinfo{journal}{Phys. Rev. B} \textbf{\bibinfo{volume}{29}},
  \bibinfo{pages}{5617} (\bibinfo{year}{1984}),
  \urlprefix\url{http://link.aps.org/doi/10.1103/PhysRevB.29.5617}.

\bibitem[{\citenamefont{Dev and Jain}(1992)}]{Dev92a}
\bibinfo{author}{\bibfnamefont{G.}~\bibnamefont{Dev}} \bibnamefont{and}
  \bibinfo{author}{\bibfnamefont{J.~K.} \bibnamefont{Jain}},
  \bibinfo{journal}{Phys. Rev. B} \textbf{\bibinfo{volume}{45}},
  \bibinfo{pages}{1223} (\bibinfo{year}{1992}),
  \urlprefix\url{http://link.aps.org/doi/10.1103/PhysRevB.45.1223}.

\bibitem[{\citenamefont{Balram et~al.}(2015{\natexlab{d}})\citenamefont{Balram,
  Wurstbauer, Wojs, Pinczuk, and Jain}}]{Balram15d}
\bibinfo{author}{\bibfnamefont{A.~C.} \bibnamefont{Balram}},
  \bibinfo{author}{\bibfnamefont{U.}~\bibnamefont{Wurstbauer}},
  \bibinfo{author}{\bibfnamefont{A.}~\bibnamefont{Wojs}},
  \bibinfo{author}{\bibfnamefont{A.}~\bibnamefont{Pinczuk}}, \bibnamefont{and}
  \bibinfo{author}{\bibfnamefont{J.~K.} \bibnamefont{Jain}},
  \bibinfo{journal}{Nat Commun} \textbf{\bibinfo{volume}{6}}
  (\bibinfo{year}{2015}{\natexlab{d}}), \bibinfo{note}{article},
  \urlprefix\url{http://dx.doi.org/10.1038/ncomms9981}.

\bibitem[{\citenamefont{Girvin et~al.}(1985)\citenamefont{Girvin, MacDonald,
  and Platzman}}]{Girvin85}
\bibinfo{author}{\bibfnamefont{S.~M.} \bibnamefont{Girvin}},
  \bibinfo{author}{\bibfnamefont{A.~H.} \bibnamefont{MacDonald}},
  \bibnamefont{and} \bibinfo{author}{\bibfnamefont{P.~M.}
  \bibnamefont{Platzman}}, \bibinfo{journal}{Phys. Rev. Lett.}
  \textbf{\bibinfo{volume}{54}}, \bibinfo{pages}{581} (\bibinfo{year}{1985}),
  \urlprefix\url{http://link.aps.org/doi/10.1103/PhysRevLett.54.581}.

\bibitem[{\citenamefont{Girvin et~al.}(1986)\citenamefont{Girvin, MacDonald,
  and Platzman}}]{Girvin86}
\bibinfo{author}{\bibfnamefont{S.~M.} \bibnamefont{Girvin}},
  \bibinfo{author}{\bibfnamefont{A.~H.} \bibnamefont{MacDonald}},
  \bibnamefont{and} \bibinfo{author}{\bibfnamefont{P.~M.}
  \bibnamefont{Platzman}}, \bibinfo{journal}{Phys. Rev. B}
  \textbf{\bibinfo{volume}{33}}, \bibinfo{pages}{2481} (\bibinfo{year}{1986}),
  \urlprefix\url{http://link.aps.org/doi/10.1103/PhysRevB.33.2481}.

\bibitem[{\citenamefont{Chang}(2003)}]{Chang03}
\bibinfo{author}{\bibfnamefont{A.~M.} \bibnamefont{Chang}},
  \bibinfo{journal}{Rev. Mod. Phys.} \textbf{\bibinfo{volume}{75}},
  \bibinfo{pages}{1449} (\bibinfo{year}{2003}),
  \urlprefix\url{http://link.aps.org/doi/10.1103/RevModPhys.75.1449}.

\bibitem[{\citenamefont{Pan et~al.}(1999)\citenamefont{Pan, Xia, Shvarts,
  Adams, Stormer, Tsui, Pfeiffer, Baldwin, and West}}]{Pan99}
\bibinfo{author}{\bibfnamefont{W.}~\bibnamefont{Pan}},
  \bibinfo{author}{\bibfnamefont{J.-S.} \bibnamefont{Xia}},
  \bibinfo{author}{\bibfnamefont{V.}~\bibnamefont{Shvarts}},
  \bibinfo{author}{\bibfnamefont{D.~E.} \bibnamefont{Adams}},
  \bibinfo{author}{\bibfnamefont{H.~L.} \bibnamefont{Stormer}},
  \bibinfo{author}{\bibfnamefont{D.~C.} \bibnamefont{Tsui}},
  \bibinfo{author}{\bibfnamefont{L.~N.} \bibnamefont{Pfeiffer}},
  \bibinfo{author}{\bibfnamefont{K.~W.} \bibnamefont{Baldwin}},
  \bibnamefont{and} \bibinfo{author}{\bibfnamefont{K.~W.} \bibnamefont{West}},
  \bibinfo{journal}{Phys. Rev. Lett.} \textbf{\bibinfo{volume}{83}},
  \bibinfo{pages}{3530} (\bibinfo{year}{1999}),
  \urlprefix\url{http://link.aps.org/doi/10.1103/PhysRevLett.83.3530}.

\bibitem[{\citenamefont{Melik-Alaverdian
  et~al.}(1997)\citenamefont{Melik-Alaverdian, Bonesteel, and
  Ortiz}}]{Melik-Alaverdian97}
\bibinfo{author}{\bibfnamefont{V.}~\bibnamefont{Melik-Alaverdian}},
  \bibinfo{author}{\bibfnamefont{N.~E.} \bibnamefont{Bonesteel}},
  \bibnamefont{and} \bibinfo{author}{\bibfnamefont{G.}~\bibnamefont{Ortiz}},
  \bibinfo{journal}{Phys. Rev. Lett.} \textbf{\bibinfo{volume}{79}},
  \bibinfo{pages}{5286} (\bibinfo{year}{1997}),
  \urlprefix\url{http://link.aps.org/doi/10.1103/PhysRevLett.79.5286}.

\bibitem[{\citenamefont{Eisenstein et~al.}(2002)\citenamefont{Eisenstein,
  Cooper, Pfeiffer, and West}}]{Eisenstein02}
\bibinfo{author}{\bibfnamefont{J.~P.} \bibnamefont{Eisenstein}},
  \bibinfo{author}{\bibfnamefont{K.~B.} \bibnamefont{Cooper}},
  \bibinfo{author}{\bibfnamefont{L.~N.} \bibnamefont{Pfeiffer}},
  \bibnamefont{and} \bibinfo{author}{\bibfnamefont{K.~W.} \bibnamefont{West}},
  \bibinfo{journal}{Phys. Rev. Lett.} \textbf{\bibinfo{volume}{88}},
  \bibinfo{pages}{076801} (\bibinfo{year}{2002}),
  \urlprefix\url{http://link.aps.org/doi/10.1103/PhysRevLett.88.076801}.

\bibitem[{\citenamefont{Kumar et~al.}(2010)\citenamefont{Kumar, Cs\'athy,
  Manfra, Pfeiffer, and West}}]{Kumar10}
\bibinfo{author}{\bibfnamefont{A.}~\bibnamefont{Kumar}},
  \bibinfo{author}{\bibfnamefont{G.~A.} \bibnamefont{Cs\'athy}},
  \bibinfo{author}{\bibfnamefont{M.~J.} \bibnamefont{Manfra}},
  \bibinfo{author}{\bibfnamefont{L.~N.} \bibnamefont{Pfeiffer}},
  \bibnamefont{and} \bibinfo{author}{\bibfnamefont{K.~W.} \bibnamefont{West}},
  \bibinfo{journal}{Phys. Rev. Lett.} \textbf{\bibinfo{volume}{105}},
  \bibinfo{pages}{246808} (\bibinfo{year}{2010}),
  \urlprefix\url{http://link.aps.org/doi/10.1103/PhysRevLett.105.246808}.

\bibitem[{\citenamefont{Zhang et~al.}(2012)\citenamefont{Zhang, Huan, Xia,
  Sullivan, Pan, Baldwin, West, Pfeiffer, and Tsui}}]{Zhang12}
\bibinfo{author}{\bibfnamefont{C.}~\bibnamefont{Zhang}},
  \bibinfo{author}{\bibfnamefont{C.}~\bibnamefont{Huan}},
  \bibinfo{author}{\bibfnamefont{J.~S.} \bibnamefont{Xia}},
  \bibinfo{author}{\bibfnamefont{N.~S.} \bibnamefont{Sullivan}},
  \bibinfo{author}{\bibfnamefont{W.}~\bibnamefont{Pan}},
  \bibinfo{author}{\bibfnamefont{K.~W.} \bibnamefont{Baldwin}},
  \bibinfo{author}{\bibfnamefont{K.~W.} \bibnamefont{West}},
  \bibinfo{author}{\bibfnamefont{L.~N.} \bibnamefont{Pfeiffer}},
  \bibnamefont{and} \bibinfo{author}{\bibfnamefont{D.~C.} \bibnamefont{Tsui}},
  \bibinfo{journal}{Phys. Rev. B} \textbf{\bibinfo{volume}{85}},
  \bibinfo{pages}{241302} (\bibinfo{year}{2012}),
  \urlprefix\url{http://link.aps.org/doi/10.1103/PhysRevB.85.241302}.

\bibitem[{\citenamefont{d'Ambrumenil and Reynolds}(1988)}]{Ambrumenil88}
\bibinfo{author}{\bibfnamefont{N.}~\bibnamefont{d'Ambrumenil}}
  \bibnamefont{and} \bibinfo{author}{\bibfnamefont{A.~M.}
  \bibnamefont{Reynolds}}, \bibinfo{journal}{Journal of Physics C: Solid State
  Physics} \textbf{\bibinfo{volume}{21}}, \bibinfo{pages}{119}
  (\bibinfo{year}{1988}),
  \urlprefix\url{http://stacks.iop.org/0022-3719/21/i=1/a=010}.

\bibitem[{\citenamefont{Balram et~al.}(2013)\citenamefont{Balram, Wu, Sreejith,
  W\'ojs, and Jain}}]{Balram13b}
\bibinfo{author}{\bibfnamefont{A.~C.} \bibnamefont{Balram}},
  \bibinfo{author}{\bibfnamefont{Y.-H.} \bibnamefont{Wu}},
  \bibinfo{author}{\bibfnamefont{G.~J.} \bibnamefont{Sreejith}},
  \bibinfo{author}{\bibfnamefont{A.}~\bibnamefont{W\'ojs}}, \bibnamefont{and}
  \bibinfo{author}{\bibfnamefont{J.~K.} \bibnamefont{Jain}},
  \bibinfo{journal}{Phys. Rev. Lett.} \textbf{\bibinfo{volume}{110}},
  \bibinfo{pages}{186801} (\bibinfo{year}{2013}),
  \urlprefix\url{http://link.aps.org/doi/10.1103/PhysRevLett.110.186801}.

\bibitem[{\citenamefont{{Jeong} et~al.}(2016)\citenamefont{{Jeong}, {Lu},
  {Hashimoto}, {Chung}, and {Park}}}]{Jeong16}
\bibinfo{author}{\bibfnamefont{J.-S.} \bibnamefont{{Jeong}}},
  \bibinfo{author}{\bibfnamefont{H.}~\bibnamefont{{Lu}}},
  \bibinfo{author}{\bibfnamefont{K.}~\bibnamefont{{Hashimoto}}},
  \bibinfo{author}{\bibfnamefont{S.~B.} \bibnamefont{{Chung}}},
  \bibnamefont{and} \bibinfo{author}{\bibfnamefont{K.}~\bibnamefont{{Park}}},
  \bibinfo{journal}{ArXiv e-prints}  (\bibinfo{year}{2016}),
  \eprint{1601.00403}.

\bibitem[{\citenamefont{Bergholtz et~al.}(2007)\citenamefont{Bergholtz,
  Hansson, Hermanns, and Karlhede}}]{Bergholtz07}
\bibinfo{author}{\bibfnamefont{E.~J.} \bibnamefont{Bergholtz}},
  \bibinfo{author}{\bibfnamefont{T.~H.} \bibnamefont{Hansson}},
  \bibinfo{author}{\bibfnamefont{M.}~\bibnamefont{Hermanns}}, \bibnamefont{and}
  \bibinfo{author}{\bibfnamefont{A.}~\bibnamefont{Karlhede}},
  \bibinfo{journal}{Phys. Rev. Lett.} \textbf{\bibinfo{volume}{99}},
  \bibinfo{pages}{256803} (\bibinfo{year}{2007}),
  \urlprefix\url{http://link.aps.org/doi/10.1103/PhysRevLett.99.256803}.

\bibitem[{\citenamefont{Bergholtz et~al.}(2008)\citenamefont{Bergholtz,
  Hansson, Hermanns, Karlhede, and Viefers}}]{Bergholtz08}
\bibinfo{author}{\bibfnamefont{E.~J.} \bibnamefont{Bergholtz}},
  \bibinfo{author}{\bibfnamefont{T.~H.} \bibnamefont{Hansson}},
  \bibinfo{author}{\bibfnamefont{M.}~\bibnamefont{Hermanns}},
  \bibinfo{author}{\bibfnamefont{A.}~\bibnamefont{Karlhede}}, \bibnamefont{and}
  \bibinfo{author}{\bibfnamefont{S.}~\bibnamefont{Viefers}},
  \bibinfo{journal}{Phys. Rev. B} \textbf{\bibinfo{volume}{77}},
  \bibinfo{pages}{165325} (\bibinfo{year}{2008}),
  \urlprefix\url{http://link.aps.org/doi/10.1103/PhysRevB.77.165325}.

\bibitem[{\citenamefont{Hansson et~al.}(2009)\citenamefont{Hansson, Hermanns,
  and Viefers}}]{Hansson09}
\bibinfo{author}{\bibfnamefont{T.~H.} \bibnamefont{Hansson}},
  \bibinfo{author}{\bibfnamefont{M.}~\bibnamefont{Hermanns}}, \bibnamefont{and}
  \bibinfo{author}{\bibfnamefont{S.}~\bibnamefont{Viefers}},
  \bibinfo{journal}{Phys. Rev. B} \textbf{\bibinfo{volume}{80}},
  \bibinfo{pages}{165330} (\bibinfo{year}{2009}),
  \urlprefix\url{http://link.aps.org/doi/10.1103/PhysRevB.80.165330}.

\bibitem[{\citenamefont{Goldman and Shayegan}(1990)}]{Goldman90a}
\bibinfo{author}{\bibfnamefont{V.}~\bibnamefont{Goldman}} \bibnamefont{and}
  \bibinfo{author}{\bibfnamefont{M.}~\bibnamefont{Shayegan}},
  \bibinfo{journal}{Surface Science} \textbf{\bibinfo{volume}{229}},
  \bibinfo{pages}{10 } (\bibinfo{year}{1990}), ISSN \bibinfo{issn}{0039-6028},
  \urlprefix\url{http://www.sciencedirect.com/science/article/pii/003960289090819T}.

\bibitem[{\citenamefont{Bonderson et~al.}(2012)\citenamefont{Bonderson,
  Feiguin, M\"oller, and Slingerland}}]{Bonderson12}
\bibinfo{author}{\bibfnamefont{P.}~\bibnamefont{Bonderson}},
  \bibinfo{author}{\bibfnamefont{A.~E.} \bibnamefont{Feiguin}},
  \bibinfo{author}{\bibfnamefont{G.}~\bibnamefont{M\"oller}}, \bibnamefont{and}
  \bibinfo{author}{\bibfnamefont{J.~K.} \bibnamefont{Slingerland}},
  \bibinfo{journal}{Phys. Rev. Lett.} \textbf{\bibinfo{volume}{108}},
  \bibinfo{pages}{036806} (\bibinfo{year}{2012}),
  \urlprefix\url{http://link.aps.org/doi/10.1103/PhysRevLett.108.036806}.

\bibitem[{\citenamefont{Sreejith et~al.}(2013)\citenamefont{Sreejith, Wu,
  W\'ojs, and Jain}}]{Sreejith13}
\bibinfo{author}{\bibfnamefont{G.~J.} \bibnamefont{Sreejith}},
  \bibinfo{author}{\bibfnamefont{Y.-H.} \bibnamefont{Wu}},
  \bibinfo{author}{\bibfnamefont{A.}~\bibnamefont{W\'ojs}}, \bibnamefont{and}
  \bibinfo{author}{\bibfnamefont{J.~K.} \bibnamefont{Jain}},
  \bibinfo{journal}{Phys. Rev. B} \textbf{\bibinfo{volume}{87}},
  \bibinfo{pages}{245125} (\bibinfo{year}{2013}),
  \urlprefix\url{http://link.aps.org/doi/10.1103/PhysRevB.87.245125}.

\bibitem[{\citenamefont{Xia et~al.}(2004)\citenamefont{Xia, Pan, Vicente,
  Adams, Sullivan, Stormer, Tsui, Pfeiffer, Baldwin, and West}}]{Xia04}
\bibinfo{author}{\bibfnamefont{J.~S.} \bibnamefont{Xia}},
  \bibinfo{author}{\bibfnamefont{W.}~\bibnamefont{Pan}},
  \bibinfo{author}{\bibfnamefont{C.~L.} \bibnamefont{Vicente}},
  \bibinfo{author}{\bibfnamefont{E.~D.} \bibnamefont{Adams}},
  \bibinfo{author}{\bibfnamefont{N.~S.} \bibnamefont{Sullivan}},
  \bibinfo{author}{\bibfnamefont{H.~L.} \bibnamefont{Stormer}},
  \bibinfo{author}{\bibfnamefont{D.~C.} \bibnamefont{Tsui}},
  \bibinfo{author}{\bibfnamefont{L.~N.} \bibnamefont{Pfeiffer}},
  \bibinfo{author}{\bibfnamefont{K.~W.} \bibnamefont{Baldwin}},
  \bibnamefont{and} \bibinfo{author}{\bibfnamefont{K.~W.} \bibnamefont{West}},
  \bibinfo{journal}{Phys. Rev. Lett.} \textbf{\bibinfo{volume}{93}},
  \bibinfo{pages}{176809} (\bibinfo{year}{2004}),
  \urlprefix\url{http://link.aps.org/doi/10.1103/PhysRevLett.93.176809}.

\bibitem[{\citenamefont{Pan et~al.}(2008)\citenamefont{Pan, Xia, Stormer, Tsui,
  Vicente, Adams, Sullivan, Pfeiffer, Baldwin, and West}}]{Pan08}
\bibinfo{author}{\bibfnamefont{W.}~\bibnamefont{Pan}},
  \bibinfo{author}{\bibfnamefont{J.~S.} \bibnamefont{Xia}},
  \bibinfo{author}{\bibfnamefont{H.~L.} \bibnamefont{Stormer}},
  \bibinfo{author}{\bibfnamefont{D.~C.} \bibnamefont{Tsui}},
  \bibinfo{author}{\bibfnamefont{C.}~\bibnamefont{Vicente}},
  \bibinfo{author}{\bibfnamefont{E.~D.} \bibnamefont{Adams}},
  \bibinfo{author}{\bibfnamefont{N.~S.} \bibnamefont{Sullivan}},
  \bibinfo{author}{\bibfnamefont{L.~N.} \bibnamefont{Pfeiffer}},
  \bibinfo{author}{\bibfnamefont{K.~W.} \bibnamefont{Baldwin}},
  \bibnamefont{and} \bibinfo{author}{\bibfnamefont{K.~W.} \bibnamefont{West}},
  \bibinfo{journal}{Phys. Rev. B} \textbf{\bibinfo{volume}{77}},
  \bibinfo{pages}{075307} (\bibinfo{year}{2008}),
  \urlprefix\url{http://link.aps.org/doi/10.1103/PhysRevB.77.075307}.

\bibitem[{\citenamefont{Bonderson and Slingerland}(2008)}]{Bonderson08}
\bibinfo{author}{\bibfnamefont{P.}~\bibnamefont{Bonderson}} \bibnamefont{and}
  \bibinfo{author}{\bibfnamefont{J.~K.} \bibnamefont{Slingerland}},
  \bibinfo{journal}{Phys. Rev. B} \textbf{\bibinfo{volume}{78}},
  \bibinfo{pages}{125323} (\bibinfo{year}{2008}),
  \urlprefix\url{http://link.aps.org/doi/10.1103/PhysRevB.78.125323}.

\bibitem[{\citenamefont{Pakrouski et~al.}(2016)\citenamefont{Pakrouski, Troyer,
  Wu, Das~Sarma, and Peterson}}]{Pakrouski16}
\bibinfo{author}{\bibfnamefont{K.}~\bibnamefont{Pakrouski}},
  \bibinfo{author}{\bibfnamefont{M.}~\bibnamefont{Troyer}},
  \bibinfo{author}{\bibfnamefont{Y.-L.} \bibnamefont{Wu}},
  \bibinfo{author}{\bibfnamefont{S.}~\bibnamefont{Das~Sarma}},
  \bibnamefont{and} \bibinfo{author}{\bibfnamefont{M.~R.}
  \bibnamefont{Peterson}}, \bibinfo{journal}{Phys. Rev. B}
  \textbf{\bibinfo{volume}{94}}, \bibinfo{pages}{075108}
  (\bibinfo{year}{2016}),
  \urlprefix\url{http://link.aps.org/doi/10.1103/PhysRevB.94.075108}.

\bibitem[{\citenamefont{{Mong} et~al.}(2015)\citenamefont{{Mong}, {Zaletel},
  {Pollmann}, and {Papi{\'c}}}}]{Mong15}
\bibinfo{author}{\bibfnamefont{R.~S.~K.} \bibnamefont{{Mong}}},
  \bibinfo{author}{\bibfnamefont{M.~P.} \bibnamefont{{Zaletel}}},
  \bibinfo{author}{\bibfnamefont{F.}~\bibnamefont{{Pollmann}}},
  \bibnamefont{and}
  \bibinfo{author}{\bibfnamefont{Z.}~\bibnamefont{{Papi{\'c}}}},
  \bibinfo{journal}{ArXiv e-prints}  (\bibinfo{year}{2015}),
  \eprint{1505.02843}.

\bibitem[{\citenamefont{Zhu et~al.}(2015)\citenamefont{Zhu, Gong, Haldane, and
  Sheng}}]{Zhu15}
\bibinfo{author}{\bibfnamefont{W.}~\bibnamefont{Zhu}},
  \bibinfo{author}{\bibfnamefont{S.~S.} \bibnamefont{Gong}},
  \bibinfo{author}{\bibfnamefont{F.~D.~M.} \bibnamefont{Haldane}},
  \bibnamefont{and} \bibinfo{author}{\bibfnamefont{D.~N.} \bibnamefont{Sheng}},
  \bibinfo{journal}{Phys. Rev. Lett.} \textbf{\bibinfo{volume}{115}},
  \bibinfo{pages}{126805} (\bibinfo{year}{2015}),
  \urlprefix\url{http://link.aps.org/doi/10.1103/PhysRevLett.115.126805}.

\bibitem[{\citenamefont{{Hutasoit} et~al.}(2016)\citenamefont{{Hutasoit},
  {Balram}, {Mukherjee}, {Mandal}, {Wojs}, {Cheianov}, and
  {Jain}}}]{Hutasoit16}
\bibinfo{author}{\bibfnamefont{J.~A.} \bibnamefont{{Hutasoit}}},
  \bibinfo{author}{\bibfnamefont{A.~C.} \bibnamefont{{Balram}}},
  \bibinfo{author}{\bibfnamefont{S.}~\bibnamefont{{Mukherjee}}},
  \bibinfo{author}{\bibfnamefont{S.~S.} \bibnamefont{{Mandal}}},
  \bibinfo{author}{\bibfnamefont{A.}~\bibnamefont{{Wojs}}},
  \bibinfo{author}{\bibfnamefont{V.}~\bibnamefont{{Cheianov}}},
  \bibnamefont{and} \bibinfo{author}{\bibfnamefont{J.~K.}
  \bibnamefont{{Jain}}}, \bibinfo{journal}{ArXiv e-prints}
  (\bibinfo{year}{2016}), \eprint{1605.07324}.

\bibitem[{\citenamefont{{Regnault} et~al.}(2016)\citenamefont{{Regnault},
  {Maciejko}, {Kivelson}, and {Sondhi}}}]{Regnault16}
\bibinfo{author}{\bibfnamefont{N.}~\bibnamefont{{Regnault}}},
  \bibinfo{author}{\bibfnamefont{J.}~\bibnamefont{{Maciejko}}},
  \bibinfo{author}{\bibfnamefont{S.~A.} \bibnamefont{{Kivelson}}},
  \bibnamefont{and} \bibinfo{author}{\bibfnamefont{S.~L.}
  \bibnamefont{{Sondhi}}}, \bibinfo{journal}{ArXiv e-prints}
  (\bibinfo{year}{2016}), \eprint{1607.02178}.

\end{thebibliography}
\bibliographystyle{apsrev}
\end{document}